\documentclass[aps, prb, superscriptaddress, amsmath, amssymb, twocolumn, longbibliography]{revtex4-2}
\usepackage{graphicx}
\usepackage{subfigure}
\usepackage{adjustbox}
\usepackage{bm}
\usepackage{ulem}
\usepackage{color}
\usepackage{braket}
\usepackage{standalone}
\usepackage{multirow}
\usepackage{tikz}
\usepackage{mathrsfs}
\usepackage{dsfont}
\usepackage{comment}
\usepackage{amsmath, amsthm}
\usepackage{amssymb}
\usepackage{amsfonts}
\usepackage{tabularx}   % For flexible column widths
\usepackage{booktabs}   % For professional horizontal lines
\usepackage{makecell}   % For multi-line cells (optional but nice)
\usepackage{array}      % For better column control% ragged-right, auto-wrap
\usepackage[a]{esvect}
\usepackage[colorlinks,bookmarks=true,citecolor=blue,linkcolor=red,urlcolor=blue]{hyperref}
\usepackage{cleveref}
\usepackage{orcidlink}

\makeatletter
\newsavebox{\@brx}
\newcommand{\llangle}[1][]{\savebox{\@brx}{\(\m@th{#1\langle}\)}%
  \mathopen{\copy\@brx\kern-0.5\wd\@brx\usebox{\@brx}}}
\newcommand{\rrangle}[1][]{\savebox{\@brx}{\(\m@th{#1\rangle}\)}%
  \mathclose{\copy\@brx\kern-0.5\wd\@brx\usebox{\@brx}}}
\makeatother

\newcommand{\lgen}{\llangle}
\newcommand{\rgen}{\rrangle}

\graphicspath{{./figures/}}
\begin{document}

\title{Additional quantum many-body scars of the spin-$1$ $XY$ model\\ with Fock-space cages and commutant algebras}

\author{Sashikanta Mohapatra\orcidlink{0009-0001-0837-9604}}
\email{sashikanta@imsc.res.in}
\affiliation{Institute of Mathematical Sciences, CIT Campus, Chennai 600113, India}
\affiliation{Homi Bhabha National Institute, Training School Complex, Anushaktinagar, Mumbai 400094, India} 

\author{Sanjay Moudgalya}
\email{sanjay.moudgalya@gmail.com}
\affiliation{Department of Physics, Technische Universit\"{a}t M\"{u}nchen (TUM), James-Franck-Str. 1, 85748 Garching, Germany}
\affiliation{Munich Center for Quantum Science and Technology (MCQST), Schellingstr. 4, 80799 M\"{u}nchen, Germany}

\author{Ajit C. Balram\orcidlink{0000-0002-8087-6015}}
\email{cb.ajit@gmail.com}
\affiliation{Institute of Mathematical Sciences, CIT Campus, Chennai 600113, India}
\affiliation{Homi Bhabha National Institute, Training School Complex, Anushaktinagar, Mumbai 400094, India} 

\date{\today}

\begin{abstract}
Quantum many-body scars (QMBS) represent a mechanism for weak ergodicity breaking, characterized by the coexistence of atypical non-thermal eigenstates within an otherwise thermalizing many-body spectrum. In this work, we revisit the spin-$1$ $XY$ model on a periodic chain and construct several new families of exact scar eigenstates embedded within its extensively degenerate manifolds that owe their origins to an interplay of $U(1)$ magnetization conservation and chiral symmetries. We go beyond previously studied towers of states and first identify a novel set of interference-protected eigenstates resembling Fock space cage states, where destructive interference confines the wave function to sparse subgraphs of the Fock space. These states exhibit subextensive entanglement entropy, and when subjected to a transverse magnetic field, form equally spaced states whose coherent superpositions display long-lived fidelity oscillations. We further reveal a simpler organizing principle behind these nonthermal states by utilizing the commutant algebra framework, specifically by demonstrating that they are simultaneous eigenstates of non-commuting local operators. Moreover, in doing so, we uncover two more novel families of exact scars: a tower of volume-entangled states, and a set of mirror-dimer states with some free local degrees of freedom. Our results illustrate the power and interplay of interference-based and algebraic mechanisms of non-ergodicity, offering systematic routes to identifying and classifying QMBS in generic many-body quantum systems.
\end{abstract}

\maketitle

\section{Introduction}
The question of how isolated quantum many-body systems reach thermal equilibrium~\cite{deutsch1991quantum, rigol2008thermalization} lies at the heart of quantum statistical mechanics.
These systems undergo unitary evolution, and understanding how thermal behavior emerges from such reversible microscopic dynamics has long been a central pursuit.
For generic interacting systems, the prevailing theoretical framework is the Eigenstate Thermalization Hypothesis (ETH)~\cite{srednicki1994chaos,kim2014testing,deutsch2018eigenstate, Luca2016ETH}, which posits that the expectation values of local observables in individual energy eigenstates at long times coincide with thermal ensemble predictions at their corresponding energies.
As a result, all eigenstates in the bulk of the spectrum exhibit thermal behavior, and the memory of initial conditions is lost at late times.
However, ETH, and more generally the notion of thermalization, fails in certain well-understood settings.
Notable examples include integrable systems~\cite{kinoshita2006quantum,calabrese2011quantum,cassidy2011generalized,vidmar2016generalized} and many-body localized phases~\cite{BASKO2006MBL,nandkishore2015many, Pal2010MBL, Schreiber2015manybody}, where the existence of an extensive set of local conserved quantities precludes thermalization, resulting in the dynamics retaining long-term memory of initial states even at finite energy densities. 
Beyond these strong-ETH-violating paradigms, recent years have seen the emergence of a novel class of systems that break thermalization in a much weaker sense.
These systems, which are otherwise chaotic and nonintegrable, host a small subset of non-thermal eigenstates embedded within a thermal spectrum, a phenomenon now known as quantum many-body scarring~\cite{Turner2018quantumscar, Turner2018weak, moudgalya2021review, serbyn2021quantum, chandran2022review}.
Quantum many-body scars (QMBS) states often occur in quasiparticle towers of states~\cite{moudgalya2021review, chandran2022review}, which exhibit anomalously low entanglement entropy (EE) compared to typical eigenstates at similar energies. However, there are many other kinds of scars, including volume-law entangled ones~\cite{langlett2022rainbow, Ivanov2025Volume, Chiba2024Exact, Mohapatra2025Exact, Mukherjee2025SymmetricTensorScars, mestyan2025crosscapstatestunableentanglement} which are nevertheless atypical in different ways.
The existence of these states in the spectrum can lead to non-thermal dynamics in the expectation values of local observables when the system is quenched from an initial state that has a high overlap with the scar states.
QMBS have been experimentally observed in Rydberg atom arrays that approximately simulate the PXP model~\cite{bernien2017probing, Jad2023Observation}. Following this, both exact~\cite{Moudgalya2018AKLT, mark2020unified, moudgalya2020large, schecter2019weak, chattopadhyay2020quantum, moudgalya2020eta, mark2020eta, lee2020exact, banerjee2021quantum, biswas2022scars, wildeboer2022quantum, Mizuta2020Exact, Tang2022Multimagnon, Surace2021Exact, Schindler2022Exact, Gotta2022Exact, Langlett2021Hilbert, Iversen2024Tower, Miao2025Exact2DGaugeScars, bhowmick2025asymmetric, bhowmick2025GZscar} and approximate QMBS~\cite{Turner2018weak, Ho2019Periodic, moudgalya2020quantum, mohapatra2023pronounced, Wang2024QMBS, Bull2019Systematic, You2022QMBSSpin1Kitaev, Desaules2023Weak, Ren2025ScarFinder, Kerschbaumer2025QMBS, Turner2018quantumscar} have been identified in a wide range of models. 

Among the burgeoning landscape of exact scar-hosting systems, those featuring extensive zero-energy modes~\cite{Turner2018quantumscar, Schecter2018ManyBody, Lin2020Quantum, Karle2021Area, udupa2023weak, Buijsman2022Number, Pietro2023Hilbert, Brighi2024Anomalous} have emerged as particularly intriguing.
In such systems, symmetries-most commonly chiral and/or spectral-reflection symmetries-enforce a macroscopically large manifold of many-body eigenstates to be pinned at exactly zero energy~\cite{Schecter2018ManyBody}.
While these zero-energy subspaces, typically located at the center of the spectrum, might be expected to host thermal states, several works~\cite{lin2019exact, Lin2020Quantum, lee2020exact, Karle2021Area, udupa2023weak, biswas2022scars, Sau2024Sublattice, Mukherjee2025SymmetricTensorScars, Ivanov2025scar, Ivanov2025Volume} have demonstrated that they can also harbor non-thermal, scar-like states that arise as specific linear combinations of the zero-modes.
These scattered examples of scars within models possessing extensive zero-modes hint at a rich underlying structure. Yet, these are hard to construct precisely because of the extensive degeneracy of the manifold of states in which they lie, which leads conventional methods of scar identification, such as anomalous entanglement entropy, to fail. 
We note that Ref.~\cite{Karle2021Area} proposed an algorithm to deal with this degeneracy and identify scars by finding the lowest entangled states in this manifold, which successfully identifies many interesting states such as the exact MPS states in the zero-energy manifold of the PXP model~\cite{lin2019exact, Ivanov2025scar}.
In fact, they also go on to conjecture that \textit{every} local Hamiltonian with such a zero-energy manifold has an area-law entangled state within that manifold, although it is not guaranteed to be an MPS with a finite bond dimension.
While such methods are certainly useful in finding exact and approximate area-law scars in this manifold, as mentioned above and as we shall show below, there exist scars in this manifold that have higher entanglement but yet fairly simple structures, such as quasiparticle towers or even simple volume-law entangled states, and it is not clear if such algorithms can be generalized to capture these states.
This gap highlights the need for systematic and analytic construction methods capable of uncovering more kinds of nonthermal states or subspaces.
Moreover, the structures of these manifolds are themselves largely unexplored in models with multiple symmetries such as $U(1)$~\cite{Nicolau2025Fragmentation} along with chiral or inversion symmetries, and there might be additional surprises hiding there.

In this work, we revisit the paradigmatic spin-$1$ $XY$ chain and focus on identifying scars within its extensively degenerate manifolds of states. We find that these degenerate manifolds are simply those that possess zero energy under the $XY$ exchange term of the Hamiltonian; hence, we will simply refer to them as zero-energy manifolds in the rest of this work. The exact scars revealed by prior studies of this model, which are two kinds of quasiparticle towers of states~\cite{schecter2019weak, chattopadhyay2020quantum}, also reside in this zero-energy manifold. Our approach here is complementary, encompasses the previous scars, and also reveals many new families of scars not previously reported. First, we show that this zero-energy manifold is protected by a combination of global $U(1)$ magnetization conservation and chiral symmetry, which enforces a bipartite structure on the many-body Fock-space graph.
This provides a rigorous lower bound on the number of zero modes in each magnetization sector, which we compute analytically, showing that it grows exponentially with system size.
This extensive degeneracy lies precisely at the center of the many-body spectrum and serves as the foundation for constructing several new families of exact scar eigenstates. 
Our first key finding is the identification of multiple novel families of zero-energy eigenstates that can be naturally interpreted as versions of the recently-studied Fock-space cage (FSC) states~\cite{Jonay2025FockSpaceCages, BenAmi2025ManyBodyCages, Tan2025InterferenceCaged}.
These eigenstates have support only on a sparse set of Fock basis configurations, where destructive interference precisely cancels all transition amplitudes to the sublattice that is complementary to where the states have support, thereby forming closed local loops (cage) in the Fock-space graph. 
This interference-protected confinement isolates them from the thermal continuum consisting of generic states, resulting in subextensive EE and establishing them as QMBS.
When a transverse magnetic field is applied, each family forms an equally spaced ladder of finite-energy states, and coherent superpositions within a ladder display long-lived fidelity oscillations, the dynamical hallmark of scarring. We also find that states studied in previous works~\cite{schecter2019weak, chattopadhyay2020quantum} also have such FSC interpretations, which we review when we discuss the states we find.
For the simplest members of these families, such as single quasiparticle states, the FSC mechanism offers a clear and intuitive picture—simple pairwise interference loops that can be directly visualized on the Fock-space graph.
However, as one moves to higher members, such as states with a higher number of quasiparticles, the interference structure becomes increasingly intricate, making a purely geometric understanding impractical.
This growing complexity suggests that these nonthermal manifolds are organized by a deeper algebraic structure not evident from the Fock-space interference picture alone.
To uncover such hidden structures, we turn to the commutant algebra framework~\cite{moudgalya2021hilbert, Moudgalya2023From_symmetries, moudgalya2022exhaustive}, which offers a systematic algebraic lens to identify and characterize QMBS.
In this framework, QMBS are identified as simultaneous eigenstates (often referred to as ``singlets"~\cite{pakrouski2020many}) of multiple non-commuting local operators, making them generalizations of stabilizer states~\cite{gottesman1997stabilizercodesquantumerror}, which are simultaneous eigenstates of products of Pauli operators.
This property inherently implies a violation of ETH as it leads to non-uniqueness in reconstructing local Hamiltonians that have such states as eigenstates~\cite{Garrison2018Does, Qi2019determininglocal,moudgalya2022exhaustive}.
We operationalize this idea by systematically reorganizing the spin-$1$ $XY$ Hamiltonian into several carefully chosen families of non-commuting operator sets, constructed through spatially clustered, antipodal, or mirror-symmetric groupings of local exchange bonds, such that their sum exactly reconstructs the original Hamiltonian.
Employing numerical methods~\cite{Moudgalya2023numerical}, we then identify the common eigenstates of each operator family, which serve as the algebraic singlets characterizing distinct nonthermal subspaces.
Through this construction, we not only uncover a simple algebraic understanding of the previously identified FSC states, but also discover two additional classes of exact eigenstates—a tower of volume-entangled states and a set of mirror-dimer states—each displaying distinctive entanglement and dynamical properties.
The remainder of this paper is organized as follows.
In Sec.~\ref{sec: model and extensive zero modes}, we introduce the spin-$1$ $XY$ model, detail its underlying symmetries, and quantify the degeneracy of its zero-energy manifold.
In Sec.~\ref{sec: Fock_space_cages}, we present several families of exact zero-energy eigenstates interpreted as FSCs, emphasizing their interference-based confinement mechanism, entanglement properties, and dynamical behavior.
In Sec.~\ref{sec: exact eigenstates from commutant algebra formalism}, we apply the commutant algebra framework to reveal the algebraic structure underlying the FSC states and to identify two additional families of exact eigenstates—the volume-entangled and mirror-dimer scars.
Finally, in Sec.~\ref{sec: conclusion and outlook}, we summarize our results and discuss their broader implications for understanding QMBS and non-ergodic dynamics. Technical derivations and supporting calculations are provided in the appendices. 

\section{The spin-$1$ $XY$ model and its extensive zero energy modes}
\label{sec: model and extensive zero modes}
We now introduce the version of the spin-$1$ $XY$ chain we study and highlight its two symmetries-a global $U(1)$ magnetization conservation and a chiral symmetry-that play a pivotal role in shaping its many-body spectrum.
Together, these symmetries induce a bipartite structure on the Fock-space graph, leading to the emergence of an extensive number of exact zero-energy eigenstates.
We analytically quantify this degeneracy and discuss its implications as a natural foundation for constructing novel families of exact QMBS.
\begin{figure}
    \centering
    \includegraphics[scale=0.38]{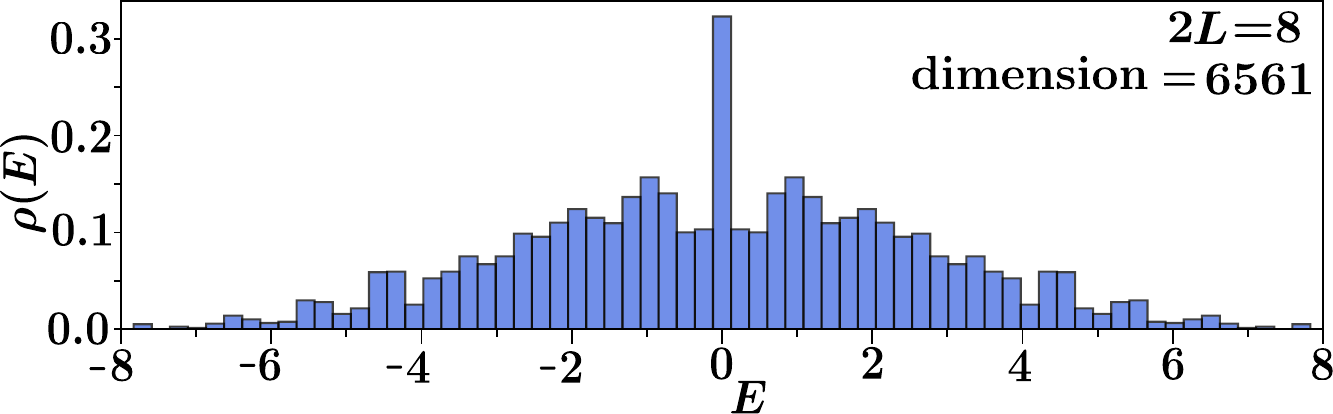}
    \caption{Many-body density of states $\rho(E)$ as a function of energy $E$, in units of the coupling strength $J$, for the $XY$ Hamiltonian $H_{XY}$ [see Eq.~\eqref{eq: full_Hamiltonian}] for $2L{=}8$ sites. The sharp central peak at $E{=}0$ reflects the extensive degeneracy of exact zero-energy eigenstates of $H_{XY}$.}
    \label{fig: DOS}
\end{figure}
\subsection{Model}
\label{sec: model}
We consider a one-dimensional spin-$1$ $XY$ chain comprising an even number of sites $N{=}2L$ with periodic boundary conditions (PBC) in a perpendicular external magnetic field, governed by the Hamiltonian $H_0$
\begin{gather}
    \label{eq: full_Hamiltonian}
    H_{0} = H_{XY} + H_{z}, \\ 
    H_{XY} = J\sum_{i=1}^{2L} ( S^x_i S^x_{i+1} + S^y_i S^y_{i+1}), \;\; H_{z} = h \sum_{i=1}^{2L} S^z_i. \nonumber
\end{gather}
Here, $S^{\alpha}_i ( \alpha {=} x, y, z )$ are spin-$1$ operators acting on local basis states $|m\rangle
_{i}{\in}\{|{-}1\rangle,|0\rangle,|1\rangle\}$ at site $i$ satisfying $S^{z}|m\rangle{=}m|m\rangle$. 
The Hamiltonian $H_0$ consists of two parts: a nearest-neighbor $XY$ exchange interaction $H_{XY}$, the strength of which is controlled by the coupling constant $J$, and a Zeeman term $H_{z}$ corresponding to a uniform magnetic field of strength $h$ along the $z$-axis. The Hamiltonian $H_0$ conserves the total magnetization $S_{\rm tot}^{z}{=}\sum_{i=1}^{2L} S^z_i$, leading to a global $U(1)$ symmetry that partitions the Hilbert space into magnetization sectors $M{\in}\{{-}2L, {-}2L{+}1, {\cdots}, 2L{-}1, 2L\}$.
Additionally, the model has reflection, and translational symmetries. 
Interestingly, the exchange term $H_{XY}$ admits an extensive number of exact zero-energy eigenstates as evident from its density of states shown in Fig.~\ref{fig: DOS}, which exhibits a sharp peak at $E{=}0$.
This high degeneracy forms the basis of our construction of analytically tractable exact eigenstates.
Our goal is to quantify this degeneracy and explicitly construct several distinct families of exact eigenstates within the zero-energy manifold of $H_{XY}$.
Throughout this work, when we use the phrase ``zero-energy eigenstates," we mean the eigenstates have zero energy with respect to $H_{XY}$.
When the Zeeman field $H_{z}$ is turned on, these states acquire finite energies under $H_{0}$ equal to $hM$ that are integer multiples of the field $h$ and, as we will show, display characteristic features of QMBS.
Even in the presence of the Zeeman field $H_{z}$, the degeneracy of states, now at finite energies, remains extensive. 
Before we proceed further, we note an important and subtle point regarding the integrability of $H_{0}$. Previous studies~\cite{Atsuhiro2003AnSU2, schecter2019weak, chattopadhyay2020quantum} established that $H_0$ exhibits Poissonian level-statistics in even magnetization sectors, indicating a symmetry, which has been identified to be a hidden non-local $SU(2)$ symmetry, which was additionally believed to lead to integrability of those sectors~\cite{chattopadhyay2020quantum}. However, once this non-local $SU(2)$ symmetry is resolved, the level statistics match those of the Gaussian Orthogonal Ensemble (GOE) of the random-matrix theory, demonstrating that the spin-$1$ $XY$ model is not integrable~\cite{odea2024levelstatisticsdetectgeneralized}. Since QMBS are typically defined only in non-integrable systems after resolving all symmetries, the eigenstates we construct are true QMBS of the spin-$1$ $XY$ model, even without the addition of any other perturbations.
Moreover, we show that our interpretation of QMBS as simultaneous eigenstates of non-commuting operators allows us to construct many kinds of perturbations $V_{\rm pert}$ that readily break the non-local symmetry while still preserving our constructed eigenstates as exact scar eigenstates of the perturbed Hamiltonian $H_{\rm scar}{=}H_{0}{+}V_{\rm pert}$, which is even more clearly non-integrable.
The particular forms of $V_{\rm pert}$ vary depending on the scar family and will be specified in the relevant sections. 
Readers might also notice that, unlike previous studies of the spin-$1$ $XY$ model in the QMBS literature~\cite{schecter2019weak}, we omit the single-ion anisotropy term $\sum_{i=1}^{2L}{(S_{i}^{z})^{2}}$ or the next-to-next-nearest-neighbor $XY$ exchange term $J_3\sum_{i=1}^{2L} ( S^x_i S^x_{i+3} {+} S^y_i S^y_{i+3})$ in the definition of the model.
This choice is deliberate, as such perturbations were included as the appropriate $V_{\rm pert}$ for stabilizing the set of previously-obtained QMBS while breaking the fine-tuned symmetries of $H_0$ (Note that the analogous next-nearest-neighbor $XY$ exchange term $J_2\sum_{i=1}^{2L} ( S^x_i S^x_{i+2} {+} S^y_i S^y_{i+2})$ does not preserve either the previously obtained scars~\cite{Halder2025Perturbations_QMBS_spin1XY} or the scars we unravel in this work.).
As we will discuss, the exact eigenstate families we construct respond differently to those perturbations, and hence are not always the appropriate $V_{\rm pert}$ for the QMBS we obtain. 
In particular, the tower of states of Sec.~\ref{subsec: Towers of equally spaced FSCs} remain exact eigenstates upon adding the anisotropy term, but are destroyed by the next-nearest-neighbor $XY$ exchange, whereas the volume-entangled and mirror-dimer families of Sec.~\ref{sec: exact eigenstates from commutant algebra formalism} exhibit the opposite behavior.
We defer the detailed discussion of these properties to the corresponding sections; here we simply note that the Hamiltonian in Eq.~\eqref{eq: full_Hamiltonian} provides a common baseline for all subsequent constructions.
\subsection{ Chiral symmetry and extensive zero modes}
\label{sec: Chiral symmetry and extensive zero modes}
In addition to the total magnetization, $H_{XY}$ also possesses a chiral symmetry that plays a central role in generating its extensive zero-energy degeneracy.
For the chains of even length $2L$ under consideration, this symmetry is generated by the unitary and Hermitian operator, whose action is non-trivial only on the even-numbered sites,
\begin{equation}
    \mathcal{\hat{C}}{=}\prod_{i=1}^{L}\mathcal{P}_{2i}^{z},~~~\text{with}~~~\mathcal{P}_{i}^{z}{=}{}{-}e^{\iota\pi S_{i}^{z}},
    \label{eq: chiral_symmetry_operator}
\end{equation}
where $\iota{=}\sqrt{{-}1}$, that satisfies the anticommutation relation $\{H_{XY},\mathcal{\hat{C}}\}{=}0$. This implies the energy spectrum of $H_{XY}$ is symmetric around zero, i.e., has particle-hole symmetry, as seen in Fig.~\ref{fig: DOS}. 

\begin{figure}
    \centering
    \includegraphics[scale=0.33]{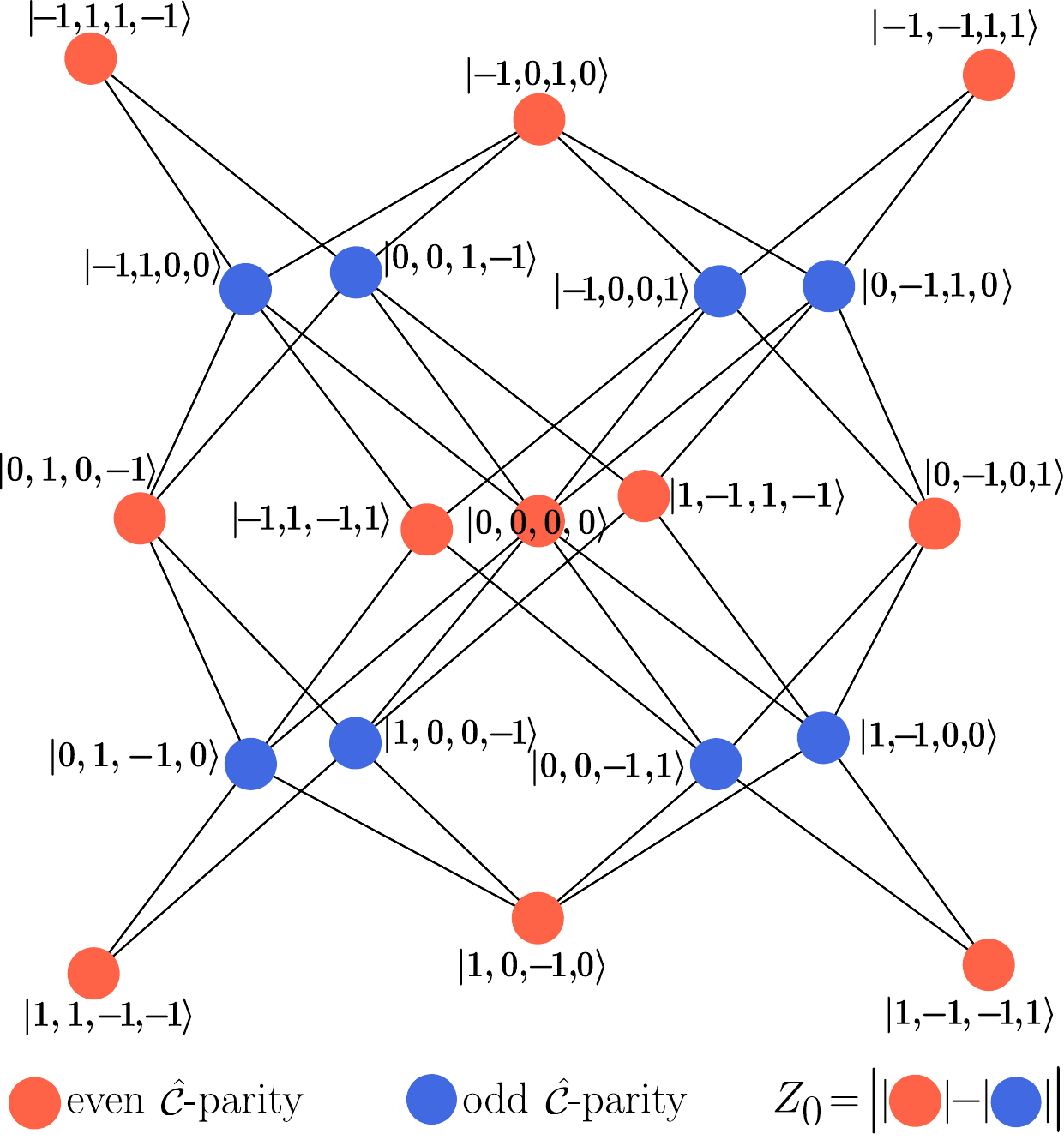}
    \caption{Fock-space graph of $H_{XY}$ [see Eq.~\eqref{eq: full_Hamiltonian}] for $2L{=}4$ sites. Basis states are colored by their $\hat{\mathcal{C}}$-parity [see Eq.~\eqref{eq: chiral_symmetry_operator} for definition of the operator $\hat{\mathcal{C}}$]: orange for sublattice $A$ (even) and blue for sublattice $B$ (odd). Edges indicate nonzero matrix elements of $H_{XY}$, connecting only opposite sublattices, revealing the bipartite nature of its Fock-space graph.}
    \label{fig: bipartite}
\end{figure}

The presence of both the global $U(1)$ and the chiral symmetries endows the Fock-space graph of $H_{XY}$ with a rich internal structure.
In this Fock-space graph of $H_{XY}$, the nodes represent the many-body basis states $\{|m_{1}, m_{2}, {\cdots}, m_{2L}\rangle\}$, while edges connect two nodes $|\alpha\rangle$ and $|\beta\rangle$ whenever $\langle\alpha|H_{XY}|\beta\rangle{\neq}0$.
It turns out that the $XY$ exchange term in Eq.~\eqref{eq: full_Hamiltonian} acts with identical strength on all nearest-neighbor bonds of the Fock-space graph, i.e., every nonzero matrix element of $H_{XY}$ is the same.
Hence, we can consider the edges of the Fock-space graph to be unweighted.
First, we can see that the Fock-space graph is bipartite.
To understand how the chiral symmetry shapes this structure, consider the action of the operator $\mathcal{P}_{i}^{z}$ on the local basis states: $\mathcal{P}_{i}^{z}|m\rangle_{i}{=}(-1)^{m{+}1}|m\rangle_{i}$, i.e., it flips the sign of the state $|0\rangle_{i}$ while leaving $|{\pm}1\rangle_{i}$ unchanged.
Therefore, the chiral operator $\mathcal{\hat{C}}$ assigns a $\mathbb{Z}_{2}$ value to each basis state based on the parity of the number of $|0\rangle$'s on its even-numbered sites.
Crucially, the action of $H_{XY}$ flips the chiral parity of any basis state, since for any computational basis state $|\Lambda\rangle$ satisfying $\mathcal{\hat{C}}|\Lambda\rangle{=}\lambda|\Lambda\rangle$ with $\lambda{=}{\pm}1$, the anticommutation of $\mathcal{\hat{C}}$ and $H_{XY}$ gives $\mathcal{\hat{C}}(H_{XY}|\Lambda\rangle){=}{-}\lambda(H_{XY}|\Lambda\rangle)$.
Thus, $H_{XY}$ connects configurations with opposite $\mathcal{\hat{C}}$-parity, implying that the Fock-space graph is bipartite: its nodes can be divided into two disjoint sublattices
\begin{itemize}
    \item sublattice $A$: basis states with $even$ number of $|0\rangle$ spins on even-numbered sites with $\mathcal{\hat{C}}$ eigenvalue ${+}1$,
    \item sublattice $B$: basis states with $odd$ number of $|0\rangle$ spins on even-numbered sites with $\mathcal{\hat{C}}$ eigenvalue ${-}1$,
\end{itemize}
and edges connect nodes belonging to different sublattices, i.e., the hopping generated by the action of $H_{XY}$ is only from a site in $A$ to a site in $B$, or vice-versa.
Next, we observe that since $\mathcal{\hat{C}}$ commutes with the total magnetization $M$, this bipartite structure of the Fock-space graph persists within each magnetization sector.
An illustrative example of this bipartite structure for a system with $2L{=}4$ sites in the $M{=}0$ sector is shown in Fig.~\ref{fig: bipartite}, where sites in sublattice $A$ ($B$) are colored orange (blue).
By ordering the basis states in a fixed magnetization sector $M$ according to their $\mathcal{\hat{C}}$-parity, $H_{XY}$ takes a block off-diagonal form:
\begin{equation}
    H_{XY}^{M}=P_{M}H_{XY}P_{M}{=}
    \begin{pmatrix}
        0 & B_{M}\\
        B_{M}^{\dagger} & 0
    \end{pmatrix},
    \label{eq: off-diagonal Hamiltonian}
\end{equation}
where $P_{M}$ is the projection operator to the total magnetization $M$ sector and $B_{M}$ is the biadjacency matrix (adjacency matrix of a bipartite graph) linking $A$ to $B$.

This characteristic off-diagonal structure of $H_{XY}^{M}$ immediately suggests the possibility of zero-energy modes in each magnetization sector.
Specifically, if $N_{A}^{M}$ and $N_{B}^{M}$ denote the number of basis states in sublattices $A$ and $B$, the dimension of the kernel of $H_{XY}^{M}$, i.e., the number of zero-energy states, must satisfy~\cite{Sutherland86, Turner2018quantumscar, Schecter2018ManyBody, moudgalya2020quantum, Jonay2025FockSpaceCages}:
\begin{equation}
    \text{dim~ker}~H_{XY}^{M}\geq Z_{M}=|N_{A}^{M}-N_{B}^{M}|.
    \label{eq: dim_kernel}
\end{equation}
This follows directly from the rank-nullity theorem, as the off-diagonal form of $H_{XY}^{M}$ restricts its rank to be at most $\text{min}(N_{A}^{M}, N_{B}^{M})$.
The number $Z_{M}$ is a rigorous lower bound on the degeneracy at zero energy.

Using a generating-function method (see Appendix~\ref{Appsec: counting zero energy modes} for details), we compute these quantities analytically.
Ultimately, the Hilbert-space dimension in a given magnetization sector $M$, denoted by $\mathcal{D}_{2L}^{M}$, is obtained as the coefficient of $x^M$ in the expansion of $(x{+}1{+}x^{-1})^{2L}$, leading to the trinomial coefficient
\begin{equation}
     \mathcal{D}_{2L}^{M}{=} \sum_{p=\max(0,M)}^{\lfloor (M{+}2L)/2 \rfloor}\binom{2L}{p}\binom{2L{-}p}{M{+}2L-2p}.
    \label{eq: U(1) sector dimension}
\end{equation}
Similarly, the lower bound on the number of zero modes is given by the coefficient of $x^{M}$ in $(x^{2}{+}1{+}x^{-2})^{L}$, which  is non-zero only for \textit{even} $M$, and given by $\mathcal{D}_{L}^{M/2}$, i.e.
\begin{equation}
    Z_{M}{=}\begin{cases}
\displaystyle
\sum_{p = \max(0,\, M/2)}^{\left\lfloor (L + M/2)/2 \right\rfloor}
\binom{L}{p} \binom{L - p}{ M/2 + L - 2p}, & \text{even}~M, \\[1.2em]
0, & \text{odd}~M.
\end{cases}
    \label{eq: Z_M_in_trinomial_expansion}
\end{equation}
Using a saddle-point approximation (see Appendix~\ref{Appsec: counting zero energy modes}) valid for any even-$M{\ll}\sqrt{L}$ in the large-$L$ limit, we find
\begin{equation}
    Z_{{\rm even}-M}{\sim}\frac{3^{L}}{\sqrt{(4/3)\pi L}} \exp\left[{-}\frac{3M^2}{16L} \right],
    \label{eq: Z_M_asymptotic_scaling}
\end{equation}
demonstrating that the bound on the number of exact zero-energy eigenstates grows exponentially with system size, with the largest contribution coming from the large $U(1)$ sectors near $M{=}0$.
This extensive degeneracy at zero energy of $H_{XY}$, which lies precisely at the middle of its spectrum, opens the possibility of constructing non-generic superpositions of the degenerate eigenstates with features reminiscent of QMBS~\cite{banerjee2021quantum}, which we explore below. 
Multiple recent works have shown that such zero-mode manifolds can host interesting eigenstates, with either area-law~\cite{lin2019exact, Ivanov2025scar} or volume-law entanglement~\cite{Chiba2024Exact, Ivanov2025Volume, Mohapatra2025Exact, Mukherjee2025SymmetricTensorScars, mestyan2025crosscapstatestunableentanglement}, demonstrating the possibility of interesting kinds of nonthermal phenomena in the manifold.
\subsection{Previously known scar eigenstates in the zero-energy manifold of $H_{XY}$}
\label{sec: Previously known scar tower}
Before proceeding to the construction of new eigenstates in this manifold, we briefly recall the known examples of exact scar eigenstates in the spin-$1$ $XY$ model.
These earlier constructions also reside in the zero-energy manifold of $H_{XY}$, and are also exclusively within its even magnetization sectors.
A prototypical example is the bimagnon tower introduced in Ref.~\cite{schecter2019weak}.
These low-entangled states, denoted by $|\mathbb{B}_{n}\rangle$, are constructed by repeated application of a bimagnon creation operator $J^{+}$ on the ferromagnetic vacuum state $|\Omega\rangle{=}\bigotimes_{i}|{-}1\rangle_{i}$, i.e.,
\begin{equation}
    |\mathbb{B}_{n}\rangle=(J^{+})^{n}|\Omega\rangle,~~~~~\text{where}~~~~~J^{+}{=}\sum_{i=1}^{2L}({-}1)^{i}(S_{i}^{+})^2.
    \label{eq: type_1 scar tower}
\end{equation}
The state $|\mathbb{B}_{n}\rangle$ lies in the even magnetization sector $M_{n}{=}2(n{-}L)$.
Each state $|\mathbb{B}_{n}\rangle$ is annihilated by $H_{XY}$, and therefore the family of states $\{|\mathbb{B}_{n}\rangle\}$ form an equi-spaced tower under the full Hamiltonian $H_{0}$.
Another family, known as bond-bimagnon states $|\mathbb{B}_{n}'\rangle$, was introduced in Ref.~\cite{schecter2019weak} and later examined in detail in Ref.~\cite{chattopadhyay2020quantum}.
These states, given by
\begin{equation} |\mathbb{B}_{n}'\rangle{=}\sum_{i_1{\neq}i_2{\cdots}{\neq}i_n}({-}1)^{\sum_{k{=}1}^{n}i_{k}}\prod_{k{=}1}^{n}S_{i_k}^{+}S_{i_k{+}1}^{+}|\Omega\rangle,
\label{eq: type_2 scar tower}
\end{equation}
also lie in the kernel of $H_{XY}$ and within even magnetization sectors $M_{n}'{=}2(n{-}L)$. 

These examples highlight the rich structure of the zero-energy manifold of $H_{XY}$ and motivate a broader search for hidden nonthermal eigenstates within it.
In the remainder of this work, we construct several previously unreported families of exact eigenstates, arising from mechanisms of destructive interference in Fock space and from algebraic organization through commutant structures, thereby substantially extending the known landscape of exact scars in the spin-$1$ $XY$ model.
\section{Novel family of Fock-space cage eigenstates}
\label{sec: Fock_space_cages}
In this section, we introduce a class of exact zero-energy eigenstates of $H_{XY}$ that are confined to compact, dynamically isolated subgraphs within its many-body Fock space.
These states, which we refer to as Fock-space cage (FSC)~\cite{Jonay2025FockSpaceCages, Nicolau2025Fragmentation, BenAmi2025ManyBodyCages, Tan2025InterferenceCaged} states, arise from local interference that prevents the spreading of wave function amplitude under the action of the $H_{XY}$, leading to perfectly Fock-space localized eigenstates (although not in real-space) in a disorder-free, translation-invariant system.
A crucial property that enables such interference-based localization is that $H_{XY}$ connects basis states with equal transition amplitudes, making its Fock-space graph effectively binary, with off-diagonal elements, when appropriately normalized, taking only values $0$ or $1$.
As a result, localization of states in Fock space can arise purely from sign cancellations rather than fine-tuned amplitudes, increasing their likelihood of occurrence.
While such interference-induced localization has recently been explored in
constrained settings~\cite{Jonay2025FockSpaceCages, Nicolau2025Fragmentation, BenAmi2025ManyBodyCages, Tan2025InterferenceCaged}, our finding reveals that a similar mechanism can emerge in a fully unconstrained model without any fine-tuned geometry.
In particular, previously reported exact scar states of the spin-$1$ $XY$ model given in Eqs.~\eqref{eq: type_1 scar tower} and \eqref{eq: type_2 scar tower} also lend themselves to an interpretation in terms of FSCs as illustrated in Appendix~\ref{Appsec: FSC_bimagnon}.
Moreover, as we demonstrate and focus on below, FSCs are effective in uncovering new scar families in the spin-$1$ $XY$ chain.
\subsection{Definition and destructive interference mechanism}
\label{ssec: construction_mechanism_FSCs}
\begin{figure}
    \centering
    \includegraphics[scale=0.58]{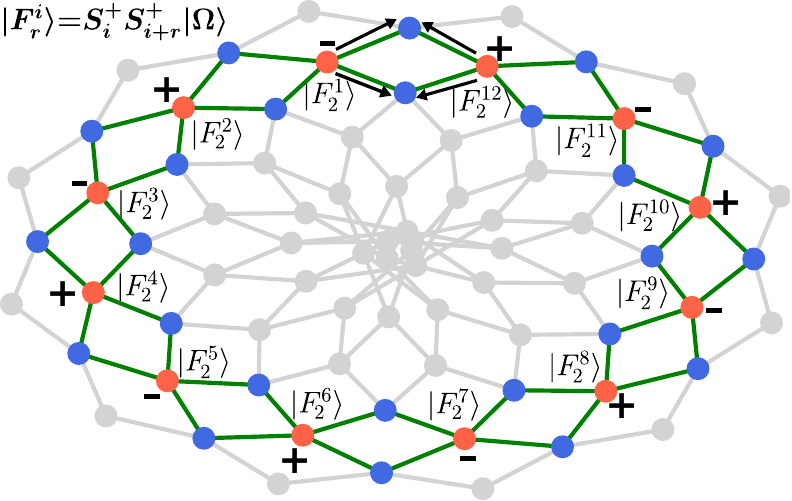}
    \caption{Schematic illustration of the Fock-space graph connectivity for the Fock-space cage state $|\Omega_2\rangle$ in a system with $2L{=}12$ sites in the magnetization sector $M{=}{-}10$. The orange nodes represent the basis states $|F_{2}^{i}\rangle$~[see Eq.~\eqref{eq: FSC_states}] that constitute the state, and their amplitudes, ${\pm}1$, are indicated by just the sign. The blue nodes are direct neighbors of the orange nodes under the Hamiltonian $H_{XY}$ [see Eq.~\eqref{eq: full_Hamiltonian}], and green edges show these connections. Destructive interference ensures that all blue nodes receive zero net amplitude from the orange nodes. Dimmed nodes represent the remaining Fock space states, which are decoupled from the cage.}
    \label{fig: FSC_example}
\end{figure}

The defining feature of an FSC state is that it has support on a sparse set of basis configurations of the Hilbert space, arranged with carefully chosen relative phases such that all virtual transitions to neighboring configurations under the action of a Hamiltonian cancel through destructive interference, making the state an exact zero-energy eigenstate.
In the Fock-space graph, this appears as a closed loop (or ``cage”) within which all nonzero amplitudes reside, dynamically isolated from the rest of the Hilbert space.
Unlike generic zero modes, whose cancellation patterns are generically delocalized over the entire Hilbert space, FSC states are compact and local in Fock space, and as we show, they display hallmark features of QMBS.
To illustrate this mechanism concretely, we consider a family of two-magnon states built by applying pairs of spin-raising operators at sites separated by a fixed distance $r$ to the ferromagnetic vacuum $|\Omega\rangle$ to obtain
\begin{equation} 
    |\Omega_{r}\rangle{=}\sum_{i=1}^{2L}({-}1)^{i}|F_{r}^{i}\rangle~~~~~\text{with}~~~~~|F_{r}^{i}\rangle{=}S_{i}^{+}S_{i{+}r}^{+}|\Omega\rangle,
    \label{eq: FSC_states}
\end{equation}
where $r{=}1,2,{\cdots}, L$ (owing to PBC, the largest separation $r_{\rm max}{=}L$)~\footnote{Note that for $r{=}1$, the state $|\Omega_{1}\rangle$ coincides with the state $|\mathbb{B}_{1}'\rangle$ defined in Eq.~\eqref{eq: type_2 scar tower}, however even in that case the tower we will build on top of it in the next section is entirely distinct.}.
The state $|\Omega_{r}\rangle$ resides in the magnetization sector $M{=}{-}2L{+}2$ and is built from a superposition of $2L$ distinct configurations $|F_{r}^{i}\rangle$ with spin raised at sites $i$ and $i{+}r$.
All of these $|F_{r}^{i}\rangle$ share the same $\mathcal{\hat{C}}$-parity $({-}1)^{r}$, so the support of $|\Omega_{r}\rangle$ lies entirely on one sublattice of the bipartite Fock-space graph defined by $H_{XY}$.
As discussed in Sec.~\ref{sec: Chiral symmetry and extensive zero modes}, the hopping action of $H_{XY}$ connects each support node $|F_{r}^{i}\rangle$ to basis states only in the complementary sublattice.
However, any such complementary-sublattice node is reached from both $|F_{r}^{i}\rangle$ and $|F_{r}^{i{+}1}\rangle$ with identical matrix-element magnitude but opposite relative sign, inherited from the $({-}1)^{i}$ alternating coefficients in $|\Omega_{r}\rangle$ [see Eq.~\eqref{eq: FSC_states}].
Their amplitude in $H_{XY}|\Omega_{r}\rangle$ therefore cancels exactly, and every would-be leakage path is blocked.
Hence $|\Omega_{r}\rangle$ is an exact zero-energy eigenstate of $H_{XY}$.
Note that even though this can be shown simply by explicitly analyzing the action of $H_{XY}$ on the states directly without any reference to the Fock-space graph, we have illustrated this in the graph language to highlight the FSC origins of these states.
A schematic depiction of this cancellation mechanism is shown in Fig.~\ref{fig: FSC_example} for $|\Omega_{1}\rangle$ in a $12$-site chain.

Computing the von Neumann EE $\mathcal{S}_{L}{=}{-}\text{Tr}_{A} (\rho_A\ln\rho_A)$, where $\rho_{A}$ is the reduced density matrix of one half of the chain, of $|\Omega_{r}\rangle$ for $r{\leq}L/2$ we obtain (see App.~\ref{Appsec: EE of Fock-space cages} for derivation):
\begin{equation}
    \mathcal{S}_{L}^{(r)}{=}\frac{L{-}r}{L}\ln\left(\frac{2L}{L{-}r}\right){+}\frac{r}{L}\ln(2L),
    \label{eq: EE_of_FSC}
\end{equation}
which, for finite $r$ in the large-$L$ limit approaches $\ln(2)$, as expected for single local quasiparticle eigenstates~\cite{Moudgalya2018AKLT, castroalvaredo2018entanglement}. On the other hand, for $r{\propto}L$, $\mathcal{S}_{L}^{(r)}{\sim}\ln(L)$, because the single quasiparticle is now non-local. Thus, the entanglement always violates the volume law and identifies the simple FSCs $|\Omega_r\rangle$ as QMBS since they too lie in the middle of the spectrum by definition, albeit in an extensive zero-energy manifold.
\subsection{Towers of equally spaced Fock-space cages}
\label{subsec: Towers of equally spaced FSCs}
\subsubsection{States}
Interestingly, the two-magnon cages introduced in Eq.~\eqref{eq: FSC_states} represent the most elementary members of a larger hierarchy of exact zero-energy eigenstates of $H_{XY}$.
For \textit{each} fixed separation $r{\in}\{1, {\cdots} L\}$, we identify an entire family of zero-energy eigenstates labeled by an integer $n{=}0, 1, {\cdots}, 2L{-}2$, denoted $|r,n\rangle$, that share the same underlying interference structure but exhibit progressively more complex amplitude patterns.
Their explicit form (up to normalization) is given by
\begin{equation}
\begin{split}
    |r,n\rangle=&\sum_{i=1}^{2L}({-}1)^{i(n{+}1)}(J_{i,r}^{+})^{n} (S_{i}^{+}S_{i{+}r}^{+})|\Omega\rangle,
\end{split}
    \label{eq: new_towers}
\end{equation}
where the operator $J_{i,r}^{+}$ is defined as
\begin{equation}
    \begin{split}
        &J_{i,r}^{+}{=}\frac{1}{2}\sum_{\substack{i{\leq}j{\leq}i{+}2L{-}1\\j{\neq}i,i{+}r}}(-1)^{f(j,i,r)} (S_{j}^{+})^2,\\ &\text{with} ~~
    f(j,i,r){=} \begin{cases}
        j{-}i &\text{for}\hspace{0.2cm}i{+}1{\leq}j{<}i{+}r\\
        j{-}i{-}1 &\text{for}\hspace{0.2cm}j{>}i{+}r
    \end{cases}. 
    \end{split}
\end{equation}
An equivalent but more instructive form of Eq.~\eqref{eq: new_towers} can be written in terms of the bimagnon scar states of Eq.~\eqref{eq: type_1 scar tower} as
\begin{equation}
    |r,n\rangle{=}\sum_{i=1}^{2L}(-1)^{i(n{+}1)}|00\rangle_{i,i{+}r}|\mathbb{B}_{n}\rangle_{i{+}1,\cdots,i{+}r{-}1,i{+}r{+}1,\cdots,i{-}1},
    \label{eq: newstates_interms_of_old}
\end{equation}
where $|00\rangle_{i,i{+}r}$ denotes a state with a magnon each at sites $i$ and $i{+}r$, and $|\mathbb{B}_{n}\rangle$ is an $n$-bimagnon background living on the remaining $2L{-}2$ sites (thus $n$ ranges from $0$ to $2L{-}2$). 
The lowest member $|r,0\rangle{\equiv}|\Omega_{r}\rangle$ corresponds to the simple FSC state discussed earlier in Eq.~\eqref{eq: FSC_states}.
\subsubsection{Revivals}
\begin{figure}
    \centering
\includegraphics[scale=0.56]{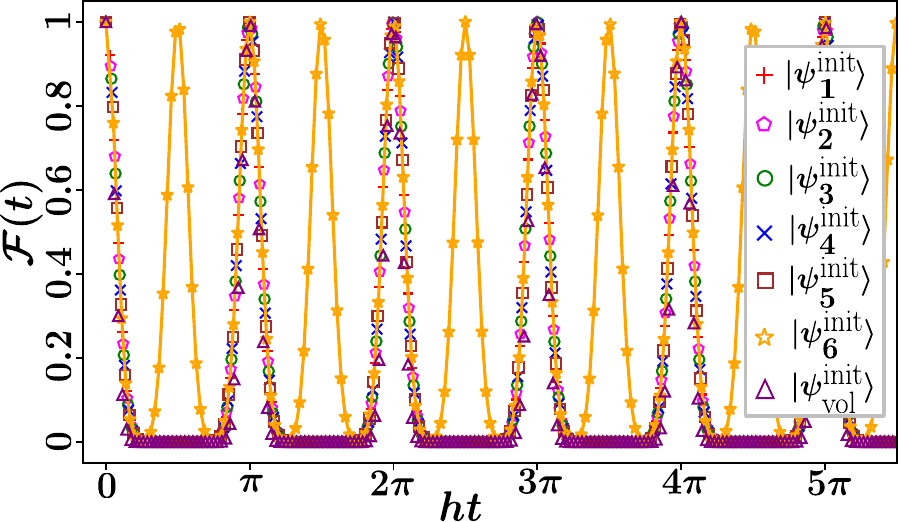}
    \caption{Fidelity $\mathcal{F}(t){=}|\langle\psi(t)|\psi(0)\rangle|^2$ for a spin-$1$ chain of size $2L{=}12$ sites, shown for initial states $|\psi_{r}^{\rm init}\rangle$ [see Eq.~\eqref{Eq: Initial_states_for_revivals}] (with $r{=}1,2,{\cdots},6$) and $|\psi_{\rm vol}^{\rm init}\rangle$~[see Eq.~\eqref{eq: EAP_state}] constructed as a coherent superposition over the scar eigenstates of the towers $|r,n\rangle$, and the volume-entangled tower $|\mathbb{V}_n\rangle$, respectively. The state $|\psi_{r}^{\rm init}\rangle$ is time-evolved with $H_{0}$ [see Eq.~\eqref{eq: full_Hamiltonian}], while $|\psi_{\rm vol}^{\rm init}\rangle$ is time-evolved with $H_{\rm vol}$ [see Eq.~\eqref{eq: volume_scar_tower}]. All cases exhibit persistent fidelity revivals with period $\pi/h$, consistent with the uniform energy spacing $2h$, except for $|\psi_{r{=}L}^{\rm init}\rangle$, which oscillates with period $\pi/(2h)$ as discussed in the main text.}
\label{fig: fidelity}
\end{figure}

The state $|r,n\rangle$ is an eigenstate of the total magnetization with magnetization $M_{r,n}{=}2(n{+}1{-}L)$.
Hence, under the full Hamiltonian $H_0$ of Eq.~\eqref{eq: full_Hamiltonian}, the state $|r,n\rangle$ acquires energy $E_{r,n}{=}2h(n{+}1{-}L)$.
Consequently for fixed $r$, the set of states $\{|r,n\rangle\}$ forms an equally spaced tower of $H_0$ with spacing $\Delta E_{r,n}{=}2h$.
For the special case $r{=}L$, the alternating phase structure of Eq.~\eqref{eq: new_towers} enforces $|L,n\rangle{=}0$ for all odd (even) values of $n$ when $L$ is even (odd), so the effective spacing of that particular tower doubles to $4h$.
Note that $|r,n\rangle$ also remains an exact eigenstate of the single-ion anisotropy term $D\sum_{i=1}^{2L}(S_i^z)^2$ under which it acquires a uniform energy shift of $(L{-}2)D$ that is independent of $n$.
The uniform energy spacing immediately implies perfect dynamical revivals from appropriately chosen initial states. A simple choice for the initial states is $|\psi_{r}^{\text{init}}\rangle{=}\sum_{n=0}^{2L-2}|r,n\rangle$ that that are coherent superpositions of states from a single tower. Their explicit form,
\begin{equation}
      |\psi_{r}^{\text{init}}\rangle
  {=}\sum_{i}({-}1)^{i}|0,0\rangle_{i,i{+}r}\bigotimes_{j{\neq}i,i{+}r}(|{-}1\rangle{+}({-}1)^{i{+}f(j,i,r)}|1\rangle)_{j},
  \label{Eq: Initial_states_for_revivals}
\end{equation}
reveals that they are inherently low entangled since they are essentially a single quasiparticle on top of a product state; in the large-$L$ limit, their half-chain EE approaches $2\ln(2)$. Under time evolution with $H_{0}$, their fidelity $\mathcal{F}_r(t){=}|\langle\psi_{r}^{\text{init}}|e^{-\iota H_{0}t}|\psi_{r}^{\text{init}}\rangle|^{2}$ exhibits perfect revivals with period $T{=}\pi/h$ [or $T{=}\pi/(2h)$ for the $r{=}L$ case due to the double energy spacing of $4h$]. These persistent oscillations from a simple initial state, as illustrated in Fig.~\ref{fig: fidelity}, are a definitive dynamical hallmark of scarring. Note that it is in principle also possible to further superpose the states in Eq.~\eqref{Eq: Initial_states_for_revivals} for different $r$ and obtain Matrix Product State initial states that exhibit similar revivals, but we will not explicitly illustrate them here.

\begin{figure}
    \centering
\includegraphics[scale=0.5]{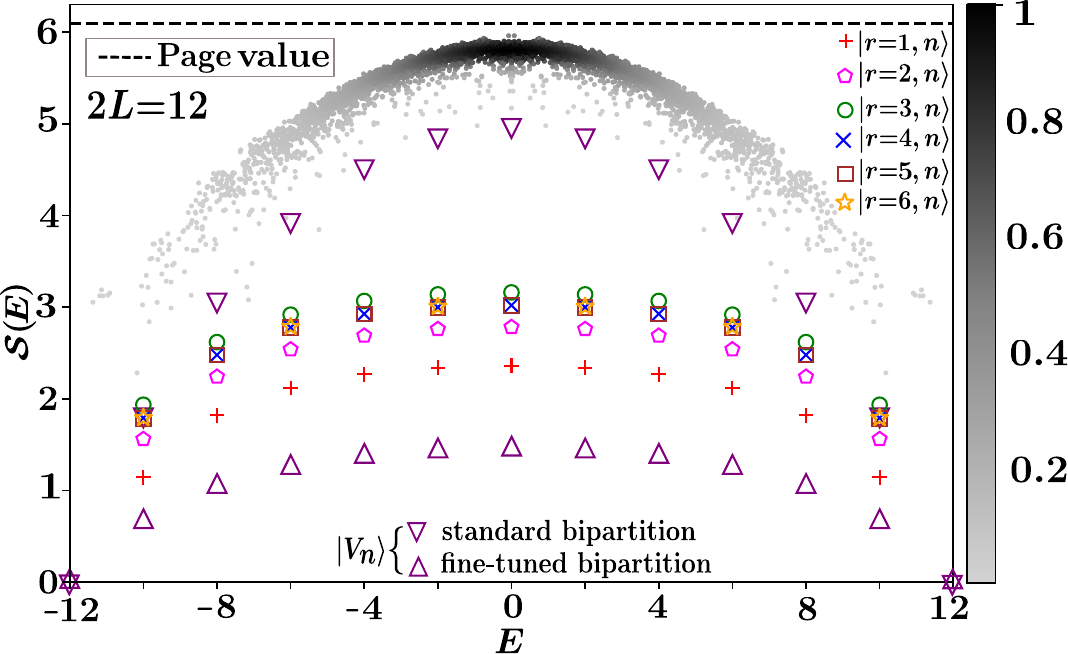}
    \caption{Bipartite entanglement entropy (EE) of all scar states constructed in this work for a spin-$1$ chain of size $2L{=}12$ sites. The plot shows EE of the tower states $\{|r,n\rangle\}$~[see Eq.~\eqref{eq: new_towers}] for $r{=}1,2,{\cdots},6$, and the special volume-entangled tower states $\{|\mathbb{V}_{n}\rangle\}$~[see Eq.~\eqref{eq: volume_scar_tower}], both for the standard bipartition with a cut at $L$, and the fine-tuned antipodal bipartition, discussed in the text. For comparison, the EE of all eigenstates of a non-integrable Hamiltonian $H_{\rm comm}^{(2)}$~[see Eq.~\eqref{eq: V_comm_def}] with $J{=}h{=}1$, $\epsilon_{i}{\in}[0,0.2]$ and periodicity $r{=}2$, is shown in the background as grey dots with its color encoding by the density of states with a given EE (darker=higher density). The black dotted line marks the Page value, the expected EE of a random state. Although the scar states lie in different magnetization sectors and are exact eigenstates of the spin-$1$ $XY$ model with different integrability-breaking perturbations, this plot provides a standard comparative representation of their subthermal EE, highlighting their nonthermal nature. }
\label{fig: EE_towers}
\end{figure}
\begin{figure}
    \centering
\includegraphics[scale=0.5]{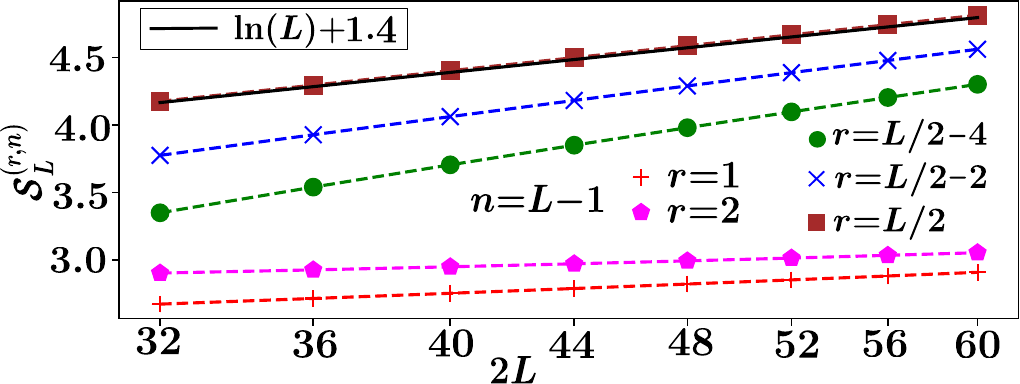}
    \caption{ Scaling of the half-chain entanglement entropy of mid-spectrum states $|r,n {=} L {-} 1\rangle$ evaluated using Eq.~\eqref{Eq: EE_r,n} as a function of system size $2L$. The $x{-}$axis is plotted on a log scale. The data shown for $r{=}1,2, L/2{-}4, L/2{-}2, L/2$ fall on straight lines $\mathcal{S}_{L}^{(r, L-1)} {=} m_r \ln(L) {+} c_r$ (indicated by dotted lines), demonstrating logarithmic growth of entanglement. For clarity, the explicit fit is shown for $r{=}L/2$, where $m_r{=}1$ and $c_r{=}1.4$, in agreement with the asymptotic analysis of Eq.~\eqref{Eq: asymptotic highest EE}, which predicts a unit slope for this case.}
\label{fig: EE_scaling_r_n}
\end{figure}
\subsubsection{Entanglement}
The entanglement properties of the states in the tower further corroborate their nonthermal character. In Fig.~\ref{fig: EE_towers} we plot the half-chain EE of all $|r,n\rangle$ for a chain of size $2L{=}12$. Across all towers, the entropy remains far below the Page value $\mathcal{S}_{\rm Page}{=}L\ln(3){-}1/2$~\cite{page1993}, forming distinct outliers when compared with the typical entanglement of chaotic eigenstates at similar energy densities.
An explicit expression for the EE follows from resolving the bipartite structure of $|r,n\rangle$ [see Appendix.~\ref{Appsec: EE of the towers states}]. For $r{\leq}L/2$, this reads
\begin{equation}
    \begin{split}
        \mathcal{S}_{L}^{(r,n)}{=}&{-}2\sum_{k{=}0}^{{\rm min}(n,L{-}2)}(\lambda_{k}^{{\rm T}_1})^2\ln (\lambda_{k}^{{\rm T}_1})^2 \\
        &{-}2r\sum_{k{=}0}^{{\rm min}(n,L{-}1)}(\lambda_{k}^{{\rm T}_2})^2\ln (\lambda_{k}^{{\rm T}_2})^2,     
    \end{split}
    \label{Eq: EE_r,n}
\end{equation}
where
\begin{equation}
    \begin{split}
        (\lambda_{k}^{{\rm T}_1})^2{=}\frac{(L{-}r)\binom{L{-}2}{k}\binom{L}{n{-}k}}{2L\binom{2L{-}2}{n}}~~{\rm and}~~(\lambda_{k}^{{\rm T}_2})^2{=}\frac{\binom{L{-}1}{k}\binom{L{-}1}{n{-}k}}{2L\binom{2L{-}2}{n}}. 
    \end{split}
\end{equation}
In Fig.~\ref{fig: EE_scaling_r_n}, we show the scaling of EE of mid-spectrum states with $n{=}L{-}1$, which possess the highest EE within each tower, for different towers.
For all towers, the data exhibit a clear linear dependence on $\ln (L)$, indicating logarithmic growth of entanglement with system size.
To make this trend explicit, focusing on the highest entangled case $r{=}L/2$, the asymptotic in $L$ analysis (given in Appendix~\ref{Appsec: Asymptotic limit highest EE state}) yields
\begin{equation}
    \mathcal{S}_{L}^{(L/2,L-1)} \sim \ln\left(L\right) + \text{constant},
    \label{Eq: asymptotic highest EE}
\end{equation}
which agrees with numerical scaling [see Fig.~\ref{fig: EE_scaling_r_n}] up to a constant offset.
This logarithmic, subthermal scaling is consistent with the physical interpretation that the states $|r,n\rangle$ are composed of an extensive number of identical quasiparticles on a background of a logarithmically entangled state $\ket{\Omega_r}$ of Eq.~\eqref{eq: FSC_states}, which are only expected to add ${\sim}\ln(L)$ entanglement~\cite{moudgalya2021review}.
This scaling contrasts sharply with the extensive ``volume-law” scaling ($\mathcal{S}_{L}{\propto}L$) and establishes $|r,n\rangle$ as genuine QMBSs.

\begin{table}[t]
\centering
\setlength{\tabcolsep}{3pt} 
\renewcommand{\arraystretch}{1.05}
%\footnotesize <--- make BOTH panels small
% ------------ Panel (a) ------------
\begin{minipage}{0.55\columnwidth}
\centering
Table I (a):\; $r{=}2,2L{=}12$ \\[0.25em]
\begin{tabular}{c c c c c c}
\hline
$n$ & $N_{\text{sup}}$ & $N_{\text{con}}$ & $\mu$ & $\mathcal{D}_{2L}^{M}$ & $\phi$ \\
\hline
0,10 & 12   & 24    & 2    & 78    & 0.46 \\
1,9  & 120  & 336   & 2.8  & 1221  & 0.37 \\
2,8  & 540  & 1848  & 3.42 & 8074  & 0.29 \\
3,7  & 1440 & 5568  & 3.86 & 28314 & 0.25 \\
4,6  & 2520 & 10416 & 4.13 & 58278 & 0.22 \\
5    & 3024 & 12768 & 4.22 & 73789 & 0.21 \\
\hline
\end{tabular}
\end{minipage}
\hfill
% ------------ Panel (b) ------------
\begin{minipage}{0.35\columnwidth}
\centering
Table I (b):\; $n{=}L{-}1$ \\[0.25em]
\begin{tabular}{c c c}
\hline
$2L$ & $\phi\,(r{=}1)$ & $\phi\,(r{=}2)$ \\
\hline
6   & 0.8 & 0.73 \\
8   & 0.48 & 0.62 \\
10   & 0.27 & 0.38 \\
12  & 0.15 &  0.22\\
14   & 0.08 & 0.12 \\
16  & 0.04 & 0.06 \\
\hline
\end{tabular}
\end{minipage}

\vspace{0.6em}
\caption{Quantification of the active Fock-space region for tower states $|r,n\rangle$. 
(a) reports the number of support nodes $N_{\mathrm{sup}}$, connected nodes $N_{\mathrm{con}}$, cancellation multiplicity $\mu{=}N_{\mathrm{con}}/N_{\mathrm{sup}}$, sector dimension $\mathcal{D}_{2L}^{M}$, and active fraction $\phi{=}(N_{\mathrm{sup}}{+}N_{\mathrm{con}})/\mathcal{D}_{2L}^{M}$ for all states of the $r{=}2$ tower with system size $2L{=}12$.
While $N_{\mathrm{sup}}$ grows polynomially with $n$ and $N_{\mathrm{con}}$ increases as a larger multiple, the active fraction remains small.
(b) shows that $\phi$ for the mid-spectrum states ($n{=}L{-}1$) in different towers (here $r{=}1,2$) decreases with system size (shown up to $2L{=}16$), strongly suggesting that it tends to zero in the thermodynamic limit. This demonstrates that even the most complicated mid-spectrum states involve interference processes confined to a vanishing fraction of the Hilbert space, and can thus be regarded as complicated FSCs.}
\label{tab: FSC_quantification}
\end{table}

\subsubsection{Fock-space cage statistics}
\label{Sec: Fock-space cage statistics}
To better understand how these states fit within the FSC perspective, we analyze three quantities for each state $|r,n\rangle$: (i) the number of support nodes $N_{\rm sup}$ — basis configurations with nonzero amplitude ; (ii) the number of connected nodes $N_{\rm con}$ — additional configurations reached under the action of $H_{XY}$ where the destructive interferences occur; and (iii) the dimension of the corresponding magnetization sector $M_{r,n}$ [see Eq.~\eqref{eq: U(1) sector dimension}].
The union of support and connected nodes defines the \textit{active region} associated with the state, and the ratio of the active-region size to the sector dimension defines the \textit{active fraction} $\phi$, which measures how much of the Hilbert space participates in the interference process.
Table~\ref{tab: FSC_quantification} summarizes these quantities for representative values of $r,{n}$ and $L$.
For a fixed system size $2L$, as one moves from the bottom of a tower towards the middle of the spectrum approaching $M{=}0$, the number of support nodes grows rapidly as $2L\binom{2L{-}2}{n}$.
The number of connected nodes grows proportionally but lacks a simple closed-form expression, reflecting the wavefunction’s increasing structural complexity. Yet, the active fraction remains small throughout [see Table~\ref{tab: FSC_quantification}(a)]. In addition, for a fixed tower member, such as a mid-spectrum state with $n{=}L{-}1$, this fraction decreases steadily with increasing system size [see Table~\ref{tab: FSC_quantification}(b)]. Together, these trends imply that the entire active region involved in interference occupies a vanishingly small portion of the Hilbert space in the thermodynamic limit. 
 
\begin{figure}
    \centering
\includegraphics[scale=0.32]{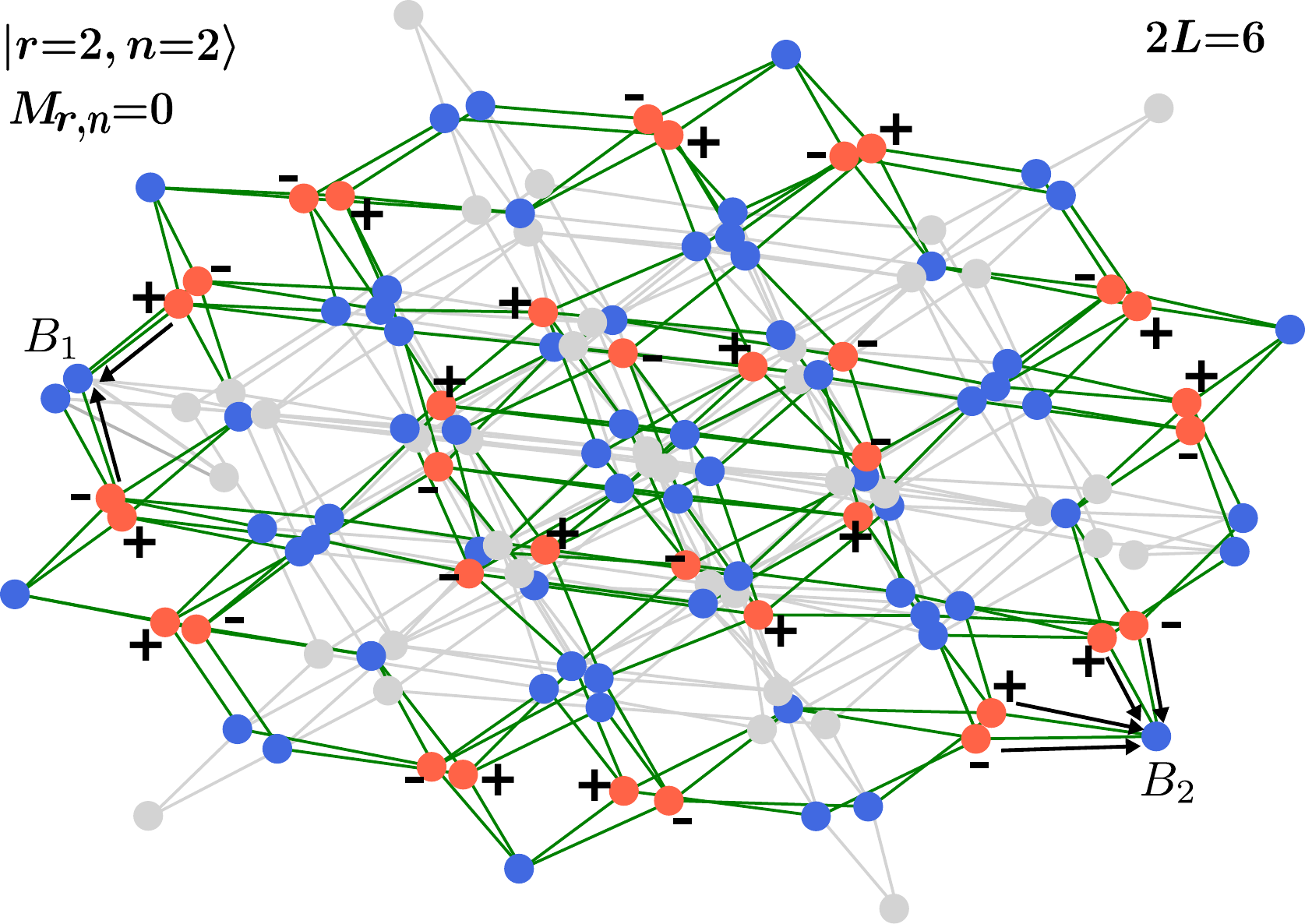}
    \caption{Fock-space representation of a mid-spectrum scar state $|r{=}2,n{=}2\rangle$ for a chain of size $2L{=}6$. Orange nodes denote basis states with nonzero amplitude ($\pm1$), while blue nodes are connected via $H_{XY}$. Although destructive interference at all blue nodes continues to confine the non-zero amplitudes, forming a complicated Fock-space cage, the cancellation network becomes highly interconnected and spatially extended, making the structure increasingly difficult to visualize. Two representative blue nodes (labelled $B_{1}$ and $B_{2}$) highlight this mechanism: ``$B_{1}$" exhibits simple pairwise cancellation, whereas ``$B_{2}$" shows multipath interference involving several $\pm$ paths.}
\label{fig: complicated_FSC_example}
\end{figure}

However, at the microscopic level, the cancellation pattern becomes increasingly intricate as $n$ grows.
For the simple two-magnon cages, i.e., $|r,n{=}0\rangle$, each connected node receives two equal and opposite contributions, producing a simple, visually transparent cancellation pattern.
For higher $n$, the number of contributing paths proliferates; connected nodes typically receive many amplitudes whose phases must coordinate for cancellation.
This cancellation multiplicity ($\mu$)—captured in Table~\ref{tab: FSC_quantification}~(a) by the rising ratio of connected to support nodes—spreads the interference network over a large, highly connected subgraph.
Direct visualization quickly becomes opaque (as shown in Fig.~\ref{fig: complicated_FSC_example} for a representative mid-tower state).
In that sense, while all $|r,n\rangle$ can be regarded as \textit{complicated} FSCs, the intuitive cage picture becomes impractical beyond the lowest states in the tower. 
We note that the previously studied bimagnon states $|\mathbb{B}_n\rangle$ and $|\mathbb{B}'_n\rangle$ of Eqs.~(\ref{eq: type_1 scar tower})~\cite{schecter2019weak} and (\ref{eq: type_2 scar tower})~\cite{chattopadhyay2020quantum} show similar statistics, which further supports their interpretation within the FSC framework (see Appendix~\ref{Appsec: FSC statistics for the bimagnon tower}).
Although algorithmic approaches have been proposed recently~\cite{Jonay2025FockSpaceCages, Tan2025InterferenceCaged} to identify FSC states systematically, they rely on iteratively searching for closed interference loops within the bipartite Fock-space graph.
Such methods work well when the cages involve only a small number of nodes, but in the cases of tower of states, both the support and the connected subgraphs rapidly increase with system size, while the number of possible loop configurations—the distinct ways to assign compensating paths that neutralize every connected node-grows exponentially.
Consequently, the search space becomes prohibitively large, and these automated techniques—though capable of detecting our states for very small chains or the simplest single quasiparticle states in the tower — do not generalize efficiently to larger system sizes and states with a higher number of quasiparticles.
These observations motivate the need for a more systematic operator-level framework to characterize these states.
In the next section, we discuss the characterization of these states in terms of commutant algebras,  which does precisely that.
This picture naturally recognizes them as scar eigenstates due to the existence of whole families of non-commuting operators under which these states remain exact eigenstates~\cite{moudgalya2022exhaustive}. 
Furthermore, it reveals additional scar families, notably those with volume-law entanglement and mirror-dimer structures, that are not immediately apparent from the FSC perspective.
\subsubsection{Connections to other unified frameworks}
Before we move on, we emphasize that most of the states $\ket{r, n}$ [similar to the $|\mathbb{B}'_n\rangle$ states of Eq.~\eqref{eq: type_2 scar tower}] appear to lie beyond the reach of systematic frameworks that have been proposed to unify different families of scar eigenstates such as Restricted Spectrum Generating Algebras (RSGA), and symmetry-based constructions like ``tunnels to towers,"~\cite{mark2020eta, ODea2020tunnelstotowers} and quasisymmetry formalisms~\cite{Ren2021Quasisymmetry}.
These frameworks are suited for describing towers of states that are generated by the action of simple ladder operators on top of some base state $\ket{\psi_0}$, e.g., of the form $(Q^\dagger)^n\ket{\psi_0}$. Among the various towers, we are only able to find that structure in the case $r{=}1$. This tower satisfies the defining condition of an order-1 RSGA [the detailed proof is given in Appendix~\ref{Appsec: RSGA_FSC}]. With $|\Omega_{1}\rangle$ as the root, successive tower states can be obtained by repeated action of the raising operator $J^{+}$ defined in Eq.~\eqref{eq: type_1 scar tower}, giving $|1,n\rangle{=}(J^{+})^{n}|\Omega_{1}\rangle$. This places the $r{=}1$ tower on the same footing as the known bimagnon scar tower given in Eq.~\eqref{eq: type_1 scar tower}. However, for $r{>}1$, we cannot identify a comparably simple ladder operator. It remains possible that an appropriate operator exists but is significantly more complicated; we leave its explicit construction to future work.
\begin{table*}[t]
\centering
\renewcommand{\arraystretch}{1.3}
\setlength{\tabcolsep}{4pt}
\begin{tabular}{|c|c|c|c|c|c|}
\hline
\# 
& \multicolumn{2}{c|}{QMBS family}
& \makecell{$M$}
& \makecell{Minimal set of non-\\commuting local operators} 
& \makecell{Bipartite EE\\characteristic} \\
\hline
\# 1 & \makecell{Bimagnon tower\\
$|\mathbb{B}_{n}\rangle$, Eq.~\eqref{eq: type_1 scar tower},\\ Ref.~\cite{schecter2019weak} \\}
& $n{=}0,1,{\cdots},2L$
  & $2(n{-}L)$ 
  & $\{h_{i,i{+}1}\}$, Eq.~\eqref{eq: exchange_bonds} 
  & \makecell{subextensive\\
  Fig. 2 of Ref.~\cite{schecter2019weak}} \\
\hline
\# 2 & \makecell{Bond-bimagnon tower\\
$|\mathbb{B}'_{n}\rangle$, Eq.~\eqref{eq: type_2 scar tower},\\ Ref.~\cite{schecter2019weak,chattopadhyay2020quantum}}
& $n{=}0,1,{\cdots},2L$
  & $2(n{-}L)$
  & -
  & \makecell{subextensive\\
  Fig. 5 of Ref.~\cite{chattopadhyay2020quantum}} \\
\hline
\# 3 & \makecell{FSC towers\\$|r,n\rangle$, Eq.~\eqref{eq: new_towers}} 
& \makecell{For each $r{\in}\{1,{\cdots},L\},$\\$n{=}0,{\cdots},2L{-}2$ 
}
  & $2(n{+}1{-}L)$
  & \makecell{(i) $2L~{\rm mod}~r{=}0$:
  $\{\mathcal{H}_{j}^{(r)}\}$,\\ Eq.~\eqref{eq: new_generators_of_local_algebra}\\
  (ii) generic $r{>}1$:
  $\{q_{i,i{+}1}\}$, \\
  $r{=}1$: $\{q_{i,i{+}3}\}$ \\Eq.~\eqref{eq: generator_generic_r}}
 & \makecell{subextensive\\
  Fig.~\ref{fig: EE_towers}} \\
\hline
\# 4 & \makecell{Volume-entangled\\tower $|\mathbb{V}_n\rangle$, Eq.~\eqref{eq: volume_scar_tower}}
& $n{=}0,1,{\cdots},2L$
  & $2(n{-}L)$
  & $\{\mathcal{H}_{j}^{(L)}\}$~\eqref{eq: new_generators_of_local_algebra}
  & \makecell{(i) standard cut: 
  \\extensive, Eqs.~\eqref{eq: EE_volumetower_standardpartition}, \eqref{eq: EE_volumetower_standardpartition2} \\ (ii) fine-tuned cut:\\ subextensive, Eqs.~\eqref{eq: EE_volumetower_athermalpartition}, \eqref{eq: EE_volumetower_athermalpartition2}    } \\
\hline
\# 5 & \makecell{Mirror-dimer\\states $|\mathbb{M}_{m,m'}^k\rangle$, Eq.~\eqref{eq: mirror_dimer_state}}
& \makecell{For each $k{\in}\{1,{\cdots},L\},$\\$m,m'{\in}\{{-}1,0,1\}$}
  & $m{+}m'$
  & $\{\mathcal{M}_{j}^{k}\}$, Eq.~\eqref{eq: mirror_local}
  & \makecell{(i) dimers cut:
  $L\ln(3)$ \\(ii) mirror-paired: 0} \\
\hline
\end{tabular}
\caption{Summary of all exact QMBS of the spin-$1$ $XY$ model, which are zero-energy eigenstates of $H_{XY}$ [see Eq.~\eqref{eq: full_Hamiltonian}]. The table lists both previously known scars (\#1, \#2) and the new families discovered in this work (\#3, \#4, \#5).  For each case, we indicate their total magnetization $M$, the number of states, the minimal set of non-commuting local operators that realize them as some of the simultaneous eigenstates (that highlight their non-thermal nature and could connect them to the commutant algebra framework), and their characteristic entanglement scaling. Together, these examples demonstrate how diverse non-thermal subspaces—ranging from frustration-free bimagnons to mirror-dimer states — can all be unified through the study of simultaneous eigenstates of appropriate minimal sets of non-commuting local operators.}
\label{Table: summary_of_all_scar_states}
\end{table*}

\section{Exact eigenstates from commutant algebras}
\label{sec: exact eigenstates from commutant algebra formalism}
In this section, we review the framework of commutant algebras~\cite{moudgalya2022exhaustive, Moudgalya2023From_symmetries,moudgalya2021hilbert}, which has been demonstrated to be a powerful tool for characterizing QMBS and associated Hamiltonians.
As we will show, this framework provides an algebraic explanation for the tower of exact eigenstates $|r,n\rangle$ discussed earlier [see Eq.~\eqref{eq: new_towers}] and, at the same time, offers a constructive route to discover new families of scar states beyond those accessible from the FSC perspective.

\begin{figure}[h]
    \centering
\includegraphics[scale=0.43]{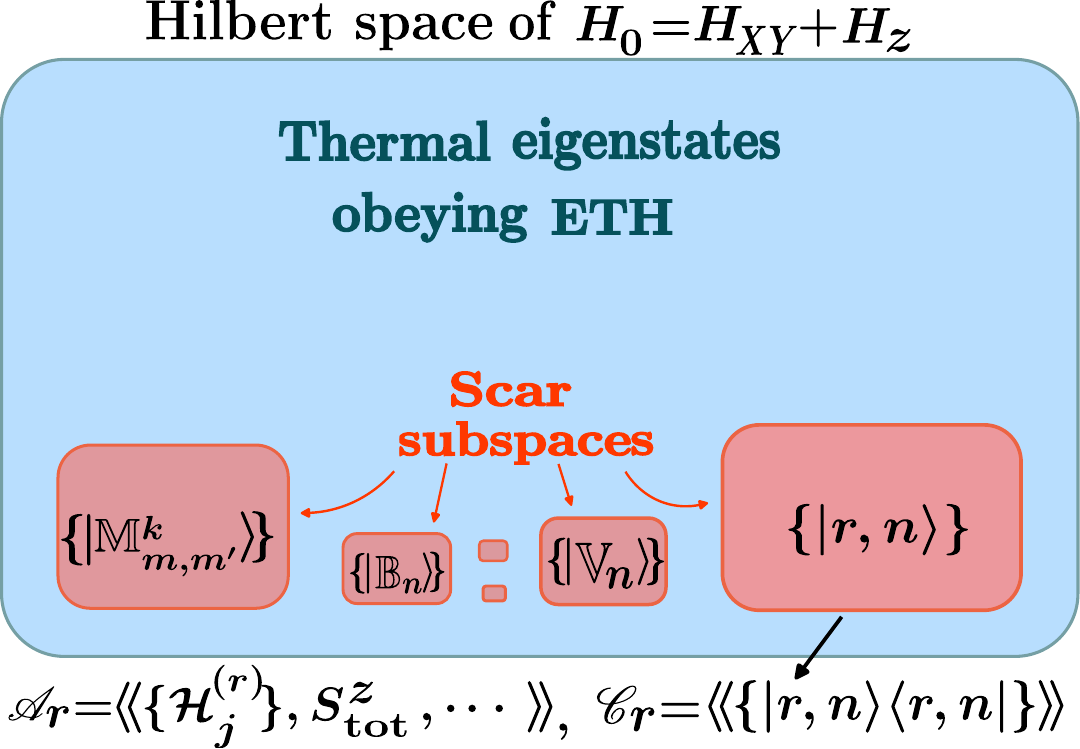}
    \caption{Schematic of the Hilbert-space structure in the presence of exact QMBS and their interpretation via the commutant algebra framework. The Hilbert space fragments into a large thermal block obeying ETH and smaller dynamically isolated scar manifolds that host exact nonthermal eigenstates. For the example shown, corresponding to the $|r,n\rangle$ tower of Eq.~\eqref{eq: new_towers}, a minimal subset of local non-commuting generators defines the working algebra $\mathscr{A}_{r}$, to which $H_{0}$ belongs, while the projectors onto the scar states span the commutant $\mathscr{C}_{r}$ and act as conserved quantities within the corresponding nonthermal block. This construction illustrates how the commutant framework explains the isolated scar subspaces embedded within an otherwise thermal many-body spectrum.
    }
    \label{fig: scar_subspace_commutant}
\end{figure}

\subsection{Commutant algebras and QMBS} 
In systems hosting exact QMBS, the Hilbert space effectively partitions into a vast ``thermal" subspace consistent with ETH and a small ``non-thermal" subspace comprising exact eigenstates that violate ETH (see Fig.~\ref{fig: scar_subspace_commutant}).
Importantly, these small subspaces are referred to as ``non-thermal" or ETH-violating only if this emergent block structure \textit{cannot} be explained by conventional global symmetries. This is because standard symmetries, such as total spin or particle number conservation, that are generally associated with local conserved quantities, merely partition the Hilbert space into sectors in which slightly modified versions of ETH that account for the local conserved quantities are still expected to hold~\cite{deutsch2018eigenstate}.
In contrast, scars form tiny atypical islands embedded within these sectors, and their existence cannot be explained by local conserved quantities, which demands a different organizing principle.
The commutant algebra framework provides precisely such a principle, offering a natural explanation for how these exceptional non-thermal subspaces can emerge from non-local symmetries that result from the intrinsic operator structure of the Hamiltonian.
\subsubsection{Key properties}
First, we consider a family of Hamiltonians of the form $H'{=}\sum_{\alpha}J_{\alpha}\hat{h}^{\alpha}$, where $\{\hat{h}^{\alpha}\}$ is a set of Hermitian local operators (either \textit{strictly local} with support on a few nearby sites on the lattice, or \textit{extensive local}, i.e., a sum of such terms) that do not commute with each other, and $J_{\alpha}$ are coupling constants.
We define two key algebraic structures:
\begin{itemize}
    \item The bond or local algebra $\mathscr{A}{=}\lgen \{\hat{h}^{\alpha}\} \rgen$ generated by arbitrary products and linear combinations of the local terms along with the identity, owing to which we call $\{\hat{h}^{\alpha}\}$ generators, and
    \item The commutant algebra $\mathscr{C}{=}\{\hat{O}:[\{\hat{h}^{\alpha}\},\hat{O}]{=}0~~\forall \alpha\}$, the set of operators that commute with every element of $\mathscr{A}$.
\end{itemize}
The algebra $\mathscr{C}$ is essentially the symmetry algebra of the family of Hamiltonians we are interested in, and $\mathscr{A}$ is the algebra of all operators with that symmetry. 
It has been shown that many kinds of conventional symmetries that appear in quantum many-body physics can be understood as commutant algebras with this structure~\cite{moudgalya2021hilbert, Moudgalya2023From_symmetries}.
Within this algebraic language, scar eigenstates are recognized as the ``singlets” of the algebra $\mathscr{A}$, i.e., they are simultaneous eigenstates of all operators in the algebra, in particular its local generators $\hat{h}_{\alpha}$~\cite{pakrouski2020many, moudgalya2022exhaustive}.
This characterization automatically implies that these states violate the conventional form of ETH, since by definition, they are eigenstates for an entire family of Hamiltonians; hence, the information of any single parent Hamiltonian is not in the state. In other words, ETH satisfying states are expected to have a unique local parent Hamiltonian that has those states as eigenstates~\cite{Garrison2018Does, Qi2019determininglocal}, and since singlets of $\mathscr{A}$ violate that property by definition, they are ETH-violating states, or QMBS (Strictly speaking, one should also examine the structure of the algebra generated by the local operators recovered from the state and ensure that it does not have other simple symmetries that would lead to a symmetry-based explanation of these states. In this work, we assume that this algebra is ``generic enough'' to not contain such symmetries. See Ref.~\cite{moudgalya2022exhaustive} for a more detailed discussion on this point.).
The projectors onto these scar states further lie within $\mathscr{C}$ and effectively act as non-local conserved quantities for the entire family of Hamiltonians spanned by $\mathscr{A}$, and can lead to block-decompositions that isolate scars from the remaining thermal states (see Fig.~\ref{fig: scar_subspace_commutant}) but cannot be explained by ordinary symmetries. More importantly for this work, this highlights the fact that \textit{scars are essentially simultaneous eigenstates of non-commuting local operators}, and re-frames the search for scars as an algebraic problem of identifying such states given a non-commuting local operator set.
\subsubsection{Illustration with the bimagnon tower of scars}
To illustrate these ideas concretely, consider the Hamiltonian $H_{XY}$ of Eq.~\eqref{eq: full_Hamiltonian} written as
\begin{equation}
\label{eq: exchange_bonds}
    H_{XY}{=}J\sum_{i=1}^{2L}h_{i,i{+}1}~~~\text{with}~~~h_{i,i{+}1}{=}S^x_i S^x_{i+1} + S^y_i S^y_{i+1}.
\end{equation}
The states of the well-known bimagnon scar tower given in Eq.~\eqref{eq: type_1 scar tower} are simply frustration-free eigenstates of $H_{XY}$ satisfying $h_{i, i{+}1}|\mathbb{B}_n\rangle {=}0~~~\forall i$. We also numerically find that the set of nearest-neighbor exchange bonds $\{h_{i, i{+}1}\}$ together with the total magnetization $S_{\rm tot}^{z}$ define a natural starting algebra $\mathscr{A}_{0}{=}\lgen\{h_{i, i{+}1}, S_{\rm tot}^{z}\}\rgen$, for which the bimagnon tower constitutes the only set of \textit{singlets}. It is easy to see that the Hamiltonian terms $\{h_{i,i+1}\}$ do not commute with each other, hence the states of the bimagnon tower are simultaneous eigenstates of a set of non-commuting local operators, which already illustrates their non-thermal character.
Note that the commutant $\mathscr{C}_0$ of this particular algebra $\mathscr{A}_0$ contains many more symmetries than just the scar projectors, which lead to more complicated block decompositions of the Hilbert space that obscure the clear separation into ``thermal" and ``non-thermal" subspaces.
Nevertheless, this shows that even though the bimagnon states might not be referred to as scars of $H_{XY}$ since they might be explained by other conventional symmetries of this particular model, many local perturbations can be constructed from the generators of $\mathscr{A}_0$ that break the extra symmetries while preserving these states, which makes them scars of the perturbed models.
Indeed, previous works studying this model~\cite{schecter2019weak, moudgalya2022exhaustive} add perturbations such as $\sum_{i=1}^{2L} ((S_i^z)^2S_{i+1}^{z}{+}S_i^z(S_{i+1}^{z})^2)$, $\sum_{i=1}^{2L}(S_{i}^{z})^2$ to make these bimagnon states as scars.
This example distills the essential insight that we use for the rest of this work: finding minimal sets of non-commuting local operators that host scar eigenstates as simultaneous eigenstates is already sufficient to characterize the states as scars, even though more work is needed to identify the exhaustive set of perturbations that preserve \textit{only} the scars and break all other symmetries~\cite{moudgalya2022exhaustive}.
\subsubsection{Beyond the bimagnon towers}
While the bimagnon tower is completely understood within this algebraic framework, the spin-$1$ $XY$ model hosts many additional exact eigenstates, such as the towers $|r,n\rangle$ introduced in Sec.~\ref{subsec: Towers of equally spaced FSCs}, that are not captured by the previously known algebra, i.e., they are not simultaneous eigenstates of the strictly local terms $h_{i,i+1}$ of Eq.~\eqref{eq: exchange_bonds}.
This motivates the search for an alternate algebraic characterization of these states.
In particular, one may ask if these states are simultaneous eigenstates of a different set of generators $\{\hat{h}_{\alpha}'\}$ such that the local algebra $\mathscr{A}'{=}\lgen \{\hat{h}_{\alpha}'\} \rgen$ would still contain $H_{XY}$, in which case the simultaneous eigenstates of the generators would be eigenstates of $H_{XY}$ as well.
While there are infinitely many choices of generators that one can consider, inspired by the bimagnon scars here we focus on generators $\{\hat{h}_{\alpha}'\}$ such that the Hamiltonian $H_{XY}$ can be simply realized as a direct sum $H_{XY}{=}\sum_{\alpha}\hat{h}_{\alpha}'$ (rather than as a complicated sums of products of them).
The simultaneous eigenstates obtained in this manner are guaranteed to be exact non-thermal eigenstates of $H_{XY}$~\cite{Garrison2018Does, Qi2019determininglocal,moudgalya2022exhaustive}. 
In the following subsection, we demonstrate how the simplest such choice—formed by periodically clustering the nearest-neighbor bonds $\{h_{i,i+1}\}$ into groups whose total sum reconstructs $H_{XY}$—naturally accounts for a large subset of the $|r,n\rangle$ states introduced earlier.
Slight modifications of this idea further yield additional families of exact eigenstates, most notably a volume-entangled tower and a set of mirror-dimer states, demonstrating the versatility of the commutant-inspired approach.
We summarize the examples and results in Table~\ref{Table: summary_of_all_scar_states}. 
Note that since these generators are no longer strictly local operators or uniform sums of such operators, this generalizes the constructions of bond algebras of Ref.~\cite{moudgalya2022exhaustive}. However, these generators are nevertheless sums of such strictly local operators; hence, all statements about their singlets not having a unique local Hamiltonian with the singlets as eigenstates remain valid, which, as we have discussed earlier, is sufficient to qualify these singlets as scars.
We also do not attempt to construct the complete algebra whose commutant contains \textit{only} the scar projectors, which would be a much more formidable task.
Instead, we explicitly build perturbations that break all simple symmetries of $H_0$ while leaving these states as exact eigenstates.
Given the form of such perturbations, it seems highly unlikely that there is any simple symmetry that would survive and explain the existence of these eigenstates. 
\subsection{Algebraic origin of the states $|r,n\rangle$} 
Having reviewed the commutant framework, we now turn to the explicit characterization of the tower states $|r,n\rangle$ introduced in Sec.~\ref{subsec: Towers of equally spaced FSCs}.
\subsubsection{Clustering approach}\label{subsubsec: clustering}
\begin{figure}
    \centering
\includegraphics[scale=0.57]{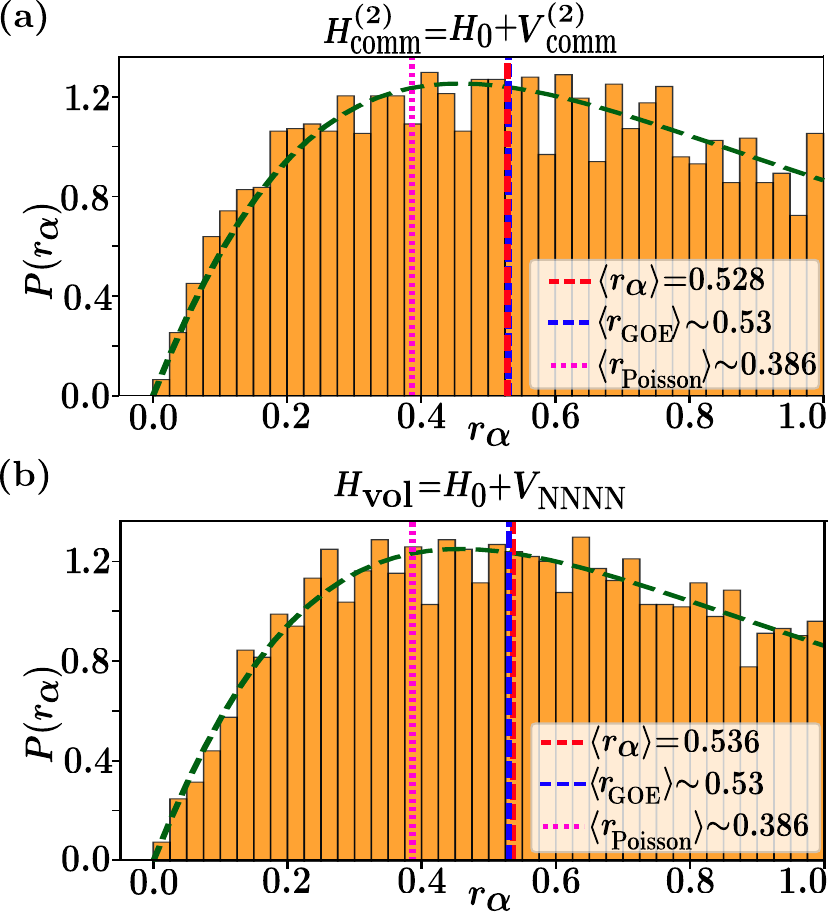}
    \caption{Level statistics showing adjacent-gap ratio distribution $P(r_{\alpha})$ for the Hamiltonian $H_{\rm scar}{=}H_{0}{+}H_{\rm pert}$. Calculations are done for a spin-$1$ chain of size $2L{=}14$ sites with periodic boundary conditions and $(J,h){=}(1,1)$ within the symmetry-resolved sector of magnetization $M{=}{-}10$, momentum $p{=}0$, and inversion parity $\mathcal{I}{=}1$ (dimension$ {=}9840$). Only the bulk (retaining eigenstates numbered, in ascending order, from 50 to 4300) of the spectrum is used, excluding extensive zero modes (a) with perturbation $V_{\rm comm}^{(2)}$~[see Eq.~\eqref{eq: V_comm_def}] using staggered periodic couplings $\epsilon_{i}{=}\epsilon_{i{+}2}{\in}[0,0.2]$, (b) with perturbation $V_{\rm NNNN}$ [see Eq.~\eqref{eq: V_NNNN}] using $J_{3}{=}0.2$. In both cases, the distribution agrees with the analytic prediction~\cite{Atas2013Distribution} from the Gaussian Orthogonal Ensemble of Random Matrix Theory (blue), with average $\langle r_{\alpha}\rangle{\approx}0.53$.
    }
    \label{fig: levelstat}
\end{figure}

Specifically, we define a set of $r$ periodic cluster operators as
\begin{equation}
    \mathcal{H}_{j}^{(r)}{=}\sum_{k{=}0}^{2L/r{-}1}h_{j{+}kr,j{+}kr{+}1},~~~j{=}1, \cdots, r,
    \label{eq: new_generators_of_local_algebra}
\end{equation}
where $r$ is a non-trivial divisor of the system size $2L$ (i.e. $2{\leq}r{\leq}L$), such that $H_{XY}{=}\sum_{j{=}1}^r\mathcal{H}_{j}^{(r)}$.
Each $\mathcal{H}_{j}^{(r)}$ is the sum of $2L/r$ non-overlapping exchange terms separated by $r$ sites, and the family of operators $\{\mathcal{H}_{j}^{(r)}\}$ provides a minimal non-commuting generator set whose direct sum yields $H_{XY}$.
We then search for simultaneous eigenstates of all $\mathcal{H}_{j}^{(r)}$.
Note that we excluded divisors $r{=}1$ and $r{=}2L$ since $r{=}1$ amounts to considering the entire Hamiltonian as single block while $r{=}2L$ involves considering each $h_{i, i{+}1}$, which simply yields the bimagnon towers as we explained previously [see Eqs.~\eqref{eq: exchange_bonds} and \eqref{eq: type_1 scar tower}]. 
Remarkably, for every allowed divisor $r$ we find that the tower states $\{|r,n\rangle\}$ of Eq.~\eqref{eq: new_towers} satisfy (see Appendix.~\ref{Appsec: annihilation_by_cluster})
\begin{equation}
\mathcal{H}_{j}^{(r)}|r,n\rangle{=}0~~~~~\forall j,~\forall n,
    \label{eq: sim_zero_energy_eigenstates}
\end{equation} 
i.e., they are common zero-energy eigenstates of all cluster operators, and therefore of $H_{XY}$.
In addition, the common kernel of $\{\mathcal{H}_{j}^{(r)}\}$ only contains the $|r,n\rangle$ tower (except for $r{=}L$, see Sec.~\ref{Sec: volume_tower}), aside from the already known bimagnon states $\ket{\mathbb{B}_n}$.
This property was confirmed numerically using a simultaneous block-diagonalization technique~\cite {Moudgalya2023numerical} up to system size $2L{=}10$.
We note, however, that being simultaneous eigenstates of such clustered operators does not by itself guarantee low entanglement; such connections can only be made when the underlying operators are strictly local~\cite{yao2022bounding}.
Understanding this structure also motivates the construction of perturbations that preserve the $\ket{r,n}$ states while apparently breaking all other symmetries of $H_{XY}$.
For instance, in the divisor $r$-case, a simple choice is a periodically modulated $XY$ interaction:
\begin{equation}
V_{\text{comm}}^{(r)}{=} \sum_{j{=}1}^{r} \epsilon_j \mathcal{H}_{j}^{(r)} = \sum_{j{=}1}^{2L} \epsilon_j \left( S_j^x S_{j+1}^x + S_j^y S_{j+1}^y \right),
\label{eq: V_comm_def}
\end{equation}
with arbitrary periodic coefficients $\epsilon_j{=}\epsilon_{j{+}r}$.
The states $\{|r,n\rangle\}$ still remain exact eigenstates of $H_{\rm comm}^{(r)}{=}H_{0}{+}V_{\text{comm}}^{(r)}$ with energy unchanged at $E_{r,n}{=}2h(n{+}1{-}L)$.
Meanwhile, the distribution of adjacent gap ratios $r_{\alpha}{=}\text{min}(\delta_{\alpha{+}1}/\delta_{\alpha},\delta_{\alpha}/\delta_{\alpha{+}1})$ with $\delta_{\alpha}{=}\mathcal{E}_{\alpha}{-}\mathcal{E}_{\alpha{-}1}$, where \{$\mathcal{E}_{\alpha}\}$ are eigenvalues of $H_{\rm comm}^{(r)}$, follows the Wigner-Dyson distribution associated with Gaussian Orthogonal Ensemble of Random Matrix Theory, confirming that the perturbed model is nonintegrable [see Fig.~\ref{fig: levelstat} (a)].
This same chaotic reference Hamiltonian was used in Fig.~\ref{fig: EE_towers} for the EE comparison, where the states $|r,n\rangle$ persist as robust low-entropy outliers in an otherwise thermal spectrum.

In summary, the periodic clustering of exchange bonds provides a minimal algebraic structure that exactly reproduces some of the $|r,n\rangle$ towers as singlets of a noncommuting operator family.
\subsubsection{Reverse-engineering approach}\label{subsubsec:reverse}
For a non-divisor $r$ ($2L~{\rm mod}~r{\neq}0$), we could not identify any simple clustered operator set whose sum reproduces $H_{XY}$ while simultaneously hosting $|r,n\rangle$ as eigenstates.
Nevertheless, to understand whether these cases still fall in the commutant framework, i.e., whether they are simultaneous eigenstates of non-commuting local operators, we follow a \textit{reverse-engineering} approach.
Starting from the states $|r,n\rangle$, we numerically search for local operators that host them as eigenstates using well-known methods~\cite{Qi2019determininglocal, chertkov2018computational, ODea2020tunnelstotowers, ren2021deformed, Moudgalya2023numerical}.
Interestingly, for all $r{>}1$ (including $2L~{\rm mod}~r{=}0$), we find strictly local annihilators that satisfy (see Appendix~\ref{Appsec: reverse engineer})
\begin{equation}
\begin{split}
     &q_{i,i{+}1}|r,n\rangle{=}0~~~\forall r{>}1,\forall i,~
     \forall n,~~~~~\text{where}\\&q_{i,i{+}1}{=}(|0,0\rangle\langle1,{-}1|{+}|0,0\rangle\langle{-}1,1|{+}\text{H.c.})_{i,i{+}1},
\end{split}
\label{eq: generator_generic_r}
\end{equation}
where H.c. denotes Hermitian conjugate. For $r{=}1$ an analogous structure holds for a next-to-next nearest neighbor operator set: $q_{i,i{+}3}|1,n\rangle{=}0~~\forall i,n$.
The existence of these local annihilators suffices to account for the fact that these are scars.

\begin{figure*}[t]
    \centering
    \includegraphics[width=\textwidth]{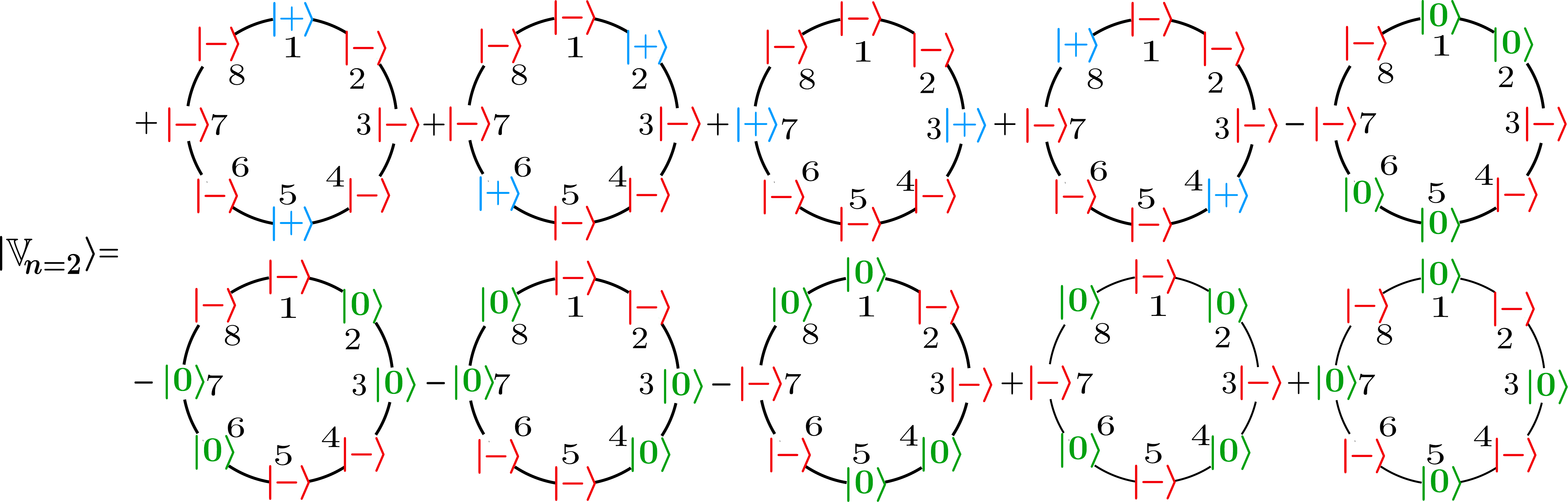}
    \caption{Illustration of a volume-law tower state $|\mathbb{V}_2\rangle$ [see Eq.~\eqref{eq: volume_scar_tower}] for a chain of size $2L{=}8$ sites. For ease of notation, the local basis states $|{-}1\rangle, |0\rangle, |1\rangle$ are simply denoted as $|{-}\rangle, |0\rangle, |{+}\rangle$. The state is a coherent superposition of configurations with two momentum $\pi$ bimagnons either located on distinct antipodal pairs or doubly occupying a single antipodal pair.}
    \label{fig: volume_scar_example}
\end{figure*}

An immediate advantage of this approach is that it naturally identifies a large class of perturbations that preserve the states $|r,n\rangle$.
Introducing such perturbations is essential for realizing these states as true QMBS in a non-integrable setting, since the parent Hamiltonian $H_{0}$ has many additional symmetries.
Since any operator belonging to the algebra $\lgen \{q_{i, i{+}1}\}\rgen$ (for $r{>}1$) or $\lgen \{q_{i, i{+}3}\}\rgen$ (for $r{=}1$) have the states $|r,n\rangle$ as eigenstates, such operators can be used directly as admissible perturbations $V_{\rm pert}$ that preserve the scars.
These range from simple nearest-neighbor sums $\sum_i q_{i,i+1}$ to multi-site constructions such as $\sum_{i} q_{i,i+1}q_{i+2,i+3}$ built from products of the local annihilators.
Thus, this reverse engineering approach provides an alternative way to recognize that the QMBS are simultaneous eigenstates of non-commuting local operators, which in turn enables the construction of a wider class of scar-preserving perturbations. In addition, identifying local operators that annihilate these states might provide a way to understand these in the Shiraishi-Mori framework of QMBS~\cite{shiraishi2017systematic}, the exploration of which we leave for future work.
\subsection{Additional family of scars: volume-entangled states}
\label{Sec: volume_tower}
In this and the next subsection, we identify two algebraic constructions that yield new families of scars: a volume-entangled tower and a set of mirror-dimer states by naturally extending the commutant algebra construction presented above to search for new states beyond the $|r,n\rangle$ towers.
Motivated by the periodic bond clustering of Hamiltonian terms that led to a characterization of many of the $\ket{r,n}$ states in Sec.~\ref{subsubsec: clustering}, we adopt alternative but symmetry-related combinations of the Hamiltonian terms to identify two further families of exact zero-energy eigenstates of $H_{XY}$: a \textit{volume-entangled tower}, which we describe in this section, and a set of \textit{mirror-dimer states}, which we describe in the next section.
\subsubsection{Structure of the states}
We first identify a distinct set of exact volume-entangled zero-energy eigenstates of $H_{XY}$, which appear when the half-chain length $L$ is even.
These states emerge as simultaneous zero modes of the set of generators obtained by pairing antipodal exchange terms separated by half the system size, i.e., 
\begin{equation}
\mathcal{H}_{j}^{(L)}{=}h_{j,j{+}1}{+}h_{j{+}L,j{+}L{+}1},\;\;j{=}1, {\cdots}, L
\end{equation}
such that $H_{XY}{=}\sum_{j=1}^{L}\mathcal{H}_{j}^{(L)}$.
The common kernel of these operators contains not only the previously discussed $|L,n\rangle$ states but also an additional family of (unnormalized) states denoted $|\mathbb{V}_{n}\rangle$ (see Appendix~\ref{Appsec: Annihilation of volume tower}):
\begin{equation}
   |\mathbb{V}_{n}\rangle{=} \sum_{1{\leq}i_1{\leq}{\cdots}{\leq}i_{n}{\leq}L} ({-}1)^{\sum_{k{=}1}^{n} i_k} \prod_{k{=}1}^{n}S_{i_k}^+S_{i_k{+}L}^{+}|\Omega\rangle,
   \label{eq: volume_scar_tower}
\end{equation}
where $n{=}0,1,{\cdots},2L$.
Note that although it may appear that these states could be generated by repeatedly applying a single operator such as $Q^{\dagger}{=}\sum_{i = 1}^L{(-1)^i S^+_i S^+_{i+L}}$, this does \textit{not} reproduce the required equal-weight structure of $|\mathbb{V}_{n}\rangle$-$(Q^{\dagger})^n$ produces configuration–dependent combinatorial coefficients. A short illustrative example to elucidate this is provided in Appendix~\ref{Appsec: volumelaw_example}.
Each $|\mathbb{V}_n\rangle$ corresponds to a configuration with $n$ ``antipodal bimagnons," dispersing/rotating with momentum $\pi$ that is obtained by acting pairs of spin-raising operators at sites $i$ and $i{+}L$ on the fully polarized vacuum $|\Omega\rangle$ [see Fig.~\ref{fig: volume_scar_example} for a schematic for $|\mathbb{V}_{n{=}2}\rangle$ for $2L{=}8$].
Note that, unlike the previously studied ``bond bimagnon" scars [Eq.~\eqref{eq: type_2 scar tower}] in Refs.~\cite {schecter2019weak,chattopadhyay2020quantum}, where bimagnons are maximally local since the magnons reside on neighboring sites, here they are \textit{maximally nonlocal} since the magnons are $L$ sites apart. 
As these states only involve even numbers of excitations, their total magnetization is $M_{n}{=}{-}2(L{-}n)$.
Thus, they are eigenstates of the full Hamiltonian $H_{0}$ with eigenvalues $E_{n}{=}{-}2h(L{-}n)$.
As before, we can directly see that the simple perturbation $V_{\rm comm}^{(L)}{=}\sum_{j{=}1}^{L} \epsilon_j \mathcal{H}_{j}^{(L)}$ (with $\epsilon_j{=}\epsilon_{j{+}L}$) annihilates all $|\mathbb{V}_{n}\rangle$.
In addition, we find that the translationally invariant next-to-next-nearest-neighbor (NNNN) $XY$ Hamiltonian considered in Ref.~\cite{schecter2019weak}:
\begin{equation}
    V_{\rm NNNN}{=}J_{3}\sum_{i=1}^{2L} ( S^x_i S^x_{i+3} + S^y_i S^y_{i+3}),
    \label{eq: V_NNNN}
\end{equation}
with arbitrary coupling $J_{3}$, also preserves the entire tower $|\mathbb{V}_{n}\rangle$.
Thus, the resulting total Hamiltonian 
\begin{equation}
\label{eq: Hamiltonian_for_volume_law_tower}
    H_{\rm vol}{=}H_{0}{+}V_{\rm NNNN}
\end{equation}
admits $|\mathbb{V}_{n}\rangle$ as a tower of exact eigenstates, with unchanged energies $E_{n}{=}{-}2h(L{-}n)$.
Moreover, the level-spacing statistics of $H_{\rm vol}$ follow the Wigner–Dyson distribution as shown in Fig.~\ref{fig: levelstat}(b), confirming that the model is non-integrable. 
\subsubsection{Entanglement properties}
\begin{figure}
    \centering
\includegraphics[scale=0.50]{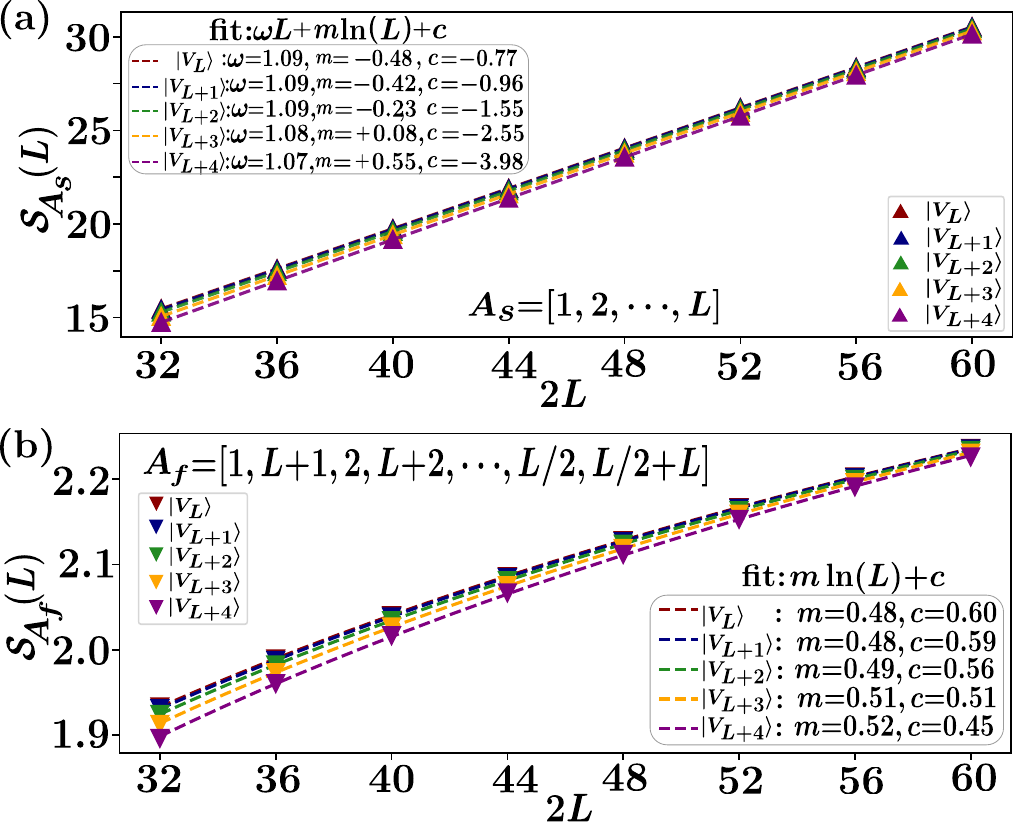}
    \caption{Scaling of bipartite entanglement entropy (EE) for the scars of the tower $|\mathbb{V}_{n}\rangle$, constructed as eigenstates of the cluster operators $\mathcal{H}_{j}^{(L)}$ for even $L$ evaluated using Eqs.~\eqref{eq: EE_volumetower_standardpartition} and \eqref{eq: EE_volumetower_athermalpartition}. (a) Under the standard bipartition $A_{s}{=}\{1, 2, {\cdots}, L\}$, the EE exhibits volume-law scaling. The dotted lines show fits of the form $\mathcal{S}_{A_s}(L){=}{\omega}L{+}m\ln(L){+}c$ (with the corresponding fit parameters listed in the legend), consistent with the large $L$ limit given in Eq.~\eqref{eq: EE_volumetower_standardpartition2}. (b) For the fine-tuned bipartition $A_{f}{=}\{1, L{+}1, 2, L{+}2, {\cdots}, L/2, L/2{+}L\}$, i.e., consisting of $L/2$ pairs of antipodal sites, the EE grows at most logarithmically with system size. The dotted lines show fits $\mathcal{S}_{A_f}(L){=}m\ln(L){+}c$, consistent with the large $L$ limit given in Eq.~\eqref{eq: EE_volumetower_athermalpartition2}. Data correspond to states $n{=}L, L{+}1,{\cdots}, L{+}4$.
    }
\label{fig: EE_scaling_volume_tower}
\end{figure}

The entanglement properties of $|\mathbb{V}_{n}\rangle$ are markedly different from those of other scar states discussed in this work.
They depend sensitively on the choice of the bipartition as illustrated in Fig.~\ref{fig: EE_towers}, where we show the bipartite EE of all $|\mathbb{V}_{n}\rangle$ states for a $2L{=}12$-site chain under a standard bipartition, where subsystem $A_{s}{=}\{1,2,{\cdots}, L\}$ consists of $L$ contiguous sites, and a fine-tuned antipodal bipartition, where subsystem $A_{f}{=}\{1, L{+}1, 2, L{+}2, {\cdots}, L/2, L/2{+}L\}$ also comprising $L$ sites consists of $L/2$ pairs of antipodal sites. Under the fine-tuned bipartition, $|\mathbb{V}_{n}\rangle$ has sub-extensive (typical scar-like) entanglement, while under the standard bipartition, $|\mathbb{V}_{n}\rangle$ (specifically mid-spectrum states of the tower) exhibits substantially higher EE than other exact scars, placing them close to the thermal continuum.  
The simplicity of the $|\mathbb{V}_{n}\rangle$ states [Eq.~\eqref{eq: volume_scar_tower}] allows for an analytical calculation of their EE (see App.~\ref{Appsec: EE_of_volume-entangled tower}).
For a standard half-chain bipartition $A_s$, the EE of $|\mathbb{V}_{n}\rangle$ is
\begin{equation}
    \mathcal{S}_{A_s}(n){=}\ln (\mathcal{D}_{L}^{M_{n}/2}),
    \label{eq: EE_volumetower_standardpartition}
\end{equation}
where $\mathcal{D}_{L}^{M}$ denotes the dimension of the spin-$1$ Hilbert space of $L$ sites with magnetization $M$ [see Eq.~\eqref{eq: U(1) sector dimension}].
Using this, we compute the EE for several representative states in the tower, specifically mid-spectrum states with $n{=}L, L{+}1, {\cdots}, L{+}4$ that reside in the largest magnetization sectors $M_{n}{=}0,2,{\cdots},8$.
These are the central states of the tower and exhibit the highest entanglement in the family.
In Fig.~\ref{fig: EE_scaling_volume_tower}(a), we plot the standard-cut EE of these states as a function of system size $2L$.
Numerical fits reveal that the EE grows linearly with system size, consistent with a volume-law–type growth but with a logarithmic correction. In fact, one can show analytically, e.g., for the mid-spectrum state $|\mathbb{V}_{n{=}L}\rangle$ (which exhibits the maximal entanglement within the tower), EE takes the asymptotic form
\begin{equation}
    \mathcal{S}_{A_s}(L)\underset{\text{large }L}{\longrightarrow}L\ln(3){-}\frac{1}{2} \ln (L) {+} {\rm constant.}
    \label{eq: EE_volumetower_standardpartition2}
\end{equation}
Notably, this scaling matches the symmetry-resolved Page-curve behavior expected for random states in the largest $U(1)$-conserving sector~\cite{Murciano2022Symmetry, lau2022page}.
In sharp contrast, for the fine-tuned bipartition $A_{f}$, we find (see App.~\ref{Appsec: EE_of_volume-entangled tower}) the EE of $|\mathbb{V}_{n}\rangle$ takes the form:
\begin{equation}
\begin{split}
    &\mathcal{S}_{A_{f}}(n){=}-\sum_{k{=}K_{\rm min}}^{K_{\rm max}}\lambda_{k}^{2} \ln \lambda_{k}^{2},\\
    &{\rm where}~~~~ \lambda_{k}{=}\frac{\sqrt{\mathcal{D}_{L/2}^{k}}\sqrt{\mathcal{D}_{L/2}^{M_n/2-k}}}{\sqrt{\mathcal{D}_{L}^{M_n/2}}}~~~{\rm and}\\&
    K_{\rm min}{=}{\rm max}\left(\frac{-L}{2},\frac{M_{n}{-}L}{2}\right),~~K_{\rm max}{=}{\rm min}\left(\frac{L}{2},\frac{M_{n}{+}L}{2}\right).
\end{split}
    \label{eq: EE_volumetower_athermalpartition}
\end{equation}
We evaluate the fine-cut EE using the same set of states with  $n{=}L, L{+}1,{\cdots}, L{+}4$. Unlike the standard bipartition—where the EE grows visibly with system size—the fine-cut EE changes only very weakly as the system size is increased, indicating a strong suppression of entanglement for this partition [see Fig.~\ref{fig: EE_scaling_volume_tower} (b)]. Similarly, we find the asymptotic limit for the mid-state
$|\mathbb{V}_{n{=}L}\rangle$
\begin{equation}
    \mathcal{S}_{A_{f}}(L)\underset{\text{large }L}{\longrightarrow}\frac{1}{2}\ln(L) {+}{\rm constant},
    \label{eq: EE_volumetower_athermalpartition2}
\end{equation}
demonstrating that the fine-tuned cut obeys subextensive EE scaling that grows at most logarithmically with system size. The numerical fits shown in Fig.~\ref{fig: EE_scaling_volume_tower}(b) are fully consistent with this predicted logarithmic scaling.

Since entanglement entropy alone is not a definitive diagnostic of ETH satisfaction [i.e., if a state has volume-law entanglement entropy that does not necessarily imply that it respects ETH (but note that the contrapositive is true, i.e., if a state has area-law entanglement entropy, that does violate ETH)], the nonthermal character of these states can also be seen at the level of observable expectation values. For generic local observables, the expectation values in these states are identical to those of typical thermal states. However, for a class of structured nonlocal, but few-body observables with support only on pairs of antipodal sites such as $S^{z}_{i}S^{z}_{i+L}$, the expectation values remain anomalous, consistent with their low entanglement under fine-tuned bipartitions. This distinguishes these states from typical ETH-respecting thermal eigenstates, which usually have high entanglement across all bipartitions.

%\addACB{The volume-entangled tower demonstrates that entanglement entropy alone is not a definitive diagnostic of ETH violation since the states in this tower are still distinct from typical thermal states. Moreover, the expectation value of few-body local observables in these states is identical to that of typical thermal states. As with our previous work on similar volume-law entangled states~\cite{Mohapatra2025Exact}, owing to the antipodal dimer structure of these states, the expectation value of non-local, but few-body observables, like $\langle S^{z}_{i}S^{z}_{i+L} \rangle$, distinguishes these states from ETH-respecting typical thermal states.}
%
 
%
\subsubsection{Oscillations from a simple but volume-law entangled initial state}
\begin{figure}
    \centering
 \includegraphics[scale=0.25]{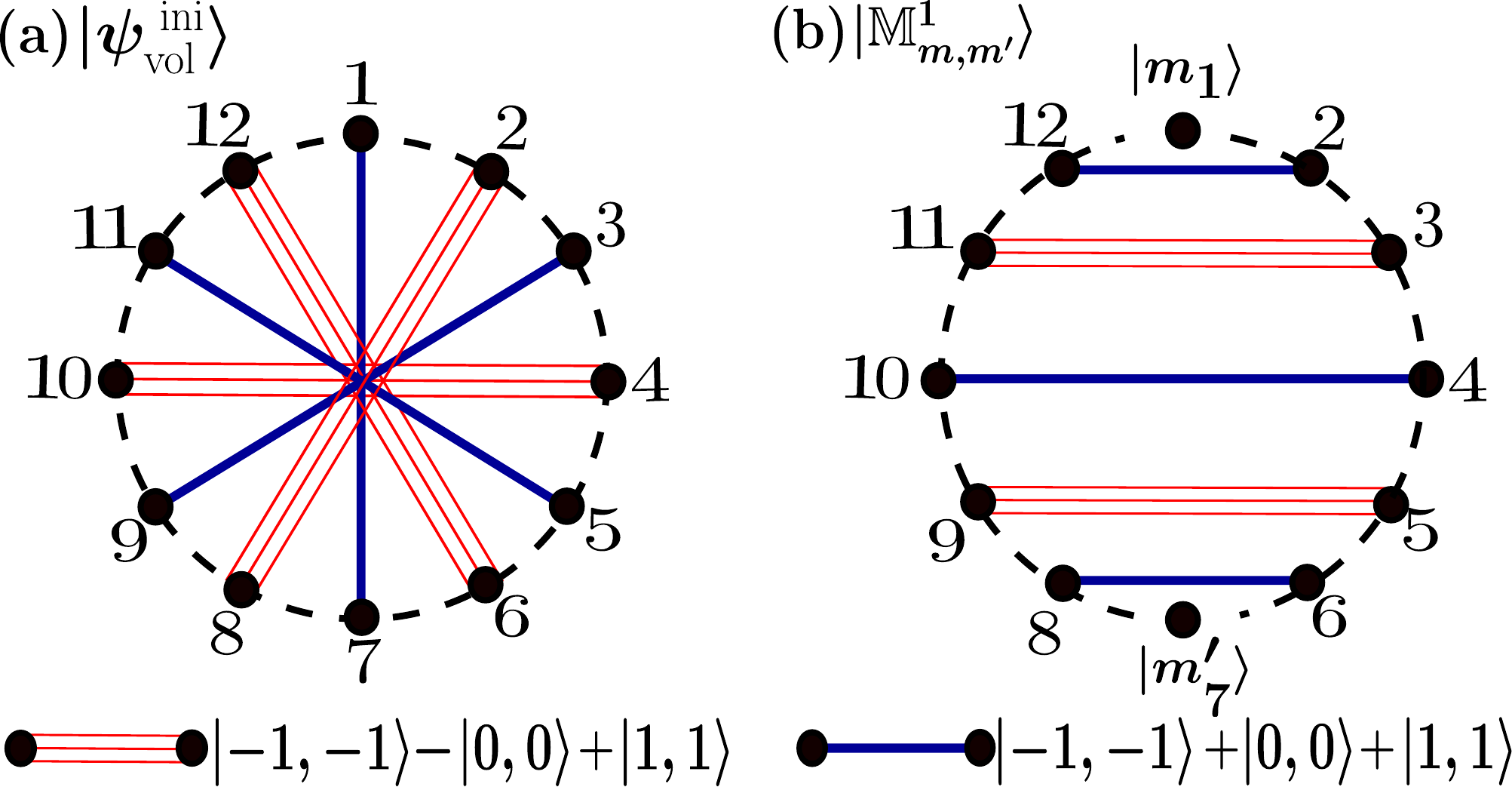}
    \caption{(a) The special initial state, $|\psi^{\rm init}_{\rm vol}\rangle{=}\sum_{n}|\mathbb{V}_{n}\rangle$, obtained by coherently superposing states in the volume-law tower $|\mathbb{V}_{n}\rangle$ defined in Eq.~\eqref{eq: volume_scar_tower}, for $2L{=}12$ sites, that exhibits long-lived fidelity oscillations [see Fig.~\ref{fig: fidelity}]. Its structure is an exact product of entangled dimers between antipodal sites, leading to volume-law entanglement under standard bipartition and vanishing entanglement under antipodal bipartition. (b) Mirror scar state $|\mathbb{M}_{m, m'}^k\rangle$ [see Eq.~\eqref{eq: mirror_dimer_state}] with $k{=}1$ for $2L{=}12$ sites, formed by pairing dimers symmetrically about a reflection axis passing through the sites $k$ and $k{+}L$. The two spins located on the symmetry axis, labeled by $m, m'$, can take any of the nine possible two-site spin-$1$ configurations.}
\label{fig: dimer_state}
\end{figure}

To probe the dynamical behavior associated with the special tower of Eq.~\eqref{eq: volume_scar_tower}, we consider the initial state $|\psi_{\rm vol}^{\rm init}\rangle{=}\sum_{n{=}0}^{2L}|\mathbb{V}_{n}\rangle$, formed by summing over all scar states in the tower. This state exhibits long-lived fidelity oscillations (see Fig.~\ref{fig: fidelity}) under time evolution with $H_{\rm vol}$ [defined in Eq.~\eqref{eq: Hamiltonian_for_volume_law_tower}], indicating robust nonthermal dynamics. Remarkably, $|\psi_{\rm vol}^{\rm init}\rangle$ can be written exactly as a product of maximally entangled dimers between antipodal site pairs [see Fig.~\ref{fig: dimer_state}(a)]:
\begin{equation}
\begin{split}   &|\psi_{\rm vol}^{\rm init}\rangle{=}\prod_{i{=}1}^{L}|\phi^{+}\rangle_{i,i{+}L}|\phi^{-}\rangle_{i{+}1,i{+}L{+}1},\\
    &{\rm where}~~~~~ |\phi^{\pm}\rangle_{i,j}{=}\left(|{-}1,{-}1\rangle{\pm}|0,0\rangle{+}|1,1\rangle \right)_{i,j}.
\end{split}
\label{eq: EAP_state}
\end{equation}
This structure is reminiscent of the entangled antipodal pair (EAP) eigenstates discussed in Refs.~\cite{Chiba2024Exact, Ivanov2025Volume, Mohapatra2025Exact}, though here it appears in a distinct context: not as an eigenstate, but as an initial state that supports long-lived oscillations.
This distinguishes the tower from previously studied scar families, where the initial states exhibiting oscillations are typically weakly entangled under all bipartitions.
In contrast, $|\psi_{\rm vol}^{\rm init}\rangle$ is highly entangled with $\mathcal{S}_{A_{s}}{=}L\ln(3)$ under the standard bipartition, but separable for the fine-tuned antipodal bipartition, where it becomes a simple product state.
This also directly explains the differences observed in the EE scaling of the constituent scar states $|\mathbb{V}_{n}\rangle$ under different bipartitions, which can be viewed as projections of $|\psi_{\rm vol}^{\rm init}\rangle$ onto fixed magnetization sectors.

\subsection{Additional family of scars: Mirror-dimer states with unconstrained spins}
We now turn to a final class of exact scar eigenstates discovered through the construction of simultaneous eigenstates of non-commuting clustered operators. These states possess a distinctive mirror-dimer structure and feature two unconstrained spins located along the reflection axis that remain completely free.
To construct them, we place the spin-$1$ chain of length $2L$ on a ring and choose a reflection axis passing through an antipodal pair of lattice sites $k$ and $k{+}L$. For each such axis, we define a family of mirror-symmetric operators
  \begin{equation}
      \mathcal{M}_{j}^{k}{=}h_{k-{j},k{-}j{+}1}{+}h_{k+{j},k{+}j{-}1},~j{\in}\{1,2,\cdots,L\},
      \label{eq: mirror_local}
  \end{equation}
where each term couples two nearest-neighbors that are mirror images with respect to the chosen reflection axis. Again the Hamiltonian $H_{XY}{\equiv}\sum_{j{=}1}^{L}\mathcal{M}_{j}^{k}$.
We find that $H_{XY}$ possesses a set of simultaneous zero modes of all the $\mathcal{M}_{j}^{k}$, which we denote by $|\mathbb{M}_{m,m'}^{k}\rangle$ (see Appendix~\ref{Appsec: annihilation_mirror}):
\begin{equation}
|\mathbb{M}_{m,m'}^k\rangle{=}\left[\prod_{\substack{j{=}1}}^{L-2} |\phi^{-}\rangle_{k{-}j,k{+}j}|\phi^{+}\rangle_{k{-}j{-}1,k{+}j{+}1}  \right]|m,m'\rangle_{k,k{+}L},
\label{eq: mirror_dimer_state}
\end{equation}
where $m,m'{\in}\{-1,0,1\}$ and $|\phi^{\pm}\rangle$ are maximally entangled spin-$1$ dimers defined previously in Eq.~\eqref{eq: EAP_state}.
Each state $|\mathbb{M}_{m, m'}^k\rangle$ therefore comprises a fully dimerized background that is symmetric about the reflection axis passing through $k$ and $k{+}L$, along with two freely configurable spins at the ends of this axis in the state $|m,m'\rangle$, which can take any of the nine possible two-site spin-$1$ configurations. A schematic of $|\mathbb{M}_{m, m'}^1\rangle$ is shown in Fig.~\ref{fig: dimer_state}(b). The total magnetization of the state $|\mathbb{M}_{m, m'}^k\rangle$ is $M_{m, m'}{=}m{+}m'$, allowing these states to appear in both even- and odd-magnetization sectors.
This does not contradict the earlier chiral-symmetry arguments, which showed that the number of zero-energy eigenstates is lower bounded by zero in the odd-magnetization sectors [see Eq.~\eqref{eq: Z_M_in_trinomial_expansion}]; states can certainly exist even if this lower bound is zero.
A representative state $|\mathbb{M}_{m,m'}^{1}\rangle$ for $2L{=}12$ is shown in Fig.~\ref{fig: dimer_state}(b). 
Furthermore, these mirror-dimer states $|\mathbb{M}_{m,m'}^k\rangle$ are also annihilated by the NNNN $XY$ interaction term $V_{\rm NNNN}$ defined in Eq.~\eqref{eq: V_NNNN}, making them exact eigenstates of a fully non-integrable Hamiltonian $H_{\rm mirror}{=}H_{0}{+}V_{\rm NNNN}$ with energy determined by magnetization as $E_{m,m'}^{k}{=}h(m{+}m')$.
The entanglement properties of these states depend strongly on the bipartition, as expected from their structure.
If the subsystem lies entirely on one side of the reflection axis, i.e., contains exactly one site from each entangled dimer, the EE grows linearly with the subsystem size, displaying volume-law scaling as each dimer cut by the bipartition contributes a fixed entropy of $\ln(3)$. In contrast, if the subsystem includes only mirror-reflected site pairs, i.e., both sites of the dimers reside in the same subsystem, the state factorizes, and its entanglement entropy vanishes.

Thus, generically, these states are non-thermal. A particularly striking feature of these states is the complete freedom of the two spins located at the reflection center. The choice of $m$ and $m'$ in Eq.~\eqref{eq: mirror_dimer_state} at sites $k$ and $k{+}L$ is arbitrary, and any configuration leaves the full state a zero-energy eigenstate of $H_{XY}$. Consequently, they are immune to local decoherence or perturbations that act only on these central spins at $k$ and $k{+}L$. It would be interesting to explore whether this unusual freedom could have practical implications.

\section{Conclusion and outlook}
\label{sec: conclusion and outlook}
In this work, we have revisited the question of exact QMBS in the spin-$1$ $XY$ chain, and focused on those residing within its extensively degenerate zero-energy manifolds that are annihilated by the $H_{XY}$ part of the Hamiltonian.
First, we traced the origin of these extensive degeneracies to the joint action of global $U(1)$ magnetization conservation and a chiral symmetry, which together enforce a bipartite structure on the model's Fock-space connectivity graph (Sec.~\ref{sec: model and extensive zero modes}).
By employing generating function methods, we quantified this degeneracy analytically and demonstrated that it grows exponentially with system size, providing a natural platform for constructing exact nonthermal eigenstates.

Partly motivated by recent works on identifying non-thermal eigenstates from quantum interference effects in Fock space, i.e., so-called Fock-Space Cage (FSC) states~\cite{Jonay2025FockSpaceCages, BenAmi2025ManyBodyCages, Tan2025InterferenceCaged}, we studied the Fock-space graph of this model to uncover an entire hierarchy of FSC eigenstates, parameterized by a positive integer separation $r$ (Sec.~\ref{sec: Fock_space_cages}), each realizing a distinct interference-protected cage structure within the zero-energy manifold.
We used this perspective to also show that previously known scars~\cite{schecter2019weak,chattopadhyay2020quantum} also admit a natural interpretation as FSC states, where destructive interference confines their support to closed loops in the Fock-space graph.
These novel quasiparticle towers of eigenstates we found are also isolated from the thermal continuum in entanglement and exhibit subextensive entanglement entropy, appearing as clear outliers well below the Page value.
Coherent superpositions of these ladder states, while not simple product states but possessing finite, low entanglement even in the thermodynamic limit, display long-lived fidelity oscillations, the defining dynamical hallmark of scarring. 
The lowest members of these families are simple-looking on the Fock-space graph and admit transparent geometric interpretations based on simple pairwise interference cancellations, which qualify them as FSC states.
However, for higher members, the interference patterns become increasingly complex and multipath in nature, making the geometric picture less intuitive and points toward a deeper organizing principle.
To uncover this hidden principle, we turned to ideas from the commutant algebra framework (Sec.~\ref{sec: exact eigenstates from commutant algebra formalism}), which motivates the understanding of QMBS as simultaneous eigenstates of multiple non-commuting local operators.

By systematically reorganizing the terms in the spin-$1$ $XY$ Hamiltonian into carefully chosen families of non-commuting operators—constructed from periodically clustered, antipodal, and mirror-symmetric combinations of exchange bonds—we identified the algebraic structure that leads to the existence of the FSC states. In fact, all the FSC states we identified can be reinterpreted in the commutant algebra framework as simultaneous eigenstates of some set of local operators (i.e., singlets of some bond algebras), but such singlets of bond algebras do not always lend themselves to a simple interpretation in terms of FSC states.
Using the commutant framework, we also uncovered two additional classes of exact eigenstates.
The first of these, arising from antipodal magnon pairings, displays a striking bipartition-dependent entanglement structure: while exhibiting volume-law scaling under standard cuts, they show sub-extensive EE under fine-tuned antipodal partitions.
The second family, consisting of mirror-dimer states, features reflection-symmetric dimer configurations with two unconstrained spins.
These unconstrained spins are completely free—any choice of their local configuration yields an exact zero-energy eigenstate of $H_{XY}$, which hints at their potential applications in quantum information processing.
Moreover, the mirror-dimer and volume-entangled states identified here consist of long-range entangled spin pairs, which are structurally similar to nonlocal Bell-pair states, for which a concrete protocol for realization in programmable quantum simulators has been recently proposed~\cite{Agarwal2023long}.
In addition, a scheme to realize volume-entangled exact scar states in the PXP model on near-term Rydberg quantum simulators was proposed in Ref.~\cite{Ivanov2025Volume}.
These works suggest that the mirror-dimer and volume-entangled states can potentially be prepared on such platforms, and it would be interesting to come up with concrete methods to realize them experimentally in future.

A concise summary of all exact scar families, both previously known and newly discovered, is presented in Table~\ref{Table: summary_of_all_scar_states}. The table lists their total magnetizations, minimal generating operator sets within the commutant algebra framework, and characteristic entanglement properties, serving as a unified map of how many kinds of nonthermal manifolds of states — from frustration-free bimagnons to highly entangled mirror-dimer states—emerge from common algebraic principles. 
In the future, it would be important to uncover other associated properties of these states, such as their stability under weak generic perturbations~\cite{lin2020slow, Li2025Dynamics}, the associated asymptotic quantum many-body scars~\cite{gotta2023asymptotic, moudgalya2023symmetries, kunimi2024proposal, ren2024quasi, kunimi2025systematic}, and potentially a complete classification of the distinct kinds of local parent Hamiltonians that could realize these as eigenstates~\cite{mark2020eta, ODea2020tunnelstotowers,  moudgalya2022exhaustive, gioia2025distinct}. Finally, it would also be interesting to explore whether there are yet additional QMBS hiding in the spin-$1$ $XY$ model that might be revealed as simultaneous eigenstates of other sets of non-commuting operators that we have not considered in this work.

Beyond the specific findings for the spin-$1$ $XY$ chain, this work underscores a broadly applicable methodology for uncovering exact scarred eigenstates in chaotic quantum systems.
The commutant algebra perspective in principle provides a versatile and model-independent route to constructing scar eigenstates as simultaneous eigenstates of non-commuting operator families, which can be searched using simple numerical methods~\cite{Moudgalya2023numerical}. While finding the appropriate set of non-commuting operators can be challenging, in this work, we have shown that simple structures can arise there too, e.g., by considering appropriate clusters of Hamiltonian terms. Indeed, it is also known that the scars of the spin-$1$ Affleck-Kennedy-Lieb-Tasaki model~\cite{AKLT_model_1987, moudgalya2018exact} can be understood as common eigenstates of non-commuting operators obtained from a clustering of alternate terms in its Hamiltonian~\cite{ODea2020tunnelstotowers, rozon2023broken}, which raises the question of whether studying such clusterings could form yet another organizational principle for the vast landscape of QMBS models and Hamiltonians.
In the future, it would be interesting to apply these ideas to a broader class of Hamiltonians, even going beyond one dimension, to see if they could prove to be a simple tool for identifying interesting QMBS.
By illustrating the interplay of geometric interference mechanisms with algebraic operator structure, this work highlights important methods for identifying, classifying, and engineering long-lived coherent dynamics in strongly interacting systems, and may prove instrumental in future efforts to design robust quantum states with long-lived coherence. 

\begin{acknowledgments}
We acknowledge useful discussions with Nicholas O'Dea, Frank Pollmann, Arnab Sen, and Maksym Serbyn. This work was undertaken on the Nandadevi supercomputer, which is maintained and supported by the High-Performance Computing Center at the Institute of Mathematical Sciences, India. S. Mohapatra is grateful to the School of Natural Sciences at the Technical University of Munich (TUM) for its hospitality in the summer of 2025. S. Moudgalya acknowledges support from the Munich Center for Quantum Science and Technology (MCQST) and the Deutsche Forschungsgemeinschaft (DFG, German Research Foundation) under Germany’s Excellence Strategy--EXC--2111--390814868.
\end{acknowledgments}

\appendix
\section{Coefficient of $x^{M}$ in the asymptotic expansion of $(x^2{+}1{+}x^{-2})^L$}
\label{Appsec: counting zero energy modes}
The coefficient of $x^M$, denoted $Z_M$, in the expansion of $(x^2+1+x^{-2})^L$ for large $L$ can be obtained by using the method of steepest descent on its integral representation. We first express $Z_M$ using Cauchy's integral formula as
\begin{equation}
    Z_M = \frac{1}{2\pi \iota} \oint_C \frac{(x^2+1+x^{-2})^L}{x^{M+1}} dx,
\end{equation}
where $\iota{=}\sqrt{{-}1}$. Choosing the unit circle, $x{=}e^{\iota\theta}$, as the contour of integration, the integral transforms into
\begin{equation}
    Z_M = \frac{1}{2\pi} \int_0^{2\pi} (1+2\cos(2\theta))^L e^{-\iota M\theta} d\theta.
\end{equation}
Assuming $M$ to be fixed, in the large-$L$ regime, the integral is dominated by the regions where the integrand's magnitude is at its maximum.
The term $(1{+}2\cos(2\theta))^L$ has maxima at $\theta{=}0$ and $\theta{=}\pi$, where it equals $3^L$.
We approximate the integral by rewriting the integrand as an exponential, $e^{f(\theta)}$, where $f(\theta){=} L\ln(1{+}2\cos(2\theta)) {-} \iota M\theta$, and expand the exponent around $\theta{=}0$ and $\theta{=}\pi$.
Near $\theta{=}0$, we Taylor expand $\cos(2\theta) {\approx} 1{-}2\theta^2$ to obtain the approximation
\begin{equation}
    f(\theta) \approx L\ln(3-4\theta^2) - \iota M\theta = L\ln (3) - \frac{4L}{3}\theta^2 - \iota M\theta.
\end{equation}
The truncation at the quadratic term is justified because the width of the steepest-descent region is $\Delta\theta{\sim}1/\sqrt{L}$. Over this region, the error introduced by the next leading order quartic term $O(L\theta^4)$ scales as $L(\Delta\theta)^4{\sim} 1/L$, which becomes negligible for large $L$. The contribution from the $\theta{=}0$ peak, denoted $I_0$, is then a Gaussian integral
\begin{equation}
   I_0 \approx \frac{3^L}{2\pi} \int_{-\infty}^{\infty} e^{-\iota M\theta - \frac{4L}{3}\theta^2} d\theta .
\end{equation}
We evaluate this integral by completing the square in the exponent, i.e., 
\begin{equation}
    -\iota M\theta - \frac{4L}{3}\theta^2 = -\frac{4L}{3}\left(\theta + \frac{3 \iota M}{8L}\right)^2 - \frac{3M^2}{16L}.
\end{equation}
Thus, the integral becomes
\begin{equation}
    I_0 \approx \frac{3^L}{2\pi} e^{-\frac{3M^2}{16L}} \int_{-\infty}^{\infty} \exp\left[-\frac{4L}{3}\left(\theta + \frac{3\iota M}{8L}\right)^2\right] d\theta.
\end{equation}
In fact, the Gaussian approximation is valid not just for a fixed $M$, but so long as $M{\ll}\sqrt{L}$. For $M{\ll}\sqrt{L}$, the shift in the saddle point, $\delta\theta{\sim}M/L$, due to the oscillatory phase $e^{\iota M\theta}$, is small compared to the width of the Gaussian, $\Delta\theta$. Specifically, the condition $\delta\theta{\ll}\Delta\theta$ translates to $M/L{\ll}1/\sqrt{L}{\implies}M{\ll}\sqrt{L}$. This guarantees that the Gaussian is the leading term in the expansion. Now using the Gaussian integral, $\int_{-\infty}^{\infty} e^{-ax^2} dx{=}\sqrt{\pi/a}$, the integral evaluates to $\sqrt{3\pi/(4L)}$. Thus, the contribution from the first peak at $\theta{=}0$ is
\begin{equation}
    I_0 \approx  \frac{3^{L+\frac{1}{2}}}{4\sqrt{\pi L}}e^{-\frac{3M^2}{16L}} .
\end{equation}
The contribution from the second peak at $\theta{=}\pi$, denoted $I_\pi$, can be related to $I_{0}$ by a change of variables $\theta{\to}\pi{-}\theta$. This leads to a factor of $({-}1)^M$ from the $e^{{-}\iota M\theta}$ term, resulting in $I_\pi{=}({-}1)^M I_0$. The total coefficient $Z_M$ is the sum of these two contributions, i.e.,
\begin{equation}
    Z_M = I_0+I_\pi \approx \frac{3^{L+\frac{1}{2}}}{4\sqrt{\pi L}}e^{-\frac{3M^2}{16L}}\left(1+(-1)^M \right).
\end{equation}
This expression is zero for odd-$M$.
For even-$M{\ll}\sqrt{L}$, the coefficient of $x^{M}$ in the asymptotic in $L$ expansion of $(x^2{+}1{+}x^{-2})^L$ is
\begin{equation}
    Z_{{\rm even}-M} \sim \frac{3^{L+\frac{1}{2}}}{2\sqrt{\pi L}}e^{-\frac{3M^2}{16L}}.
\end{equation}
\section{Additional details for Sec.~\ref{sec: Fock_space_cages} }
\label{Appsec: FSC}
Here we provide detailed derivations of a few results of Sec.~\ref{sec: Fock_space_cages} and present some additional figures. 
\subsection{Fock space cage structure of the bimagnon tower}
\label{Appsec: FSC_bimagnon}
Here we show that the bimagnon scar tower [Eq.~\eqref{eq: type_1 scar tower}] of the spin-$1$ $XY$ chain originally identified in Ref.~\cite{schecter2019weak} also admits a natural interpretation as FSC states: each tower state is a zero-energy eigenstate of $H_{XY}$ whose support forms a compact, interference-stabilized subgraph of the Fock space.
This comparison highlights that FSCs offer a unifying perspective on both previously known and newly identified scars.
\subsubsection{FSC graphs for low- and mid-tower bimagnon states}
\label{Appsec: FSC graphs for bimagnon states}
\begin{figure}
    \centering
    \includegraphics[scale=0.25]{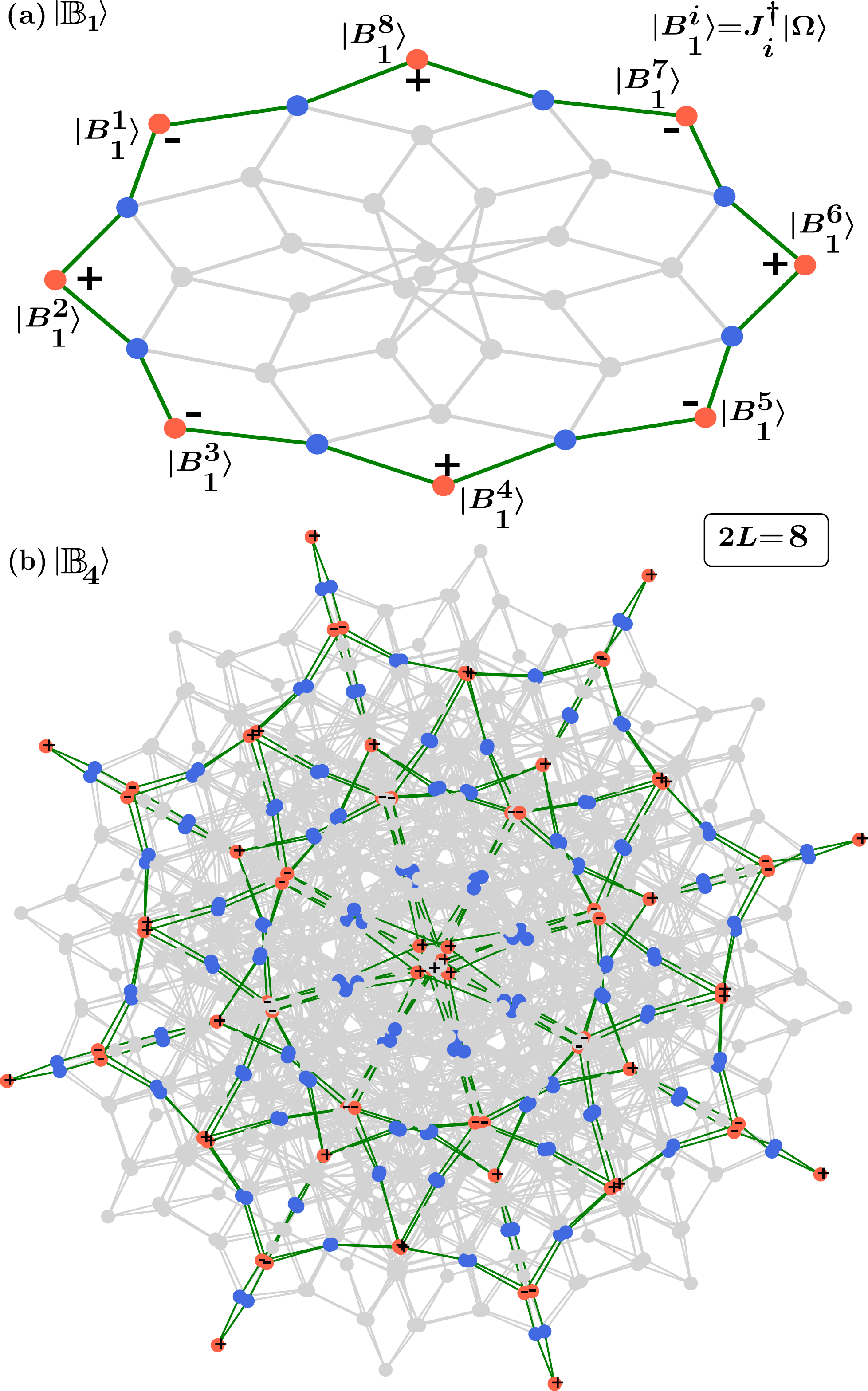}
    \caption{Fock-space cage structure for bimagnon states in a chain of size $2L=8$. (a) Simple cage for the lowest state $|\mathbb{B}_1\rangle$: the support nodes (orange) form a small set, and each connected node (blue) receives exactly two opposite-sign paths, producing a transparent two-path cancellation pattern. (b) Mid-tower state $|\mathbb{B}_4\rangle$: the active region expands, blue nodes receive many interfering contributions, and the subgraph becomes highly connected, illustrating how FSC structure grows increasingly intricate with $n$.}
    \label{Appfig: bimagnon_FSC_example}
\end{figure}

To illustrate the FSC structure concretely, Fig.~\ref{Appfig: bimagnon_FSC_example}(a) shows the Fock-space graph of the lowest nontrivial bimagnon state $|\mathbb{B}_1\rangle$ for a chain of size 
$2L{=}8$. Each connected node receives exactly two opposite-sign contributions from its neighbors, producing a clean two-path cancellation pattern characteristic of a simple, easily visualized cage.

In contrast, Fig.~\ref{Appfig: bimagnon_FSC_example}(b) displays a representative mid-tower state $|\mathbb{B}_4\rangle$ for the same system size.
Here, individual connected nodes typically receive many interfering contributions through multiple paths, and the active region becomes a densely connected subgraph. Thus, exactly as in the $|r,n\rangle$ towers discussed in Sec.~\ref{subsec: Towers of equally spaced FSCs}, the intuitive cage picture still applies, but it rapidly becomes visually opaque as one moves toward the middle of the tower and the interference network grows in complexity.

\subsubsection{FSC statistics for the bimagnon tower}
\label{Appsec: FSC statistics for the bimagnon tower}

To quantify this progression of complexity of the FSC, we compute the same diagnostics used in the main text (Sec.~\ref{Sec: Fock-space cage statistics}):
\begin{itemize}
    \item the number of support nodes $N_{\rm sup}$,
    \item the number of connected nodes $N_{\rm con}$,
    \item cancellation multiplicity $\mu{=}N_{\rm con}/N_{\rm sup}$, and
    \item the active fraction $\phi{=}(N_{\rm sup}{+}N_{\rm con})/\mathcal{D}_{2L}^{M}$
\end{itemize}
\begin{table}[t]
\centering
\setlength{\tabcolsep}{3pt} 
\renewcommand{\arraystretch}{1.05}
%\footnotesize <--- make BOTH panels small
% ------------ Panel (a) ------------
\begin{minipage}{0.55\columnwidth}
\centering
Table III (a):\; $2L{=}12$ \\[0.25em]
\begin{tabular}{c c c c c c}
\hline
$n$ & $N_{\text{sup}}$ & $N_{\text{con}}$ & $\mu$ & $\mathcal{D}_{2L}^{M}$ & $\phi$ \\
\hline
0,12 & 1   & 0    & 0    & 1    & 1 \\
1,11 & 12  & 12   & 1  & 78  & 0.31 \\
2,10  & 66  & 120  & 1.82 & 1221  & 0.15 \\
3,9  & 220 & 540  & 2.45 & 8074 & 0.09 \\
4,8  & 495 & 1440 & 2.9 & 28314 & 0.06 \\
5,7    & 792 & 2520 & 3.18 & 58278 & 0.057 \\
6    & 924 & 3024 & 3.27 & 73789 & 0.053 \\
\hline
\end{tabular}
\end{minipage}
\hfill
% ------------ Panel (b) ------------
\begin{minipage}{0.35\columnwidth}
\centering
Table III (b):\; $n{=}L$ \\[0.25em]
\begin{tabular}{c c c}
\hline
$2L$ & $\phi$ ($\mathbb{B}_{L}\rangle$) \\
\hline
6   & 0.398  \\
8   & 0.208 \\
10   & 0.106  \\
12  & 0.053 \\
14   & 0.026  \\
16  & 0.013 \\
18  & 0.006  \\
\hline
\end{tabular}
\end{minipage}

\vspace{0.6em}
\caption{FSC diagnostics for the bimagnon tower $|\mathbb{B}_{n}\rangle$: the number of support nodes $N_{\rm sup}$, connected nodes $N_{\rm con}$, cancellation multiplicity $\mu{=}N_{\rm con}/N_{\rm sup}$, and active fraction $\phi{=}(N_{\rm sup}+N_{\rm con})/\mathcal{D}_{2L}^{M}$. (a) reports for fixed system size $2L{=}12$ with increase in $n$, the active fraction $\phi$ remain small. (b) shows for fixed $n{=}L$, with an increase in system size, again $\phi$ decreases, strongly suggesting that it tends to zero in the thermodynamic limit. This demonstrates
that even the most complicated mid-spectrum states involve
interference processes confined to a vanishing fraction of the
Hilbert space, and can thus be regarded as complicated FSCs.}
\label{Apptab: bimagnon_FSC_quantification}
\end{table}
Table~\ref{Apptab: bimagnon_FSC_quantification} summarizes representative data for the bimagnon tower.
As one moves upward in the bimagnon tower [Table~\ref{Apptab: bimagnon_FSC_quantification}(a)], both $N_{\rm sup}$ and $N_{\rm con}$ grow rapidly, and the cancellation multiplicity $\mu$ increases steadily.
This mirrors the behavior observed for the $|r,n\rangle$ family [Table~\ref{tab: FSC_quantification}(a)] and reflects the proliferation of interfering paths required to maintain destructive cancellation at higher excitation numbers.
Despite this growth, the active fraction $\phi$ remains small for all $n$, indicating that only a tiny portion of the magnetization sector participates in the interference network. 
For the mid-tower state $|\mathbb{B}_{L}\rangle$, Table~\ref{Apptab: bimagnon_FSC_quantification}(b) shows that $\phi$ decreases rapidly with system size, vanishing in the thermodynamic limit. This behavior parallels the FSC statistics of the $|r,n\rangle$ states [Table~\ref{tab: FSC_quantification}(b)] and confirms that the bimagnon tower also realizes bona fide FSCs: localized in Fock space through coordinated sign cancellations across a sparse region of the Hilbert space.

Although both families realize increasingly “complicated’’ FSCs at higher excitation numbers $n$, the bimagnon tower is comparatively simpler. For the same magnetization sector, the cancellation multiplicity $\mu$ for $|r,n\rangle$ states is significantly larger (see Table~\ref{tab: FSC_quantification}), and their active fractions are higher at comparable excitation numbers. Thus, while the bimagnon scars fit cleanly within the FSC paradigm, the $|r,n\rangle$ towers exhibit richer and more highly connected interference structures—one of the key motivations for introducing the commutant algebra perspective in the main text. 

Similar statistics can be found for the ``bond-bimagnon" tower of Eq.~\eqref{eq: type_2 scar tower} as briefly summarized in Table~\ref{Apptab: bondbimagnon_FSC_quantification}.

\begin{table}[t]
\centering
\setlength{\tabcolsep}{3pt} 
\renewcommand{\arraystretch}{1.05}
%\footnotesize <--- make BOTH panels small
% ------------ Panel (a) ------------
\centering
Table IV:\; $2L{=}12$ \\[0.25em]
\begin{tabular}{c c c c c c}
\hline
$n$ & $N_{\text{sup}}$ & $N_{\text{con}}$ & $\mu$ & $\mathcal{D}_{2L}^{M}$ & $\phi$ \\
\hline
0,12 & 1   & 0    & 0   & 1    & 1 \\
1,11 & 12  & 24   & 2  & 78  & 0.46 \\
2,10  & 66  & 240  & 3.63 & 1221  & 0.25 \\
3,9  & 220 & 1080  & 4.9 & 8074 & 0.16 \\
4,8  & 495 & 2880 & 5.82 & 28314 & 0.12 \\
5,7    & 792 & 5010 & 6.32 & 58278 & 0.01 \\
6    & 923 & 5976 & 6.47 & 73789 & 0.09 \\
\hline
\end{tabular}
\caption{FSC diagnostics for the bond-bimagnon tower $|\mathbb{B}'_{n}\rangle$ of Eq.~\eqref{eq: type_2 scar tower} for system size $2L{=}12$.}
\label{Apptab: bondbimagnon_FSC_quantification}
\end{table}

\subsection{Entanglement entropy of $|\Omega_{r}\rangle$}
\label{Appsec: EE of Fock-space cages}
For convenience, we restate the normalized $|\Omega_{r}\rangle$ states:
\begin{equation} 
    |\Omega_{r}\rangle{=}\frac{1}{\sqrt{2L}}\sum_{i}^{2L}({-}1)^{i}|F_{r}^{i}\rangle~~\text{with}~~|F_{r}^{i}\rangle{=}S_{i}^{+}S_{i{+}r}^{+}|\Omega\rangle,
    \label{appeq: FSC_states}
\end{equation}
where $|\Omega\rangle{=}\bigotimes_{i}^{2L}|{-}1\rangle_{i}$. 
We now derive the bipartite EE of the FSC state $|\Omega_{r}\rangle$ with $r{\leq}L/2$ across a bipartition of the chain into two halves $A{=}\{1,{\cdots}, L\}$ and $B{=}\{L{+}1,{\cdots}, 2L\}$.
The derivation proceeds by writing the FSC state in Schmidt form. To do that, we decompose the sum over $i$ in Eq.~\eqref{appeq: FSC_states} into four disjoint regions depending on the locations of the magnons:
\begin{equation}
\begin{split}
     |\Omega_{r}\rangle{=}\underbrace{\frac{1}{\sqrt{2L}}\sum_{1{\leq}i{\leq}L{-}r}({-}1)^{i}S_{i}^{+}S_{i{+}r}^{+}|\Omega\rangle}_{{\rm T}_{1}} {+}\\
     \underbrace{\frac{1}{\sqrt{2L}}\sum_{L{-}r{<}i{\leq}L}({-}1)^{i}S_{i}^{+}S_{i{+}r}^{+}
     |\Omega\rangle}_{{\rm T}_{2}}{+}\\
     \underbrace{\frac{1}{\sqrt{2L}}\sum_{L{<}i{\leq}2L{-}r}({-}1)^{i}S_{i}^{+}S_{i{+}r}^{+}|\Omega\rangle}_{{\rm T}_{3}}{+}\\
     \underbrace{\frac{1}{\sqrt{2L}}\sum_{2L{-}r{<}i{\leq}2L}({-}1)^{i}S_{i}^{+}S_{i{+}r}^{+}|\Omega\rangle}_{{\rm T}_{4}},
\end{split}
\end{equation}
where ${\rm T}_{1}$ contains $L{-}r$ basis terms with both magnons located in subsystem $A$. Simlarly, ${\rm T}_{3}$ involves $L{-}r$ terms with magnons entrilely in $B$. Each cross terms ${\rm T}_{3}$ and ${\rm T}_{4}$ involves $r$ configurations where one magnon lies in $A$ and the other in $B$. Now simplifying ${\rm T}_{1}$:
\begin{equation}
{\rm T}_{1}{=}\frac{\sqrt{L-{r}}}{\sqrt{2L}}\underbrace{\sum_{1{\leq}i{\leq}L{-}r}\frac{({-}1)^{i}}{\sqrt{L{-}r}}S_{i}^{+}S_{i{+}r}^{+}|\Omega\rangle_{A}}_{|\alpha\rangle_{A}}
    {\otimes}|\Omega\rangle_{B}
%     &\implies {\rm T}_{1}{=}\frac{\sqrt{L-{r}}}{\sqrt{2L}}|\alpha\rangle_{A}{\otimes}|\Omega\rangle_{B},
%\end{split}
\label{appeq: T_1}
\end{equation}
where $|\alpha\rangle_{A}$ is a normalized state in subsystem $A$ and we have used $|\Omega\rangle{=}\bigotimes_{j{\in}A}|{-}1\rangle_{j}\bigotimes_{k{\in}B}|{-}1\rangle_{k}{=}   |\Omega\rangle_{A}{\otimes}|\Omega\rangle_{B}$. Similarly ${\rm T}_{3}$ can be simplified as:
\begin{equation}
%\begin{split}
    {\rm T}_{3}{=}\frac{\sqrt{L-{r}}}{\sqrt{2L}}
    |\Omega\rangle_{A}{\otimes}\underbrace{\sum_{L{<}i{\leq}2L{-}r}\frac{({-}1)^{i}}{\sqrt{L{-}r}}S_{i}^{+}S_{i{+}r}^{+}|\Omega\rangle_{B}}_{|\alpha\rangle_{B}}
     %&\implies {\rm T}_{3}{=}\frac{\sqrt{L-{r}}}{\sqrt{2L}}|\Omega\rangle_{A}{\otimes}|\alpha\rangle_{B}.
%\end{split}
\label{appeq: T_3}
\end{equation}
The cross-terms contribution becomes:
\begin{align}
    {\rm T}_{2}{=}\frac{1}{\sqrt{2L}}\sum_{L{-}r{<}i{\leq}L}\underbrace{({-}1)^{i}S_{i}  ^{+}|\Omega\rangle_{A}}_{|\beta_{i}\rangle_{A}} {\otimes} \underbrace{S_{i{+}r}^{+}
     |\Omega\rangle_{B}}_{|\gamma_{i}\rangle_{B}
     }\notag    \\
     %\implies {\rm T}_{2}{=}\frac{1}{\sqrt{2L}}\sum_{L{-}r{<}i{\leq}L}|\beta_{i}\rangle_{A}{\otimes}  |\gamma_{i}\rangle_{B}\notag    \\
     {\rm T}_{4}{=}\frac{1}{\sqrt{2L}}\sum_{2L{-}r{<}i{<}2L}\underbrace{S_{i{+}r}^{+}
     |\Omega\rangle_{A}}_{|\gamma_{i}\rangle_{A}} {\otimes} \underbrace{ ({-}1)^{i}S_{i}  ^{+}|\Omega\rangle_{B}}_{|\beta_{i}\rangle_{B}
     }%\notag    \\
     %\implies {\rm T}_{4}{=}\frac{1}{\sqrt{2L}}\sum_{L{-}r{<}i{\leq}L}|\gamma_{i}\rangle_{A}{\otimes}  |\beta_{i}\rangle_{B}
     \label{appeq: T_4}
\end{align}
Collecting all terms from Eqs.~\eqref{appeq: T_1}, \eqref{appeq: T_3} and \eqref{appeq: T_4} we write:
\begin{gather}
    |\Omega_{r}\rangle{=}\frac{\sqrt{L-{r}}}{\sqrt{2L}}|\alpha\rangle_{A}{\otimes}|\Omega\rangle_{B}{+}\frac{\sqrt{L-{r}}}{\sqrt{2L}}|\Omega\rangle_{A}{\otimes}|\alpha\rangle_{B}{+} \nonumber \\
    \frac{1}{\sqrt{2L}}\sum_{L{-}r{<}i{\leq}L}|\beta_{i}\rangle_{A}{\otimes}  |\gamma_{i}\rangle_{B}{+}\frac{1}{\sqrt{2L}}\sum_{L{-}r{<}i{\leq}L}|\gamma_{i}\rangle_{A}{\otimes}  |\beta_{i}\rangle_{B}.
\label{appeq: FSC_Schmidt_form}
\end{gather}
For $r{\leq}L/2$, the states $|\Omega\rangle, |\alpha\rangle, \{|\beta_{i}\rangle\}$ and $\{|\gamma_{i}\rangle\}$ are already mutually orthogonal and normalized. In contrast, additional orthogonalization would be necessary for $r{>}L/2$, which we do not pursue in this work.
Thus, Eq.~\eqref{appeq: FSC_Schmidt_form} represents the Schmidt form for $r{\leq}L/2$ with the nonzero Schmidt coefficients:
\begin{itemize}
    \item Two terms with coefficient $\lambda_{1}{=}\lambda_{2}{=}\sqrt{L-{r}}/\sqrt{2L}$
    \item $2r$ terms from the two sums) with coefficient $\lambda_{k}{=}1/\sqrt{2L}$ for $k{=}3,{\cdots}2{+}2r$.
\end{itemize}
The von Neumann EE is given by $\mathcal{S}_{L}^{r}{=} \sum_{k}\lambda_{k}^2\ln\lambda_{k}^2$.
Substituting the Schmidt coefficients, we get:
\begin{equation}
    \begin{split}
       \mathcal{S}_{L}^{r}{=} \frac{L{-}r}{L}\ln\left(\frac{2L}{L{-}r}\right){+}\frac{r}{L}\ln(2L) 
    \end{split}
\end{equation}
This completes the derivation of Eq.~\eqref{eq: EE_of_FSC} from the main text.
\subsection{EE of the tower states $|r,n\rangle$}
\label{Appsec: EE of the towers states}
For convenience, we restate the normalized version of the scar states $|r,n\rangle$:
\begin{equation}
\begin{split}
        |r,n\rangle{=}\frac{1}{\sqrt{\mathcal{N}}}\sum_{i=1}^{2L}&(-1)^{i(n{+}1)}|00\rangle_{i,i{+}r}{\otimes}\\&|\mathbb{B}_{n}\rangle_{i{+}1,\cdots,i{+}r{-}1,i{+}r{+}1,\cdots,i{-}1},\\
\end{split}
    \label{Appeq: newstates_interms_of_old}
\end{equation}
where
\begin{equation}
    \mathcal{N}{=}2L{\times}C_{n}^{2L{-}2}.
    \label{Appeq: normalization_new_towers}
\end{equation}
Here $C_{a}^{b}{=}\binom{b}{a}$ denotes the number of configurations with $a$ bimagnons in $b$ sites.
To compute the bipartite EE of a generic state $|r,n\rangle$ with $r{\leq}L/2$, we partition the chain into two equal halves $A{=}\{1,{\cdots}, L\}$ and $B{=}\{L{+}1,{\cdots}, 2L\}$.
Following the same approach as used above (in App.~\ref{Appsec: EE of Fock-space cages}) for the simple $n{=}0$ case, we divide the sum over $i$ in Eq.~\eqref{Appeq: newstates_interms_of_old} into four disjoint regions depending on whether the two spin-$0$ (magnons) sites of the pair $|0,0\rangle_{i, i{+}r}$ lie entirely within $A$, within $B$, or straddle the bipartition:
\begin{equation}
\begin{split}
     \sum_{i}{=}\underbrace{\sum_{1{\leq}i{\leq}L{-}r}}_{{\rm T}_{1}}{+}
     \underbrace{\sum_{L{-}r{<}i{\leq}L}
     }_{{\rm T}_{2}}{+}     \underbrace{\sum_{L{<}i{\leq}2L{-}r}}_{{\rm T}_{3}}{+}     \underbrace{\sum_{2L{-}r{<}i{\leq}2L}}_{{\rm T}_{4}},
\end{split}
\end{equation}
Here ${\rm T}_{1}$ (${\rm T}_{3}$) contains basis terms where both magnons lie entirely within subsystem $A$ ($B$), while ${\rm T}_{2}$ and ${\rm T}_{4}$ correspond to configurations where one magnon lies in $A$ and the other in $B$.\\
The first contribution,
\begin{equation}
\begin{split}
    {{\rm T}}_{1}{=}\sum_{1{\leq}i{\leq}L{-}r}\frac{1}{\sqrt{\mathcal{N}}}(-1)^{i(n{+}1)}|00\rangle_{i,i{+}r}{\otimes}\\|\mathbb{B}_{n}\rangle_{i{+}1,\cdots,i{+}r{-}1,i{+}r{+}1,\cdots,i{-}1},
\end{split}
\end{equation}
involves both spin-$0$ sites within subsystem $A$. We recall that the background state $|\mathbb{B}_n\rangle$ is a uniform superposition of all configurations containing $n$ bimagnons, which we can decompose into contributions with $k$-bimagnons located in $A$ (among the remaining $L{-}2$ sites) and $n{-}k$ in $B$:
\begin{eqnarray}
     |\mathbb{B}_n\rangle{=}\sum_{k{=}0}^{{\rm min}(n,L{-}2)}&&\sqrt{C_{k}^{L{-}2}C_{n-k}^{L}}\left(\frac{1}{\sqrt{C_{k}^{L{-}2}}}|\mathbb{B}_{k}\rangle_{A-[i,i{-}r]}\right)  \nonumber \\   
     &&{\otimes}\left(\frac{1}{\sqrt{C_{n-k}^{L}}}|\mathbb{B}_{k}\rangle_{B}\right),
\end{eqnarray}
where the states in the parenthesis $(\cdots)$.
Substituting this into ${\rm T}_{1}$, we obtain:
\begin{eqnarray}
 {\rm T}_{1}{=}\sum_{k{=}0}^{{\rm min}(n,L{-}2)}\lambda_{k}^{{\rm T}_{1}}
    &&\left(\frac{1}{\sqrt{(L{-}r)C_{k}^{L{-}2}}}|\alpha_{k}\rangle_{A}\right) \nonumber \\
    &&{\otimes}\left(\frac{1}{\sqrt{C_{n-k}^{L}}}|\mathbb{B}_{k}\rangle_{B}\right),
\end{eqnarray}
where states in the parenthesis $(\cdots)$ are normalized, and
\begin{align}
|\alpha_{k}\rangle_{A}&{=}\sum_{1{\leq}i{\leq}L{-}r}(-1)^{i(n{+}1)}|00\rangle_{i,i{+}r}|\mathbb{B}_{k}\rangle_{A-[i,i{-}r]}\nonumber \\
\lambda_{k}^{{\rm T}_{1}}&{=}\frac{\sqrt{(L{-}r) C_{k}^{L{-}2}C_{n-k}^{L}}}{\sqrt{\mathcal{N}}}.  
\end{align}
This expression represents the Schmidt decomposition of the term ${\rm T}_{1}$, and thus ${\rm T}_{1}$ contributes Schmidt coefficients $\lambda_{k}^{{\rm T}_{1}}$ for $k{=}0,{\cdots, }{\rm min}(n,L{-}2)$. 
By reflection symmetry, the term ${\rm T}_{3}$ gives an equivalent contribution, but with the roles of subsystems $A$ and $B$ interchanged.
Explicitly,
\begin{equation}
\begin{split}
    {\rm T}_{3}&{=}\sum_{L{<}i{\leq}2L{-}r}\frac{(-1)^{i(n{+}1)}}{\sqrt{\mathcal{N}}}|00\rangle_{i,i{+}r}{\otimes}|\mathbb{B}_{n}\rangle_{i{+}1,\cdots,i{-}1}\\
    &{=}\sum_{k{=}0}^{{\rm min}(n,L{-}2)}\lambda_{k}^{{\rm T}_{3}}
    \left(\frac{|\mathbb{B}_{k}\rangle_{A}}{\sqrt{C_{n-k}^{L}}}\right){\otimes}\left(\frac{|\alpha_{k}\rangle_{B}}{\sqrt{(L{-}r)C_{k}^{L-2}}}\right).
\end{split}
\end{equation}
Because both the number of terms and their weightings are identical to ${\rm T}_{1}$, the Schmidt coefficients are the same:
\begin{equation}
    \lambda_{k}^{{\rm T}_{3}}{=}\lambda_{k}^{{\rm T}_{1}},
\end{equation}
for $k{=}0,{\cdots}, {\rm min}(n,L{-}2)$.
The term ${\rm T}_2$ includes configurations where $|0\rangle_{i}$ lies in subsystem $A$ and its partner $|0\rangle_{i{+}r}$ lies in subsystem $B$:
\begin{align}
{\rm T}_2 =
\sum_{L{-}r<i\le L}
&\frac{(-1)^{i(n{+}1)}}{\sqrt{\mathcal N}}
|00\rangle_{i,i{+}r}
\otimes\nonumber \\
&|\mathbb B_n\rangle_{i{+}1,\cdots,i{+}r{-}1,i{+}r{+}1,\cdots,i{-}1}.
\end{align}
We expand the background state $|\mathbb{B}_n\rangle$ into configurations with $k$ bimagnons in $A$ and $n{-}k$ in $B$:
\begin{align}
    |\mathbb B_n\rangle{=}
\sum_{k=0}^{\min(n,L{-}1)}
&\sqrt{C_k^{L{-}1}C_{n-k}^{L{-}1}}
 \\
&\left(
\frac{|\mathbb B_k\rangle_{A-[i]}}{\sqrt{C_k^{L{-}1}}}
\right)
\otimes\left(
\frac{|\mathbb B_{n-k}\rangle_{B-[i{+}r]}}{\sqrt{C_{n-k}^{L{-}1}}}
\right) \nonumber,
\end{align}
where the states in the parenthesis $(\cdots)$ are normalized. Here, the available sites in $A$ and $B$ are each $L{-}1$, since one site in each subsystem is occupied by a $|0\rangle$.
Substituting this expansion into ${\rm T}_{2}$, we obtain
\begin{equation}
{\rm T}_2{=}\sum_{L{-}r<i\le L}
\sum_{k=0}^{\min(n,L{-}1)}
\lambda_k^{{\rm T}_2}
\left(
\frac{|\beta_k^i\rangle_A}{\sqrt{C_k^{L{-}1}}}
\right)
\otimes
\left(
\frac{|\gamma_{n-k}^i\rangle_B}{\sqrt{C_{n-k}^{L{-}1}}}
\right),
\label{Appeq: T2 term}
\end{equation}
where states in the parenthesis $(\cdots)$ are normalized, and we have
\begin{align}
    |\beta_k^i\rangle_A&{=}(-1)^{i(n{+}1)}|0\rangle_i
|\mathbb B_k\rangle_{A-[i]}, \nonumber\\
|\gamma_{n-k}^{i}\rangle_B&{=}|0\rangle_{i{+}r}
|\mathbb B_{n-k}\rangle_{B-[i{+}r]}\nonumber \\
\lambda_k^{{\rm T}_2} &{=}
\frac{\sqrt{C_k^{L{-}1}C_{n-k}^{L{-}1}}}{\sqrt{\mathcal N}}.
\end{align}
Each $i$ labels an orthogonal set of configurations $\{|\beta_k^i\rangle_A\}$ and $\{|\gamma_{n-k}^{i}\rangle_B\}$ for $r{\leq}L/2$, whereas additional orthogonalization would be necessary for $r > L/2$, which we do not pursue.
Thus, the term ${\rm T}_{2}$ yields $r$ degenerate Schmidt coefficients $\lambda_k^{{\rm T}_2}$ for each $k{=}0,{\cdots},\min(n,L{-}1)$.
Similarly, the term ${\rm T}_{4}$ contributes an identical set by symmetry. Collecting all contributions:
\begin{itemize}
    \item ${\rm T}_{1}$: coefficients $\lambda_{k}^{{\rm T}_1}$ for $k{=}0,{\cdots}, {\rm min}(n,L{-}2)$
    \item ${\rm T}_{2}$: $r$ degenerate $\lambda_{k}^{{\rm T}_2}$ for each $k{=}0,{\cdots},{\rm min}(n,L{-}1)$
    \item ${\rm T}_3$ same as ${\rm T}_{1}$
    \item ${\rm T}_{4}$ same as ${\rm T}_{2}$
\end{itemize}
Hence, the von Neumann EE of the state $|r,n\rangle$ for $r{\leq}L/2$ is given by
\begin{equation}
    \begin{split}
        &\mathcal{S}_{L}^{(r,n)}{=}{-}2\sum_{k{=}0}^{{\rm min}(n,L{-}2)}(\lambda_{k}^{{\rm T}_1})^2\ln (\lambda_{k}^{{\rm T}_1})^2 \\
        &{-}2r\sum_{k{=}0}^{{\rm min}(n,L{-}1)}(\lambda_{k}^{{\rm T}_2})^2\ln (\lambda_{k}^{{\rm T}_2})^2.       
    \end{split}
    \label{Appeq: B22}
\end{equation}
Note that the normalization condition is given by $\sum_{k{=}0}^{{\rm min}(n,L{-}2)}2(\lambda_{k}^{{\rm T}_1})^2{+}\sum_{k{=}0}^{{\rm min}(n,L{-}1)}2r(\lambda_{k}^{{\rm T}_2})^2{=}1$, which we can verify is satisfied.
\subsubsection{Large $L$ asymptotic limit}
\label{Appsec: Asymptotic limit highest EE state}
We now analyze the asymptotic behavior of the entanglement entropy in the tower $r{=}L/2$, where exactly half of the paired configurations in $|r,n\rangle$ straddle the bipartition, producing the largest bipartite entanglement among all towers at fixed $n$.
We focus on the mid-spectrum state with $n{=}L{-}1$, which lies in the magnetization sector $M_{r,n}{=}0$, and evaluate its entanglement entropy in the large $L$ limit.
The EE can be conveniently expressed through the normalized probabilities $p_k^{(1)} = 2(\lambda_k^{{\rm T}_1})^2$ and $p_k^{(2)} = 2r(\lambda_k^{{\rm T}_2})^2$, 
which satisfy $\sum_{k}{(p_k^{(1)}{+}p_k^{(2)})}{=}1$.
For $(r,n) {=} (L/2,L-1)$, the explicit forms are
\begin{equation}
    p_k^{(1)}{=}\frac{\binom{L{-}2}{k}\binom{L}{k+1}}{2\binom{2L{-}2}{L-1}},~~~
    p_k^{(2)}{=}\frac{\binom{L{-}1}{k}^2}{2\binom{2L{-}2}{L{-}1}}.
    \label{Appeq: B19}
\end{equation}
For large integers $a$, Stirling’s approximation gives the central-binomial form
\begin{equation}
    \binom{a}{b} \simeq \frac{2^{a}}{\sqrt{\pi a/2}}\; \exp\!\left[-\frac{2(b-a/2)^2}{a}\right]
    \label{Appeq: B20}
\end{equation}
which is valid near $b\simeq a/2$.
The exponent in Eq.~\eqref{Appeq: B20} describes a Gaussian envelope centered at $b{=}a/2$, with variance $\sigma_a^2{=}a/4$.
Applying this to each binomial coefficient in Eq.~\eqref{Appeq: B19}:
\begin{equation}
    \begin{split}
        \binom{L-2}{k}
&\simeq \frac{2^{L-2}}{\sqrt{\pi(L-2)/2}}
\exp\!\left[-\frac{2(k-(L-2)/2)^2}{L-2}\right],\\
\binom{L}{k+1}
&\simeq \frac{2^{L}}{\sqrt{\pi L/2}}
\exp\!\left[-\frac{2((k+1)-L/2)^2}{L}\right],\\
\binom{2L-2}{L-1}
&\simeq \frac{2^{2L-2}}{\sqrt{\pi (2L-2)/2}}.
    \end{split}
\end{equation}
Combining these for large $L$, 
\begin{equation}
    p_k^{(1)} \simeq \frac{1}{\sqrt{\pi L}} \exp\!\left[-\frac{4(k-L/2)^2}{L}\right],
\end{equation}
a Gaussian centered at $k{\simeq}L/2$ with variance $\sigma^2{=}L/4$. A similar expansion for $p_k^{(2)}$ gives
\begin{equation}
    p_k^{(2)} \simeq \frac{1}{\sqrt{\pi L}} \exp\!\left[-\frac{4(k-L/2)^2}{L}\right].
\end{equation}
 As a consistency check for this approximation, we verify the normalization in the thermodynamic limit by replacing the discrete sum over $k$ with a continuous integral
\begin{equation}
\begin{split}
    &\int p_k^{(i)}dk = \frac{1}{\sqrt{\pi L}} \int e^{-4x^2/L} dx = \frac{1}{2}\\
    &{\implies}\int (p_k^{(1)}{+}p_k^{(2)})dk = 1
\end{split}
\end{equation}
From Eq.~\eqref{Appeq: B22}, the total EE decomposes as
\begin{equation}
    \begin{split}
        &\mathcal{S}_{L}^{(r,n)}{=}\mathcal{S}_{{\rm T}_1}{+}\mathcal{S}_{{\rm T}_2},~~~~\mathcal{S}_{{\rm T}_i} = -\!\sum_k p_k^{(i)}\ln\!\frac{p_k^{(i)}}{c_i},
    \end{split}
    \label{Appeq: B25}
\end{equation}
with constants $c_1{=}2$ and $c_2{=}2r$. In the continuum limit, we then get
\begin{equation}
    \mathcal{S}_{{\rm T}_i} \approx -\!\int p^{(i)}(x)\ln p^{(i)}(x)\,dx + \tfrac{1}{2}\ln c_i.
    \label{Appeq: B26}
\end{equation}
Using the explicit Gaussian form of $p_k^{(i)}$ and the identity $\int e^{-4x^2/L}dx{=}\sqrt{\pi L}/2$, one directly obtains
\begin{equation}
    \begin{split}
       \int p^{(i)}\ln p^{(i)}dx {=} {-} \frac{1}{2}\ln(\pi L)-\frac{1}{2}. 
    \end{split}
    \label{Appeq: B27}
\end{equation}
Substituting this in Eq.~\eqref{Appeq: B26}, we get
\begin{equation}
    \begin{split}
        &\mathcal{S}_{{\rm T}_1} \approx \frac{1}{4}\ln L +\text{constant.},\\
        &\mathcal{S}_{{\rm T}_2} \approx  \frac{3}{4}\ln L +\text{constant.}.
    \end{split}
\end{equation}
Here we keep only the
$L$-dependent terms; the additive constants are absorbed into the generic ``constant” term. Adding both contributions gives the asymptotic behaviour of EE of $|r{=}L/2,n{=}L{-}1\rangle$ as
\begin{equation}
    \mathcal{S}_{L}^{(L/2,L-1)} \approx \ln\left(L\right) + \text{constant.}
    \label{Appeq: B33}
\end{equation}
Although we have not kept track of the overall additive constant, the coefficient of the logarithm is exact, as we can verify in Fig.~\ref{fig: EE_scaling_r_n} of the main text, plotting the exact EE obtained using Eq.~\eqref{Appeq: B22} against system size $2L$ on a log-scaled $x$-axis yields a straight line.
A fit of the form $\mathcal{S}_{L}^{(L/2,L-1)}{=}m \ln(L){+}c$ gives $m{=}1$, in precise agreement with the asymptotic prediction of Eq.~\eqref{Appeq: B33}.

\subsection{Restricted spectrum generating algebra of order-1 for $r{=}1$ tower}
\label{Appsec: RSGA_FSC}
The tower of states $\{|r{=}1,n\rangle\}$ admits a simple algebraic description.
We show that the root state $|\Omega_1\rangle$ of the tower satisfies the defining conditions of order-$1$ RSGA: 
\begin{equation}
\begin{aligned}
&\text{(i)}~~ H_{0} |\Omega_1\rangle {=} E_1 |\Omega_1\rangle, \\
&\text{(ii)}~~ [H_{0}, J^+] |\Omega_1\rangle {=} 2h J^+ |\Omega_1\rangle, \\
&\text{(iii)}~~ [[H_{0}, J^+], J^+] {=} 0,
\end{aligned}
\label{eq: RSGA_conditions}
\end{equation}
where $E_1$ is the energy of $|\Omega_1\rangle$ and  $J^{+}$ is defined in Eq.~\eqref{eq: type_1 scar tower}.
(i) $|\Omega_1\rangle$ is an eigenstate of $H_{0}$: As shown in Sec.~\ref{sec: Fock_space_cages}, the destructive interference mechanism ensures $H_{XY} |\Omega_1\rangle {=} 0$, and the Zeeman term acts diagonally $H_z |\Omega_1\rangle {=} 2h\, |\Omega_1\rangle$, since each configuration $S_i^+ S_{i+1}^+ |\Omega\rangle$ carries magnetization $M {=} {-}L {+} 2$, the total energy is $E_1 {=} 2h$, satisfying condition (i).

(ii) Commutator with $J^+$: Refs.~\cite{moudgalya2020eta, mark2020unified} showed that
\begin{equation}
\begin{split}
    [H_z, J^+] {=} &2hJ^{+},~~~~~~~[H_{XY}, J^+] {=}4J \sum_{i} (-1)^i f_{i,i{+}1}
\end{split}
\label{Appeq: FSC_RSGA_2}
\end{equation}
with $f_{i,j} {=} (|0,1\rangle\langle {-}1,0|{-}|1,0\rangle\langle 0,{-}1|)_{i,j}$ acting on two-site spin configurations.
Now observe that each term in $|\Omega_1\rangle$ consists of a single pair of $|0, 0\rangle$ embedded in a sea of $|{-}1\rangle$ states. The operators $h_{i,j}$ act nontrivially only on $|{-}1,0\rangle$ or $|0,{-}1\rangle$ configurations. Therefore,
\begin{align}
    &\sum_{i} (-1)^i f_{i,i{+}1}|\Omega_{1}\rangle{=} \notag\\  &\sum_{i} (-1)^{2i{-}1} f_{i,i{+}1} S_{i{-}1}^+ S_{i}^+|\Omega\rangle{+} \sum_{i}(-1)^{2i{+}1} f_{i,i{+}1} S_{i{+}1}^+ S_{i{+}2}^+|\Omega\rangle\notag\\ 
    &{=}\sum_{i} (-1)^{2i{-}1} {-}|{-}1,{\cdots},{-}1,0_{i{-}1},1_{i},0_{i{+}1},{-}1,{\cdots},{-}1\rangle{+}\notag\\& \sum_{i} (-1)^{2i{+}1} |{-}1,{\cdots},{-}1,0_{i},1_{i{+}1},0_{i{+}2},{-}1,{\cdots},{-}1\rangle
    \label{Appeq: FSC_RSGA_3}
\end{align}
Replacing $i$ by $i{+}1$ in the first sum of Eq.~\eqref{Appeq: FSC_RSGA_3}, we get $\sum_{i} (-1)^i f_{i,i{+}1}|\Omega_{1}\rangle{=}0$, implying
\begin{equation}
[H_{XY}, J^+] |\Omega_1\rangle {=} 0.
\label{Appeq: FSC_RSGA_4}
\end{equation}
Thus, from Eqs.~\eqref{Appeq: FSC_RSGA_2}
 and \eqref{Appeq: FSC_RSGA_4} we get 
\begin{equation}
[H_{0}, J^+] |\Omega_1\rangle {=} 2hJ^+ |\Omega_1\rangle,
\label{Appeq: FSC_RSGA_6}
\end{equation}
which satisfies condition (ii).

(iii) Nested commutator vanishes: 
Using Eq.~\eqref{Appeq: FSC_RSGA_2}, we get 
\begin{equation}
    [[H_{z},J^{+}],J^{+}]{=}2h[J^{+},J^{+}]{=}0
    \label{Appeq: FSC_RSGA_7}
\end{equation}
Now for the $H_{XY}$ term,
\begin{equation}
\begin{split}
     [H_{XY}, J^+]{=}& 4J\sum_{i}({-}1)^{i}[f_{i,i{+}1},J^{+}]\\
     {=}&[f_{i,i+1},(S_{i}^{+})^{2}] {-}[f_{i,i+1},(S_{i+1}^{+})^{2}]
\end{split}
\end{equation}
The operator $f_{i,i+1}$ acts non-trivially only on the two-site states $\{|0,{-}1\rangle,|{-}1,0\rangle\}$, while $(S_{i}^{+})^{2}$ or $(S_{i+1}^{+})^{2}$ map only the local $|{-}1\rangle$ to $|1\rangle$.
Consequently, $f_{i,i+1}(S_{i}^{+})^{2}$ and $f_{i,i+1}(S_{i+1}^{+})^{2}$ annihilate all two-site basis states.
Conversely, $f_{i,i+1}$ outputs only $\ket{0,1}$ or $\ket{1,0}$ from any of the two-site states, on which $(S_{i}^{+})^{2}$ or $(S_{i+1}^{+})^{2}$ act trivially. Hence $[f_{i,i+1},(S_{i}^{+})^{2}] {=} [f_{i,i+1},(S_{i+1}^{+})^{2}] {=} 0$, and therefore 
\begin{equation}
   [[H_{XY},J^{+}],J^{+}]{=} 0 
   \label{Appeq: FSC_RSGA_8}
\end{equation}
Combining Eqs.~\eqref{Appeq: FSC_RSGA_7} and \eqref{Appeq: FSC_RSGA_8}:
\begin{equation}
    [[H_{0},J^{+}],J^{+}]{=} 0 ,
\end{equation}
verifying condition (iii).
These properties ensure that the tower of states
\begin{equation}
    |\Omega_1^{n}\rangle{=}(J^+)^n|\Omega_1\rangle,
    \label{eq: RSGA tower}
\end{equation}
are all exact eigenstates of $H_{0}$, with equidistant energies $2nh{+}E_1$, while remaining entirely within the zero-energy manifold of $H_{XY}$.

\section{Additional details for Sec.~\ref{sec: exact eigenstates from commutant algebra formalism} }
\label{Appsec: Commutant}
\subsection{Annihilation of $|r,n\rangle$ by cluster operators $\mathcal{H}_{j}^{(r)}$}
\label{Appsec: annihilation_by_cluster}
In this Appendix, we analytically show that for a fixed $r$ which is a non-trivial divisor of the system size $2L$ (i.e. $2{\leq}r{\leq}L$)
\begin{equation}
\mathcal{H}_{j}^{(r)}|r,n\rangle{=}0~~{\forall}j{=}1,{\cdots}r,~~\forall n{=}0,{\cdots},2L{-}2
\end{equation}
For convenience, we recall the cluster operators
\begin{equation}
    \mathcal{H}_{j}^{(r)}{=}\sum_{k{=}0}^{2L/r{-}1}h_{j{+}kr,j{+}kr{+}1},
    \label{Appeq: new_generators_of_local_algebra}
\end{equation} 
 and for brevity, rewrite the state $|r,n\rangle$ as
\begin{equation}
 \begin{split}     &|r,n\rangle{=}\sum_{i=1}^{2L}|\psi\rangle_{i,i+r},~~~{\rm where}\\
     &|\psi\rangle_{i,i+r}{=}(-1)^{i(n{+}1)}|00\rangle_{i,i{+}r}|\mathbb{B}_{n}\rangle_{i{+}1,\cdots,i{+}r{-}1,i{+}r{+}1,\cdots,i{-}1},
 \end{split}   
    \label{Appeq: newstates_interms_of_old_r_n}
\end{equation}
Using that the bimagnon tower satisfies $h_{j,j+1}|\mathbb{B}_n\rangle{=}0$~\cite{schecter2019weak}, the only non-trivial action of an exchange term $h_{j,j+1}$ on $|\psi\rangle_{i,i+r}$ occurs when the bond touches one of the two sites carrying $|0\rangle$, i.e. when $j{\in}\{i{-}1, i, i{+}r{-}1, i{+}r\}$ (indices defined with mod $2L)$.
Therefore, for a fixed $j$, each term $h_{j{+}kr,j{+}kr{+}1}$ in $\mathcal{H}_{j}^{(r)}$ only acts on the four of the $\{|\psi\rangle_{j,j+r}\}$ whose $|0,0\rangle$ pair sits adjacent to that bond.
Thus,
\begin{equation}
    \begin{split}
&\mathcal{H}_{j}^{(r)}|r,n\rangle{=}\sum_{k{=}0}^{2L/r{-}1}h_{j{+}kr,j{+}kr{+}1}\sum_i |\psi\rangle_{i,i+r}\\   &{=}\sum_{k{=}0}^{2L/r{-}1}h_{j{+}kr,j{+}kr{+}1}\Big(|\psi\rangle_{j+kr+1,j+(k+1)r+1)}+\\
&|\psi\rangle_{j+kr,j+(k+1)r}+|\psi\rangle_{j+(k-1)r+1,j+kr+1)}\\
&+|\psi\rangle_{j+(k{-}1)r,j+kr}\Big).
    \end{split}
\end{equation}
Relabeling $k$ by $k{+}1$ in the third and fourth sum (using PBC), one finds
\begin{equation}
    \begin{split}
\mathcal{H}_{j}^{(r)}|r,n\rangle&{=}\sum_{k{=}0}^{2L/r{-}1}\Big(h_{j{+}kr,j{+}kr{+}1}|\psi\rangle_{j+kr+1,j+(k+1)r+1)}\\
&{+}h_{j{+}kr,j{+}kr{+}1}|\psi\rangle_{j+kr,j+(k+1)r}
\\&+h_{j{+}(k{+}1)r,j{+}(k{+}1)r{+}1}|\psi\rangle_{j+kr+1,j+(k{+}1)r+1)}\\
&+h_{j{+}(k{+}1)r,j{+}(k{+}1)r{+}1}|\psi\rangle_{j+kr,j+(k{+}1)r}\Big)
    \end{split}
    \label{Appeq: C5}
\end{equation}
Now we use two elementary identities (as we will show in App.~\ref{Appsec: proof_imp_lemma} shortly below)
\begin{eqnarray}
    h_{i,i+1}|\psi\rangle_{i,i+r}{=}{-} h_{i+r,i+r+1}|\psi\rangle_{i+1,i+r+1}
    \label{Appeq: imp_lemma1}
    \\
    h_{i,i+1}|\psi\rangle_{i+1,i+r+1}{=}{-} h_{i+r,i+r+1}|\psi\rangle_{i,i+r}
    \label{Appeq: imp_lemma2}
\end{eqnarray}
Using these identities for $i{=}j{+}kr$, we can see that the four terms in Eqs.~\eqref{Appeq: C5} cancel in pairs for each $k$, yielding
 \begin{equation}
     \mathcal{H}_{j}^{(r)}|r,n\rangle{=}0.
 \end{equation}
\subsubsection{Proofs of Eqs.~\eqref{Appeq: imp_lemma1} and \eqref{Appeq: imp_lemma2}}
\label{Appsec: proof_imp_lemma}
We first prove Eq.~\eqref{Appeq: imp_lemma1}. Recall that the bimagnon state $|\mathbb{B}_n\rangle$ [Eq.~\eqref{eq: type_1 scar tower}] is a superposition of all configurations of $n$ ``$|1\rangle$"s on a background of ``$|{-}1\rangle$”s. For convenience, we can write the background $|\mathbb{B}_n\rangle$ in $|\psi_{i,i{+}r}\rangle$ [Eq.~\ref{Appeq: newstates_interms_of_old_r_n}] as
\begin{equation}
\begin{split}
    &|\mathbb{B}_n\rangle_{i{+}1,\cdots,i{+}r{-}1,i{+}r{+}1,\cdots,i}\\
    &{=}|1\rangle_{i{+}1}|\mathbb{B}_{n-1}\rangle_{i{+}2,\cdots,i{+}r{-}1,i{+}r{+}1,\cdots,i}\\&{+}|{-}1\rangle_{i{+}1}|\mathbb{B}_n\rangle_{i{+}2,\cdots,i{+}r{-}1,i{+}r{+}1,\cdots,i}\\    &{=}\sum_{\alpha \in \{1,{-}1\}}|\alpha\rangle_{i{+}1}|\Lambda_{\alpha}\rangle_{i{+}2,\cdots,i{+}r{-}1,i{+}r{+}1,\cdots,i},
\end{split}
\end{equation}
where $|\Lambda_{1}\rangle{=}|\mathbb{B}_{n-1}\rangle$ ($|\Lambda_{{-}1}\rangle{=}|\mathbb{B}_{n}\rangle$) is a compact notation introduced purely to simplify expressions.
Within $|\psi\rangle_{i,i{+}r}$, the only trivial action of $h_{i,i{+}1}$ is on the two-site factor at $(i,i{+}1)$, namely
\begin{equation}
    h_{i,i{+}1}|0\rangle_{i}|\alpha\rangle_{i{+}1}{=}|\alpha\rangle_{i}|0\rangle_{i+1},~~~\alpha\in\{1,{-}1\}.
\end{equation}
Therefore,
\begin{equation}
    \begin{split}
        h_{i,i+1}|\psi\rangle_{i,i+r}&{=}(-1)^{i(n+1)}|0,0\rangle_{i+1, i{+}r}{\otimes}\\
        &\sum_{\alpha \in \{1,-1\}}|\Lambda_{\alpha}\rangle_{i{+}2,\cdots,i{+}r{-}1,i{+}r{+}1,\cdots,i{-}1}|\alpha\rangle_{i}.
    \end{split}
\end{equation}
We recognize the tail of the state as a one-site right translation of $|\mathbb{B}_{n}\rangle$:
\begin{equation}
    \begin{split}
        &\sum_{\alpha \in \{1,-1\}}\Big(|\Lambda_{\alpha}\rangle|\alpha\rangle\Big)_{i{+}2,\cdots,i{+}r{-}1,i{+}r{+}1,\cdots,i}\\&{=}T\sum_{\alpha \in \{1,-1\}}\Big(|\alpha\rangle|\Lambda_{\alpha}\rangle\Big)_{i{+}2,\cdots,i{+}r{-}1,i{+}r{+}1,\cdots,i}\\
        &{=}T|\mathbb{B}_{n}\rangle_{i{+}2,\cdots,i{+}r{-}1,i{+}r{+}1,\cdots,i},\\
    \end{split}
\end{equation}
where $T$ is the single-site translation operator. Using the known translation property $T|\mathbb{B}_n\rangle{=}(-1)^{n}|\mathbb{B}_n\rangle$, we obtain
\begin{equation}
    \begin{split}
        h_{i,i+1}|\psi\rangle_{i,i+r}&{=}\sum_{i}(-1)^{i(n+1)+n}|0,0\rangle_{i+1,i+r}{\otimes}\\
        &|\mathbb{B}_{n}\rangle_{i{+}2,\cdots,i{+}r{-}1,i{+}r{+}1,\cdots,i}
    \end{split}
    \label{Appeq: C14}
\end{equation}
Next consider the R.H.S of Eq.~\eqref{Appeq: imp_lemma1}. There the background $|\mathbb{B}_n\rangle$ in $|\psi\rangle_{i+1,i+r+1}$ can be written as 
\begin{equation}
    \begin{split}
        &|\mathbb{B}_n\rangle_{i{+}2,{\cdots},i+r,i+r+2,{\cdots},i}{=}\\
        &
        \sum_{k{=}0}^{n-1}|\mathbb{B}_k\rangle_{i{+}2,{\cdots},i+r-1}|1\rangle_{i+r}|\mathbb{B}_{n-k-1}\rangle_{i{+}r+2,{\cdots},i}\\&{+}\sum_{k{=}0}^{n}|\mathbb{B}_k\rangle_{i{+}2,{\cdots},i+r-1}|{-}1\rangle_{i+r}|\mathbb{B}_{n-k}\rangle_{i{+}r+2,{\cdots},i}.
    \end{split}
    \label{Appeq: C15}
\end{equation}
Now within $|\psi\rangle_{i+1,i+r+1}$, the non-trivial action of $h_{i+r,i+r+1}$ is 
\begin{equation}    h_{i+r,i+r+1}|\alpha\rangle_{i{+}r}|0\rangle_{i+r+1}{=}|0\rangle_{i{+}r}|\alpha\rangle_{i+r+1},~~~\alpha\in\{1,{-}1\}.
\end{equation}
Hence,
\begin{equation}
    \begin{split}
        &h_{i+r,i+r+1}|\psi\rangle_{i+1,i+r+1}{=}(-1)^{(i+1)(n{+}1)}|0,0\rangle_{i{+}1,i{+}r}{\otimes}\\&
        \Bigg(\sum_{k{=}0}^{n-1}|\mathbb{B}_k\rangle_{i{+}2,{\cdots},i+r-1}|1\rangle_{i+r+1}|\mathbb{B}_{n-k-1}\rangle_{i{+}r+2,{\cdots},i}\\&{+}\sum_{k{=}0}^{n}|\mathbb{B}_k\rangle_{i{+}2,{\cdots},i+r-1}|{-}1\rangle_{i+r+1}|\mathbb{B}_{n-k}\rangle_{i{+}r+2,{\cdots},i}\Bigg)
    \end{split}
    \label{Appeq: C17}
\end{equation}
Using Eqs.~\eqref{Appeq: C15} and \eqref{Appeq: C17}, one arrives at
\begin{equation}
    \begin{split}
        &h_{i+r,i+r+1}|\psi\rangle_{i+1,i+r+1}{=}\\&-(-1)^{i(n+1){+}n}|0,0\rangle_{i{+}1,i{+}r}
        |\mathbb{B}_{n}\rangle_{i{+}2,\cdots,i{+}r{-}1,i{+}r{+}1,\cdots,i}
    \end{split}
    \label{Appeq: C18}
\end{equation}
Comparing Eqs.~\eqref{Appeq: C14} and\eqref{Appeq: C18} proves Eq.~\eqref{Appeq: imp_lemma1}. Following a similar line of reasoning, one can establish Eq.~\eqref{Appeq: imp_lemma2}.

\subsection{Annihilation of $|r,n\rangle$ by $q_{i,i+1}$ for $r > 1$}
\label{Appsec: reverse engineer}
We show that for all $r{>}1$ the states $|r,n\rangle$ satisfy $q_{i,i{+}1}|r,n\rangle {=} 0~~\forall\, i,\;\forall\, n.$ For convenience restating the state
\begin{equation}
|r,n\rangle{=}\sum_{k=1}^{2L}({-}1)^{k(n+1)}
\,|00\rangle_{k,k+r}\,
|\mathbb{B}_{n}\rangle_{\rm all~other~sites.},
\end{equation}
Since every basis configuration appearing in $|\mathbb{B}_{n}\rangle$ contains only spins $\pm 1$, the only sites that may carry spin $0$ in $|r,n\rangle$ are the two sites $k$ and $k{+}r$. The local operator
\begin{equation}
q_{i,i+1}
=|0,0\rangle\langle 1,-1|
+|0,0\rangle\langle -1,1|
+\text{h.c.},
\end{equation}
acts non-trivially only on the three local configurations  
$|1,{-}1\rangle$, $|{-}1,1\rangle$, and $|0,0\rangle$ on the sites $(i,i+1)$.  
For all $r{>}1$ the configuration $|0,0\rangle_{i,i{+}1}$ never appears in $|r,n\rangle$, as the zero pair occupies sites separated by distance $r$.  
Hence only $|1,{-}1\rangle$ and $|{-}1,1\rangle$ need to be considered.
These two configurations arise only from those terms in the sum over $k$ for which the zero pair $(k,k{+}r)$ does not overlap with the sites $(i, i{+}1)$.
For such terms, the sites $(i,i{+}1)$ lies entirely inside $|\mathbb{B}_n\rangle$, so its local configuration depends only on $|\mathbb{B}_n\rangle$.
Because $|\mathbb{B}_n\rangle$ [Eq.~\eqref{eq: type_1 scar tower}] consists momentum $\pi$ bimagnons, for any configuration $|\gamma\rangle{=}|L\rangle|1,-1\rangle|R\rangle$ that appears in $|\mathbb{B}_n\rangle$, there is a corresponding configuration $|\gamma'\rangle{=}|L\rangle|{-}1,1\rangle|R\rangle$ that also appears with opposite sign (here $|L\rangle$ and $|R\rangle$ are some fixed configurations of $|1\rangle$ and $|{-}1\rangle$).
Since $q_{i,i+1}|1,{-}1\rangle{=}q_{i,i+1}|{-}1,1\rangle{=}|0,0\rangle$, $q_{i,i+1}$ annihilates the pair of configurations $|\gamma\rangle$ and $|\gamma'\rangle$ implying $q_{i,i+1}|\mathbb{B}_{n}\rangle{=}0$.

Since these are precisely the only configurations on which $q_{i,i+1}$ acts, it follows that
\begin{equation}
    q_{i,i+1}|r,n\rangle = 0
\qquad \forall\, i,\ \forall\, n,\ \forall\, r>1.
\end{equation}
Thus, every $|r,n\rangle$ with $r>1$ is a simultaneous zero-eigenvector of the strictly local operators $q_{i,i+1}$.

Following a similar line of reasoning one can show $q_{i,i+3}|r{=}1,n\rangle{=}0~~\forall i,~\forall n$.

\subsection{Additional details for volume entangled tower $|\mathbb{V}_n\rangle$}
\label{Sec: Additional details for volume entangled tower}
In this appendix, we provide supplementary derivations supporting the construction and properties of the volume-entangled tower introduced in the main text.

\subsubsection{Absence of a single raising operator for the volume-entangled tower}
\label{Appsec: volumelaw_example}
For convenience, we recall the definition of the states:
\begin{equation}
   |\mathbb{V}_{n}\rangle{=} \sum_{1{\leq}i_1{\leq}{\cdots}{\leq}i_{n}{\leq}L} ({-}1)^{\sum_{k{=}1}^{n} i_k} \prod_{k{=}1}^{n}S_{i_k}^+S_{i_k{+}L}^{+}|\Omega\rangle
   \label{Appeq: volume_scar_tower}
\end{equation}
It may be tempting to attempt to generate the volume-entangled tower using a single collective operator
\begin{equation}
    Q^\dagger = \sum_{i=1}^{L} (-1)^i\, S_i^{+} S_{i+L}^{+},
\end{equation}
and to define the tower states as
\begin{equation}
    |\mathbb{V}_n\rangle \propto (Q^\dagger)^{n} |\Omega\rangle.
\end{equation}
However, this construction does \textit{not} produce the equal-weight superposition that appears in Eq.~\eqref{eq: volume_scar_tower}.  
The reason is that powers of $Q^\dagger$ generate combinatorial multiplicities whenever indices overlap or repeat, resulting in non-uniform amplitudes across different configurations.

To illustrate this explicitly, consider the smallest nontrivial example with $2L = 4$.  In this case,
\begin{equation}
    Q^\dagger = S_1^{+} S_{3}^{+} - S_{2}^{+} S_{4}^{+},
\end{equation}
and one finds
\begin{equation}
\begin{split}
    (Q^\dagger)^2
    &= \bigl(S_1^{+} S_{3}^{+}\bigr)^2 
       - 2\, S_1^{+} S_{3}^{+} S_{2}^{+} S_{4}^{+}
       + \bigl(S_2^{+} S_{4}^{+}\bigr)^2 .
\end{split}
\label{eq:Q_sq_example}
\end{equation}
The three resulting configurations---two ``squared'' terms and one mixed term---appear with coefficients $1$, $-2$, and $1$, respectively.  
In contrast, the exact volume-entangled state \(|\mathbb{V}_2\rangle\) requires all three configurations to enter with the \textit{same} amplitude (up to overall signs), as enforced by the combinatorial definition in Eq.~\eqref{eq: volume_scar_tower}.

This discrepancy reflects a general feature: products of a single collective operator $Q^\dagger$ inevitably encode multiplicity factors arising from repeated index choices, producing a nonuniform distribution of amplitudes.  
Thus, the explicit combinatorial construction in Eq.~\eqref{eq: volume_scar_tower} is essential, as it guarantees uniform weights over all allowed configurations—something that cannot be achieved by taking powers of a single raising operator.

\subsubsection{Annihilation of $|\mathbb{V}_n\rangle$ by $\mathcal{H}_{j}^{(r)}$}
\label{Appsec: Annihilation of volume tower}
We now show that the state $|\mathbb{V}_n\rangle$ is annihilated by sum of antipodal exchange terms $\mathcal{H}_{j}^{(r)}{=}h_{j,j{+}1}{+}h_{j{+}L,j{+}L{+}1}$, i.e.,
\begin{equation}    \mathcal{H}_{j}^{(r)}|\mathbb{V}_n\rangle{=}0~~\forall j {\in}1,{\cdots}, L.
\label{Appeq: volume_annihilation}
\end{equation}
It is useful to recall that the coherent superposition of the states $|\mathbb{V}_n\rangle$ produces a highly entangled EAP state~\cite{Ivanov2025Volume, Chiba2024Exact, Mohapatra2025Exact, mestyan2025crosscapstatestunableentanglement} that factorizes into a product of antipodal dimers
\begin{equation}
\begin{split}   &\sum_{n}|\mathbb{V}_{n}\rangle{=}|\psi_{\rm vol}^{\rm init}\rangle{=}\prod_{\substack{i{=}1 \\ i{=}{\rm odd}}}^{L}|\phi^{+}\rangle_{i,i{+}L}|\phi^{-}\rangle_{i{+}1,i{+}L{+}1}
\end{split}
\label{Appeq: EAP_state}
\end{equation}
where
\begin{equation}
 \begin{split}    
|\phi^{\pm}\rangle_{j,k}{=}\left(|{-}1,{-}1\rangle{\pm}|0,0\rangle{+}|1,1\rangle \right)_{j,k}.
 \end{split} 
\end{equation}
\label{Appeq: EAP_state2}
Thus, projecting this state into a fixed total-magnetization sector yields precisely the states $|\mathbb{V}_n\rangle$. Since each exchange term $h_{i,i+1}$ preserves the total magnetization, it suffices to verify
\begin{equation}
    (h_{j,j{+}1}{+}h_{j{+}L,j{+}L{+}1})|\psi_{\rm vol}^{\rm init}\rangle{=}0.
\end{equation}
For the action of $h_{j,j{+}1}$, only the dimers touching the sites $j$ and $j{+}1$ in Eq.~\eqref{Appeq: EAP_state} are affected, so we isolate the local two-dimer state
\begin{equation}    |\Psi_j\rangle{=}|\phi^{+}\rangle_{j,j{+}L}|\phi^{-}\rangle_{j{+}1,j{+}L{+}1}.
\end{equation}
To evaluate the action of $h_{j,j{+}1}$, we now use the raising–lowering representation
\begin{equation}
    h_{j,j{+}1}{=}S_{i}^{+}S_{i+1}^{-}{+}S_{i}^{-}S_{i+1}^{-}
    \label{Appeq: ladder_rep}
\end{equation}
The ladder operators act on the dimer states as
\begin{equation}
    \begin{split}
        &S_j^+|\phi^{\pm}\rangle_{j,k}= (\,|0,-1\rangle \pm |1,0\rangle\,)_{j,k},\\
        &S_j^-|\phi^{\pm}\rangle_{j,k}= (\,{\pm}|{-}1,0\rangle + |0,1\rangle\,)_{j,k},\\
        &S_k^+|\phi^{\pm}\rangle_{j,k} = (\,|{-}1,0\rangle \pm |0,1\rangle\,)_{j,k},\\
        &S_k^-|\phi^{\pm}\rangle_{j,k} = (\,\pm|0,{-}1\rangle + |1,0\rangle\,)_{j,k},
    \end{split}
\end{equation}
Comparing these terms reveals the key identities
\begin{equation}
\begin{split}
    S_j^+|\phi^{\pm}\rangle_{j,k} = \pm S_k^-|\phi^{\pm}\rangle_{j,k}, \\
S_j^-|\phi^{\pm}\rangle_{j,k} = \pm S_k^+|\phi^{\pm}\rangle_{j,k}.
\end{split}
\label{Appeq: transfer}
\end{equation}
Using these relations, the first part of $h_{j,j+1}$ [Eq.~\eqref{Appeq: ladder_rep}] gives
\begin{align}
    S_j^+ S_{j+1}^- |\Psi_j\rangle
&{=} \big(S_j^+|\phi^{+}\rangle_{j,j+L}\big)
\big(S_{j+1}^-|\phi^{-}\rangle_{j+1,j+L+1}\big) \nonumber \\
&={-}S_{j+L}^- S_{j+L+1}^+ |\Psi_j\rangle,
\end{align}
and similarly,
\begin{equation}
S_j^- S_{j+1}^+ |\Psi_j\rangle
= -
S_{j+L}^+ S_{j+L+1}^- |\Psi_j\rangle.
\end{equation}
Combining these contributions 
%yields the key relation
%\begin{equation}
%h_{j,j+1}|\Psi_j\rangle
%=
%h_{j+L,j+L+1}|\Psi_j\rangle,
%\end{equation}
%which 
immediately implies
\begin{equation}
\big(h_{j,i+1} + h_{j+L,j+L+1}\big)|\Psi_j\rangle = 0.
\end{equation}
Because all other dimers in $|\psi_{\rm vol}^{\rm init}\rangle$ are unaffected by the local operator, the same cancellation holds for the full product state:
\begin{equation}
\big(h_{j,j+1}+h_{j+L,j+L+1}\big)|\psi_{\rm vol}^{\rm init}\rangle = 0,
\qquad \forall\, j.
\end{equation}
Since projection into a fixed magnetization sector commutes with all $h_{j,j+1}$, this property is inherited by all states in the tower, giving
\begin{equation}
\big(h_{j,j+1}+h_{j+L,j+L+1}\big)|\mathbb{V}_n\rangle = 0,
\qquad \forall\,n,\;\forall\,j.
\end{equation}

\subsubsection{Entanglement entropy of volume-entangled tower $|\mathbb{V}_{n}\rangle$}
\label{Appsec: EE_of_volume-entangled tower}
Before evaluating the entanglement entropy, it is helpful to rewrite the states $|\mathbb{V}_{n}\rangle$ in a form that makes their bipartite structure explicit. As discussed in the previous appendix section, the entire volume-entangled tower originates from an Entangled Antipodal-Pair (EAP) state $|\psi_{\rm vol}^{\rm init}\rangle$, which is a product of dimers connecting site $i$ with site $i{+}L$. In our earlier work~\cite{Mohapatra2025Exact}, we showed that for a real Hamiltonian with chiral symmetry $\hat{\mathcal{C}}$, this state can be written in the form
 \begin{equation}
|\psi_{\rm vol}^{\rm init}\rangle{=} \frac{1}{\sqrt{\mathcal{D}_{L}}}\sum_{|\vec{b'}\rangle{\in}\mathcal{D}_{L}}\hat{\mathcal{C}}|\vec{b'}\rangle_{1,{\cdots},L}{\otimes}|\vec{b'}\rangle_{L{+}1,{\cdots},2L},
\label{Appeq: generic_EAP}
\end{equation}
where $\mathcal{D}_{L}$ denote the full Hilbert space of $L$ site and $\vec{b'}$ runs over a basis of this space. The state $|\psi_{\rm vol}^{\rm init}\rangle$ is exactly of this form.

Since projecting this state into a fixed total-magnetization sector produces precisely $|\mathbb{V}_{n}\rangle$, it can be rewritten in the normalized form:

 \begin{equation}
|\mathbb{V}_{n}\rangle{=} \frac{1}{\sqrt{\mathcal{D}_{L}^{M_n/2}}}\sum_{|\vec{b}\rangle{\in}\mathcal{D}_{L}^{M_n/2}}\hat{\mathcal{C}}|\vec{b}\rangle_{1,{\cdots},L}{\otimes}|\vec{b}\rangle_{L{+}1,{\cdots},2L},
\label{Appeq: volume_scar_tower_rewrite}
\end{equation}
where $\vec{b}$ runs over all spin-$1$ basis configurations with magnetization $M_n/2$ and
$\mathcal{D}_{L}^{M_n/2}$ denote this Hilbert space dimension. Eq.~\eqref{Appeq: volume_scar_tower_rewrite} represents the Schmidt decomposition of $|\mathbb{V}_{n}\rangle$ for the standard bipartition $A_{s}{=}\{1,{\cdots},L\},~~B_{s}{=}\{L{+}1,{\cdots},2L\}$) with $\mathcal{D}_{L}^{M_n/2}$, with all non-zero Schmidt coefficients, $\lambda_{k}$, equal to $1/{\sqrt{\mathcal{D}_{L}^{M_n/2}}}$. Thus, the von Neumann EE  of $|\mathbb{V}_{n}\rangle$ for this bipartition is
\begin{equation}
    \mathcal{S}_{A_s}(n){=} \sum_{k}\lambda_{k}^2\ln\lambda_{k}^2{=}\ln (\mathcal{D}_{L}^{M_{n}/2}).
    \label{Appeq: EE_volume_tower_standardpartition}
\end{equation}

We now evaluate the $\mathcal{S}_{A_{s}}(n{=}L)$ for the mid-spectrum state $|\mathbb{V}_{n{=}L}\rangle$ (lying in the magnetization sector $M_{L}{=}0)$ in the large $L$ limit.
We recall that the trinomial coefficient $\mathcal{D}_{a}^{b}$ (as defined in Eq.~\eqref{eq: U(1) sector dimension}) represents the coefficient of $x^{b}$ in the expansion of $(1{+}x{+}x^{{-}1})^{a}$. For large values of $a$, following a similar derivation as done in App.~\ref{Appsec: counting zero energy modes}, we get the approximate Gaussian form of $\mathcal{D}_{a}^{b}$:
\begin{equation}
    \mathcal{D}_{a}^{b}{\simeq}\frac{3^{a}}{\sqrt{4\pi a/3}}\exp\left(-\frac{3b^2}{4a}\right).
    \label{Appeq: U(1) sector dimension asymptotic}
\end{equation}
Hence, in the asymptotic limit $\mathcal{D}_{L}^{0}$ takes the form:
\begin{equation}
    \mathcal{D}_{L}^{0}{=}\frac{3^{L{+}1/2}}{2\sqrt{\pi L}},
\end{equation}
and substituting this form in Eq.~\eqref{Appeq: EE_volume_tower_standardpartition}, we get 
\begin{equation}
    \mathcal{S}_{A_s}(L){\simeq}L\ln 3{-}\frac{1}{2} \ln \left(\frac{4\pi L}{3}\right).
\end{equation}
Now consider a fine-tuned bipartition $A_{f}{=}\{1, L{+}1, 2, L{+}2, {\cdots}, L/2, 3L/2\}$ and $B_{f}{=}\{L/2+1, L{+}L/2{+}1, {\cdots}, L, 2L\}$. To calculate the EE, we bipartition the state as:
\begin{align}
    &|\mathbb{V}_{n}\rangle{=} \frac{1}{\sqrt{\mathcal{D}_{L}^{M_n/2}}}\sum_{k{=}K_{\rm min}}^{K_{\rm max}} \notag\\
    &\Bigg[\left(\sum_{|\vec{b}\rangle{\in}\mathcal{D}_{L/2}^{k}}\hat{\mathcal{C}}|\vec{b}\rangle_{1,{\cdots},L/2}{\otimes}|\vec{b}\rangle_{L{+}1,{\cdots},3L/2}\right){\otimes} \notag \\    &\left(\sum_{|\vec{b'}\rangle{\in}\mathcal{D}_{L/2}^{M_{n}/2-k}}\hat{\mathcal{C}}|\vec{b'}\rangle_{L/2{+}1,{\cdots},L}{\otimes}|\vec{b'}\rangle_{3L/2{+}1,{\cdots},2L}\right)\Bigg].\label{Appeq: volume_scar_tower_rewrite2}
\end{align}
where $K_{\rm min}{=}{\rm max}({-}L/2,(M_{n}{-}L)/2)$, and $K_{\rm max}{=}{\rm min}(L/2,M_{n}{+}L/2)$.
For convenience, we write $\hat{\mathcal{C}}|\vec{b}\rangle_{1,{\cdots},L/2}{\otimes}|\vec{b}\rangle_{L{+}1,{\cdots},3L/2}{=}|\vec{b},\vec{b}\rangle_{A_f}$ and $\hat{\mathcal{C}}|\vec{b'}\rangle_{L/2{+}1,{\cdots},L}{\otimes}|\vec{b'}\rangle_{3L/2{+}1,{\cdots},2L}{=}|\vec{b'},\vec{b'}\rangle_{B_f}$. Normalizing and simplifying Eq.~\eqref{Appeq: volume_scar_tower_rewrite2} we deduce the Schmidt form of $|\mathbb{V}_{n}\rangle$ for the fine-tuned bipartition:
\begin{align}
    |\mathbb{V}_{n}\rangle{=}&\sum_{k{=}K_{\rm min}}^{K_{\rm max}}\Bigg[\frac{\sqrt{\mathcal{D}_{L/2}^{k}}\sqrt{\mathcal{D}_{L/2}^{M_n/2-k}}}{\sqrt{\mathcal{D}_{L}^{M_n/2}}}\\
    &\left(\sum_{|\vec{b}\rangle{\in}\mathcal{D}_{L/2}^{k}}\frac{1}{\sqrt{\mathcal{D}_{L/2}^{k}}}|\vec{b},\vec{b}\rangle_{A_f}\right){\otimes}\notag \\
    &\left(\frac{1}{\sqrt{\mathcal{D}_{L/2}^{M_n/2-k}}}\sum_{|\vec{b'}\rangle{\in}\mathcal{D}_{L/2}^{M_{n}/2-k}}|\vec{b'},\vec{b'}\rangle_{B_f}\right) \Bigg]
    \label{Appeq: Schmidt for volume tower athermal partition}
\end{align}
where the states in the parenthesis $(\cdots)$ are normalized.
Thus, the EE is given by
\begin{align}
\mathcal{S}_{A_{f}}(n)&{=}-\sum_{k{=}K_{\rm min}}^{K_{\rm max}}\lambda_{k}^{2} \ln \lambda_{k}^{2}\nonumber \\
    \lambda_{k}&{=}\frac{\sqrt{\mathcal{D}_{L/2}^{k}}\sqrt{\mathcal{D}_{L/2}^{M_n/2-k}}}{\sqrt{\mathcal{D}_{L}^{M_n/2}}}.
\label{Appeq: EE_volumetower_athermalpartition}
\end{align}
We now evaluate the $\mathcal{S}_{A_{f}}(L)$ for the mid-spectrum state $|\mathbb{V}_{L}\rangle$ (lying in the magnetization sector $M_{L}{=}0)$ in the large $L$ limit.
In this case, the Schmidt coefficients take the form
\begin{equation}
    \lambda_{k}^{2}{=}\frac{(\mathcal{D}_{L/2}^{k})^2}{\mathcal{D}_{L}^{0}}~~~{\rm as}~~~\mathcal{D}_{L/2}^{k}{=}\mathcal{D}_{L/2}^{-k}.
\end{equation}
From Eq.~\eqref{Appeq: U(1) sector dimension asymptotic}, substituting the asymptotic forms of the trinomial coefficients $\mathcal{D}_{a}^{b}$ we find
\begin{equation}
\lambda_{k}^{2}{\simeq}\sqrt{\frac{3}{\pi L}}\exp\left(-\frac{3k^2}{L}\right),
\label{Appeq: Schmidt asymptotic for volume tower athermal partition}
\end{equation}
which is a normalized Gaussian in $k$ with variance $\sigma_{k}^{2}{=}L/6$.
In the large $L$ limit, the discrete sum over $k$ can be replaced by a continuous integral.
As a consistency check, note that the above large $L$ analysis maintains the normalization $\sum_{k}\lambda_k^{2}{=}1$:
\begin{equation}
    \int_{-\infty}^{\infty}\lambda_{k}^{2}dk{=}\sqrt{\frac{3}{\pi L}}\int_{-\infty}^{\infty}\exp\left(-\frac{3k^2}{L}\right){=}\sqrt{\frac{3}{\pi L}}\sqrt{\frac{\pi L}{3}}{=}1.
\end{equation}
Now in Eq.~\eqref{Appeq: EE_volumetower_athermalpartition}, substituting the form of $\lambda_{k}$ from Eq.~\eqref{Appeq: Schmidt asymptotic for volume tower athermal partition} we obtain,
\begin{equation}
    \mathcal{S}_{A_{f}}(L){\simeq}-\int_{-\infty}^{\infty} \lambda_{k}^{2} \ln \lambda_{k}^{2}dk~{\simeq}~\frac{1}{2}\ln L{+}{\rm constant},
\end{equation}
where we keep only the $L$-dependent terms; the additive constants are absorbed into the generic ``constant” term. Thus, the entanglement entropy under the fine-tuned bipartition grows logarithmically with system size, with a predicted coefficient $1/2$. We verify this scaling numerically by computing the EE directly from the exact expression in Eq.~\eqref{eq: EE_volumetower_athermalpartition} and fitting the data to a logarithmic form. As shown in Fig.~\ref{fig: EE_scaling_volume_tower}(b) for data up to system size $2L{=}60$, the fitted coefficient $m{\approx}0.48$, in excellent agreement with the analytic coefficient $1/2$.
\subsection{Annihilation of $|\mathbb{M}_{m,m'}^{k}\rangle$ by $\mathcal{M}{j}^{k}$}
\label{Appsec: annihilation_mirror}
We now show that the mirror-dimer eigenstates
\begin{equation}
|\mathbb{M}_{m,m'}^k\rangle
= \left[
\prod_{j=1}^{L-2}
|\phi^{-}\rangle_{k-j,k+j}\,
|\phi^{+}\rangle_{k-j-1,k+j+1}
\right]
|m,m'\rangle_{k,k+L},
\end{equation}
are annihilated by the mirror-symmetric local generators
\begin{equation}
\mathcal{M}_{j}^{k}
= h_{k-j,k-j+1} + h_{k+j,k+j-1},
\qquad
j \in \{1,2,\cdots,L\}.
\end{equation}

It is convenient to distinguish two cases:\\
\textbf{(i) Bulk bonds: $2{\leq}j{\leq}L{-}2$:} For these values of $j$, the two exchange terms $(k{-}j,k{-}j{+}1)$ and $(k{+}j,k{+}j{-}1)$ act entirely within the dimer network and do not touch the central sites $k$ or $k{+}L$. The local piece of the wavefunction on the four sites involved can be written as 
\begin{gather} 
|\Psi_j^k\rangle=|\phi^{-}\rangle_{\ell,p}\,|\phi^{+}\rangle_{\ell',p'}, \\
(\ell,p)=(k-j,k+j),\; (\ell',p')=(k-j-1,k+j+1).\nonumber
\end{gather}
Using the raising-lowering representation of the spins and the relations in Eq.~\eqref{Appeq: transfer}, one finds
\begin{align}
S_{k-j}^+ S_{k-j+1}^- |\Psi_j^k\rangle
&= -\, S_{k+j}^- S_{k+j-1}^+ |\Psi_j^k\rangle, \\
S_{k-j}^- S_{k-j+1}^+ |\Psi_j^k\rangle
&= -\, S_{k+j}^+ S_{k+j-1}^- |\Psi_j^k\rangle.
\end{align}
Combining these contributions,
\begin{equation}
h_{k-j,k-j+1}|\Psi_j^k\rangle
= -\, h_{k+j,k+j-1}|\Psi_j^k\rangle,
\end{equation}
and therefore for $2 \le j \le L-2$
\begin{equation}
\mathcal{M}_{j}^{k}|\Psi_j^k\rangle
=
\bigl(h_{k-j,k-j+1}+h_{k+j,k+j-1}\bigr)|\Psi_j^k\rangle
= 0.
\end{equation}
Since all other dimers and the central spins are simply spectators, the same cancellation holds for the full state $|\mathbb{M}_{m, m'}^{k}\rangle$.\\
\textbf{(ii) Edge bonds $j{=}1$ and $j{=}L{-}1$:} For $j{=}1$, the operator
$\mathcal{M}_{1}^{k}{=}h_{k{-}1,k}{+}h_{k{+}1,k}$
acts on the state on sites $(k{-}1,k,k{+}1)$
\begin{equation}
|\Xi_1^k\rangle
=
|\phi^{-}\rangle_{k-1,k+1}\,
|m\rangle_k,
\end{equation}
with the state at all other sites factored out.
Using Eq.~\eqref{Appeq: transfer}, one can move the action of the spin raising and lowering operators $S_{k-1}^{\pm}$ and $S_{k+1}^{\mp}$ across the $|\phi^{-}\rangle_{k{-}1,k{+}1}$ dimer and show that the contributions from the two bonds cancel pairwise:
\begin{equation}
h_{k-1,k}|\Xi_1^k\rangle
= -\, h_{k+1,k}|\Xi_1^k\rangle.
\end{equation}
Hence
\begin{equation}
\mathcal{M}_{1}^{k}|\mathbb{M}_{m,m'}^{k}\rangle = 0\;\;\forall\;\;m,m'.
\end{equation}
An analogous argument applies to the bond pair at the other side of the ring, $j{=}L{-}1$, where
\begin{equation}
\mathcal{M}_{L-1}^{k}
= h_{k+L,k+L+1} + h_{k+L,k+L-1}
\end{equation}
acts on bonds touching sites $k{+}L{+}1$ and $k{+}L{-}1$ and the unconstrained spin at $k{+}L$ (note using PBC $k{+}L{\equiv}k{-}L$).
Using the same identities~\eqref{Appeq: transfer} one finds
\begin{equation}
\mathcal{M}_{L-1}^{k}|\mathbb{M}_{m,m'}^{k}\rangle = 0.
\end{equation}
Putting both cases together, we conclude that
\begin{equation}
\mathcal{M}_{j}^{k}\,|\mathbb{M}_{m,m'}^{k}\rangle = 0,
\qquad
\forall\, j \in \{1,\dots,L\},\;\forall\, m,m',
\end{equation}
i.e., each mirror-dimer state $|\mathbb{M}_{m,m'}^{k}\rangle$ is a common zero mode of all mirror generators $\{\mathcal{M}_{j}^{k}\}$.

\bibliography{references}

\end{document}